\documentclass[pdflatex,sn-mathphys-num]{sn-jnl}

\DeclareUnicodeCharacter{03B1}{\ensuremath{\alpha}} 
\usepackage[utf8]{inputenc}
\usepackage{graphicx}%
\usepackage{multirow}%
\usepackage{amsmath,amssymb,amsfonts}%
\usepackage{amsthm}%
\usepackage{mathrsfs}%
\usepackage[title]{appendix}%
\usepackage{xcolor}%
\usepackage{textcomp}%
\usepackage{manyfoot}%
\usepackage{booktabs}%
\usepackage{algorithm}%
\usepackage{algorithmicx}%
\usepackage{algpseudocode}%
\usepackage{listings}%
\usepackage[version=4]{mhchem}
\usepackage{chemist}
\usepackage{chemfig}
\usepackage{xcolor}
\definecolor{reviewerblue}{RGB}{0,70,140}
\definecolor{responsegreen}{RGB}{0,120,60}
\definecolor{changeorange}{RGB}{180,90,0}
 
\usepackage{amsmath}
\usepackage{amssymb}

\usepackage[utf8]{inputenc}
\usepackage{newunicodechar}
\newunicodechar{χ}{\ensuremath{\chi}}

\theoremstyle{thmstyleone}%
\newcommand{\cplx}[1]{\mathcal{#1}}
\theoremstyle{thmstyletwo}%

\theoremstyle{thmstylethree}%

\begin{document}

\title[Article Title]{\textbf{Revealing Light-Driven Dynamics at Nanostructured Solid–Liquid Interfaces with In-Situ SHG}}


\author{\fnm{Tarique} \sur{Anwar}}
\author{\fnm{Diana} \sur{Dall'Aglio}}

\author{\fnm{Milad} \sur{Sabzehparvar}}
\author{\fnm{Giulia} \sur{Tagliabue}}
\affil{\orgdiv{\text{Laboratory of Nanoscience for Energy Technologies (LNET), STI} }\newline  \orgname{\text{École Polytechnique Fédérale de Lausanne} }, \orgaddress{\postcode{1015}, \state{VD}, \country{Switzerland}}}


\abstract{Light and heat are key drivers of interfacial chemistry at solid–liquid boundaries, governing processes central to sustainable energy conversion, including photoelectrochemical and hydrovoltaic systems. However, non-invasive probing of light-induced interfacial dynamics remains challenging due to the inherently weak and spatially complex nature of interfacial optical signals. Here, we introduce a nanophotonic platform that amplifies second harmonic generation (SHG) from nanostructured solid–liquid interfaces by over two orders of magnitude, enabling real-time, all-optical access to interfacial processes. Crucially, we develop a rigorous overlap-integral formalism that provides a general and quantitative framework for describing SHG in nanostructured geometries. By explicitly accounting for spatially inhomogeneous electromagnetic fields, this approach connects the nonlinear response to geometry-dependent near-field distributions, and reveals new degrees of freedom—namely, independent control of attenuation and phase—that are inaccessible in planar systems. As a result, the relative contributions of surface and electric-field-induced responses can be deterministically controlled through nanophotonic design. Using in-situ SHG at silicon–oxide–electrolyte interfaces, we can resolve subtle spectral shifts ($\sim$1.3~nm) with electrolyte concentration, indicating coupling between electrical double layer (EDL) potential and semiconductor electronic polarizability. Furthermore, under controlled optical excitation, we observe reversible, intensity-dependent modulation of the interfacial susceptibility arising from two mechanisms: at low intensities, the SH signal decreases consistent with photocharging, whereas at higher intensities it increases due to photothermal effects. Together, these results establish nanophotonic-enhanced SHG as a quantitative and tunable probe of interfacial phenomena, providing a unified framework for linking optical response, electrostatics, and geometry. This work opens new avenues for actively controlling interfacial charge, potential, and chemical equilibria with light, with broad implications for energy conversion, catalysis, and nanophotonic device engineering.}

\keywords{Second Harmonic Generation, Nanostructured Silicon, Surface Charge, Photocharging, Photothermal}



\maketitle

\section{Introduction}\label{introduction_section}
In the face of an escalating energy crisis, the prospect of converting and storing energy from renewable natural resources, such as water and sunlight, without reliance on external mechanized inputs is highly appealing. Hydrovoltaic, photoelectrochemical systems, and light-enhanced blue energy technologies represent a key opportunity in energy conversion, offering a pathway to sustainable energy solutions. At the core of these systems lies the formation of a charged interface when liquid closely interacts with a solid surface, resulting in the development of an electrical double layer (EDL) \cite{li_electrokinetics_2004, memming_semiconductor_2015, understanding_EDL_2022}. This EDL consists of a Stern layer firmly anchored to the solid surface, accompanied by a diffusion layer enriched with counter-ions. These create substantial interfacial electrical fields that govern the selectivity of essential photochemical processes, such as CO$_2$ reduction \cite{noauthor_how_nodate}, surface charges in hydrovoltaic \cite{wang_hydrovoltaic_2022}, and blue energy devices \cite{graf_light-enhanced_2019} for enhancing ionic conductance, and the double-layer capacity of electrochemical supercapacitors \cite{burt_review_2014}. Under low electrolyte concentration at planar solid–liquid interfaces, the equilibrium (spatial) distribution of ions is well captured by classical continuum models, including Debye–Hückel and Gouy–Chapman–Stern theories  \cite{debye_zur_1923, stern_zur_1924, helmholtz_studien_1879}. However, for practical relevance, the deployment of hydrovoltaic and blue energy devices operating at high ionic concentrations is particularly advantageous given the vast availability of saline water. Additionally, most practical electrochemical reactions typically involve high concentrations and surface charges induced by external electrical potential. Furthermore, owing to advances in micro- and nanotechnology, all the aforementioned systems have demonstrated enhanced performance with nanostructured electrodes, and therefore, planar electrodes are seldom used. Importantly, these systems predominantly utilize metal and semiconducting materials; thus, interfacial electrical fields are highly dependent on the electronic properties of the solid electrode \cite{nagy_surface_1993-1, zwaschka_imaging_2020, zbazant_unified_2023}, and can be substantially modulated under light stimuli through the photocharging and thermal effects of light \cite{graf_light-enhanced_2019, li_kinetic_2021, anwar_enhancing_2026}, resulting in high tunability of charged species near the aqueous interface. Yet, a thorough understanding of these nanostructured interfaces and their light-driven dynamics is still lacking.

Notably, in these systems, the surface charge is not constant because interfacial chemical equilibrium between the liquid and solid phases regulates it through the dissociation of surface groups in the presence of local ionic species. The surface charges can be controlled either passively via changes to the bulk electrolyte environment (ionic strength and pH) \cite{kosmulski_ph-dependent_2009, behrens_charge_2001} or actively via heat and light stimuli \cite{azam_silica_2020, joutsuka_electrolyte_2018, noauthor_charge_2015, memming_semiconductor_2015}. Light, in particular, is typically used to power these devices sustainably, and it is expected to exert a multifaceted impact on interfacial chemical equilibria. In semiconductors, for instance, photon absorption generates electron–hole pairs, and subsequent charge trapping at the interface can significantly boost surface charge \cite{monch_semiconductor_2001}. Concurrently, light can induce local heating through photothermal effects \cite{cui_photothermal_2023}, influencing both the equilibrium distribution of interfacial charges and the kinetics of electrochemical reactions. Furthermore, any temperature increase can modify the optical properties of nanostructured electrodes via the thermo-optical effect, thereby altering the optical near field and absorption \cite{naef_light-driven_2023}. Thus, while being convoluted, light simultaneously perturbs the interfacial electrostatic landscape and the reaction kinetics, highlighting its central role in governing both electrokinetic and (photo)electrochemical processes. However, going beyond device-scale measurements, in-situ monitoring of light-driven dynamic changes in surface charge and potential, as well as local temperature and optical properties, remains challenging due to the need for highly sensitive probes that do not disrupt the local environment while providing accurate measurements of interfacial fields. As these interfaces are not perfectly sharp, techniques such as atomic force microscopy (AFM) and other tip-based methods often interfere with the surrounding area \cite{collins_probing_2014, kobayashi_nanoscale_2010}, producing a convolution of the tip and surface that perturbs the surrounding region and causes spatial averaging. Overall, achieving a microscopic understanding of light-driven charging and thermal effects at nanostructured interfaces is essential for the rational design of light-driven electrochemical systems, yet significant methodological advances remain needed \cite{cortes_experimental_2022}.

Second harmonic generation (SHG) \cite{noauthor_phase-referenced_nodate, corn_optical_SHG, yan_new_1998}, a nonlinear optical spectroscopy technique, provides a unique means to probe interfacial effects without the need for invasive probes. Due to the inherent symmetry-breaking selection rule, which for centrosymmetric materials occurs at the interface \cite{dadap_second-harmonic_1997, nagy_surface_1993, mizrahi_phenomenological_1988, guyot-sionnest_local_1987, falasconi_bulk_2001, lu_temperature-dependent_2008}, SHG has been repeatably established as an exceptional tool for examining surface charge, potential, and molecular organization across a diverse array of solid–liquid systems \cite{mifflin_second_2005, corn_optical_SHG, dreier_surface_2018, ma_new_2021}. SHG is governed by the effective nonlinear susceptibility, which encodes both the surface's intrinsic properties and its electrostatic environment. Many SHG studies aimed at probing the EDL have predominantly focused on insulating oxide--electrolyte interfaces, where the intrinsic surface contribution is relatively weak, and the response is often dominated by electric-field-induced third-order contributions \cite{corn_optical_SHG, kulyuk_second-harmonic_1991}. These contributions can arise from both the interfacial liquid and the solid, with recent studies indicating that the $\chi^{3}$ response of the solid may be significant \cite{noauthor_charge-induced_2021}. While SHG has also been widely applied to semiconductor- and metal--electrolyte interfaces, these systems generally exhibit more substantial intrinsic surface and field-induced responses \cite{zhao_contactless_2024} that are strongly coupled to the interfacial charge and potential, leading to qualitatively different SHG signals \cite{corn_optical_SHG, guyot-sionnest_electronic_1990}. Importantly, practical electrochemical and energy-conversion devices rely on micro- and nanostructured electrodes, where light–matter interactions can be engineered through nanophotonic design. Such nanostructuring enables strong control over light–matter interactions, resulting in spatially inhomogeneous fields and enhanced local optical responses at the interface. Consequently, the measured SHG signal arises from a complex interplay among intrinsic material properties, interfacial electrostatics, and geometry-dependent electromagnetic  near-fields, fundamentally differing from that at planar interfaces. Despite these opportunities, interpreting SHG responses from nanostructured interfaces remains challenging. In particular, the absence of a quantitative framework that directly links the measured SHG signal to the underlying microscopic distributions of fields, charges, and potentials limits the extraction of physically meaningful information. Consequently, while SHG provides a powerful route to probe interfacial electrostatics, its application to nanostructured interfaces demands both enhanced sensitivity and a rigorous theoretical description. Addressing these challenges is essential for leveraging SHG as a quantitative tool for in-situ and operando studies of interfacial processes in next-generation energy devices.

This work introduces a nanophotonic-enhanced, \textit{in situ} SHG platform that enables real-time probing and quantitative understanding of how light and heat independently and jointly modulate nanostructured solid–liquid interfaces. By employing a periodic array of silicon nanodisks immersed in an aqueous electrolyte, forming a semiconductor–oxide–liquid interface, we achieve more than a 200-fold enhancement in SH signal, uniquely enabling the detection of subtle changes in interfacial susceptibility despite its intrinsically small magnitude. Crucially, we develop a rigorous overlap-integral formalism that, for the first time, establishes a quantitative framework for describing the effective nonlinear susceptibility at nanostructured solid-liquid interfaces and for extracting interfacial information from SHG measurements. This framework reveals that, beyond signal enhancement, nanostructuring enables independent control of attenuation and phase, thereby providing deterministic tuning of the nonlinear interaction volume and phase interference between surface and EDL contributions. Experimentally, enhanced SHG enables us to resolve an SH spectral shift of approximately 1.3~nm with changes in electrolyte concentration, suggesting coupling between the EDL potential and the electric polarization of the solid. Furthermore, by introducing a controlled optical pump, we observe reversible, intensity-dependent modulation of the interfacial susceptibility $(\chi_{\text{eff}}^2)$ arising from two distinct mechanisms, which produce opposite sign in the effective susceptibility change. At lower intensities, the observed decrease in the SH signal (corresponding to a $\approx 5\%$ reduction in susceptibility) is consistent with photocharging effects. In contrast, at higher intensities, the SH signal increases (by $\approx 7\%$), consistent with photothermal contributions. Notably, the transition-threshold pump intensity can be tuned by nanodisk geometry, enabling the unambiguous separation of true photothermal effects from inevitable thermo-optical modulation in optically resonant systems. Beyond providing quantitative, time-resolved signatures of these processes, our approach constitutes a sensitive diagnostic tool that can be extended to monitor and manipulate light-driven electrochemical and catalytic phenomena at aqueous interfaces. Together, these capabilities enable active control of interfacial chemistry,  and electrostatics with light, opening new directions for photoelectrochemistry, hydrovoltaic, and nanophotonic device engineering.

\section{Results}
\subsection{Boosting SHG Sensitivity with Nanophotonically-informed Nanostructuring }\label{Nanostructuring_SHG_enhancement}

\begin{figure}
    \centering
    \includegraphics[width=0.9\linewidth]{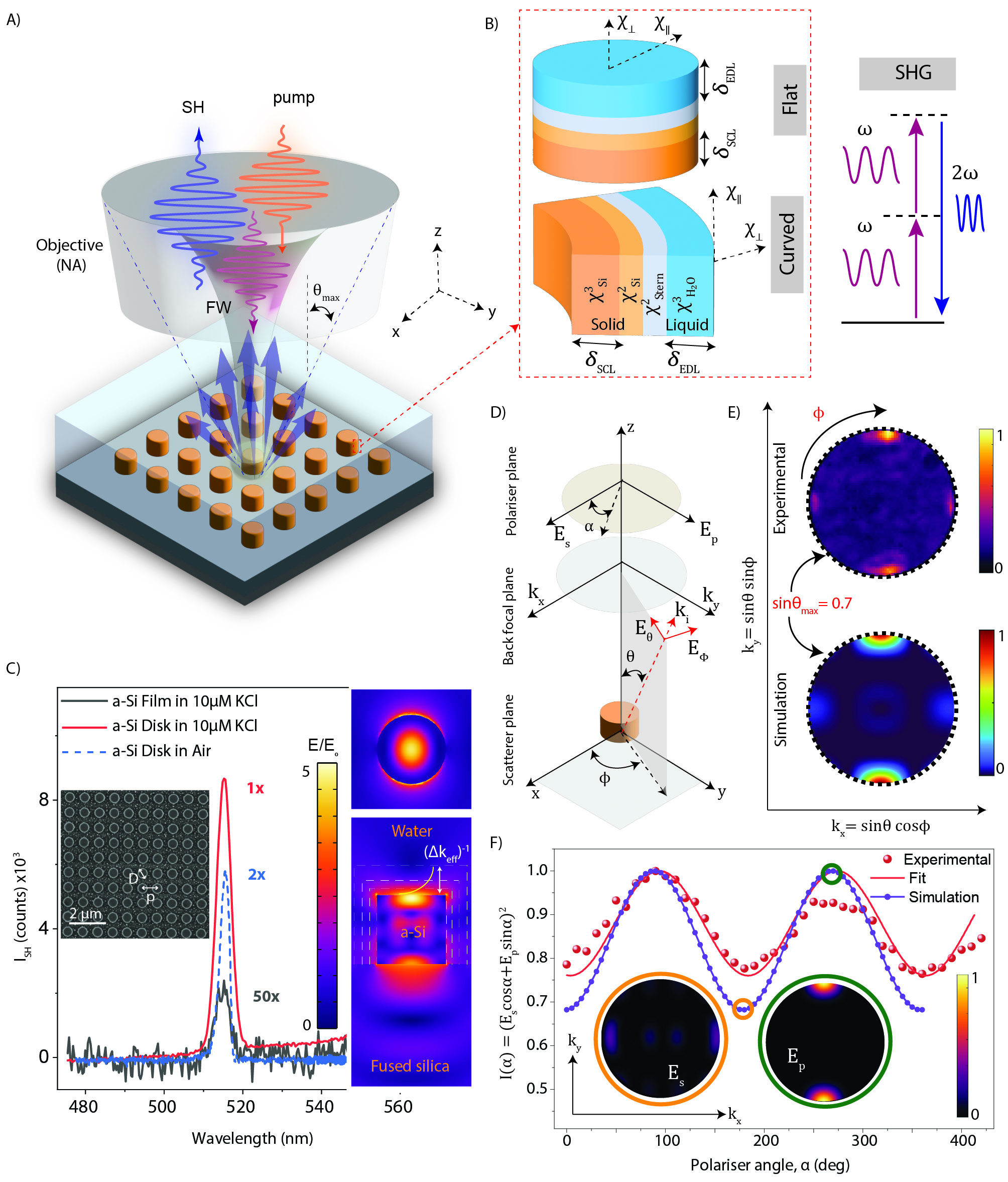}
 \caption{\textbf{Nanophotonic Enhancement and Control of Interfacial Second-Harmonic Generation at Solid–Liquid Interfaces.}
(A) Schematic of the SHG experiment showing illumination of the nanostructured sample and collection of the emitted SH signal through an objective with numerical aperture defining a maximum collection angle $\theta_{\text{max}}$. The periodic Si nanodisk array is immersed in the electrolyte and excited by a fundamental-wave (FW) beam and a pump beam. The FW probes interfacial nonlinear polarization via SHG, while the pump modulates interfacial charge density, potential, and local temperature.
(B) Schematic of the semiconductor--electrolyte interface, illustrating that the total nonlinear susceptibility arises from four distinct contributions: two intrinsic surface second-order responses (solid and liquid sides) and two field-induced third-order (EFISH) contributions associated with the space-charge layer (SCL) in silicon and the electric double layer (EDL) in the electrolyte. The relevant susceptibility components are expressed in terms of normal $(\perp)$ and tangential $(\parallel)$ tensor elements for both planar and curved nanodisk surfaces. The inset illustrates the SHG process driven by an incident FW field at frequency $\omega$.
(C) SH intensity from the nanodisk array (red) compared to a planar silicon film (black) under identical conditions, demonstrating an enhancement of $\sim 200\times$. The dashed blue curve shows the SH response of the nanodisk array in the absence of electrolyte, highlighting the critical role of the solid--liquid interface. \textit{Right inset:} simulated FW electric-field enhancement ($\sim 5\times$) in the nanodisk array, leading to a strong increase in SH intensity due to its quartic dependence ($I_{\mathrm{SH}} \propto |E(\omega)|^4$). The dashed contours denote surfaces at a distance $r_{\text{n}}$ normal to the nanodisk, and the white curve shows the near field decay characterized by an effective decay length $(\Delta k_{\mathrm{eff}})^{-1}$. \textit{Left inset:} scanning electron microscopy (SEM) image of the nanodisk array.
(D)Schematically showing the SH emission at an angle $\theta$ and $ \phi$ in a spherical coordinate. The nanodisk is located at the scatterer plane, and the electric components $E_{\theta}$ and $E_{\phi}$, and the corresponding propagation direction $k_{\text{i}}$. The back focal plane with the corresponding wave-vector in the x and y directions. The polarizer plane with the S and P components and the angle $\alpha$ at which the SH fields are detected. 
(E) The back focal plane image of the 510 nm disk (top), experimentally detected with an objective of NA=0.7, and (bottom) from SHG simulation, shows the angular emission of the SH fields. 
(F) SH intensity as a function of polarizer angle showing the mixing of two different emitted polarizations. Corresponding back focal plane images showing the 90-degree rotation of the angular emission pattern.}
    \label{fig:main_figure_1}
\end{figure}

Second harmonic (SH) generation entails a nonlinear optical process in which two photons of frequency $\omega$ interact with materials, leading to the generation of a single photon at the frequency of $2\omega$
(Fig. \ref{fig:main_figure_1} A). In our system (Fig.  \ref{fig:main_figure_2} A and SI \ref{experimental_setup}, Fig. \ref{fig:ch4_operando_SHG_setup}), optical excitation is driven by a fundamental-wave (FW) pulsed laser beam centered at 1030 nm (focused beam and linearly polarized), which leads to the coherent generation of a SH beam centered around 515 nm. An additional laser, centered at 633 nm (also focused and linearly polarized), is used for light-driven modulation. The SH intensity $I_{\text{SH}}(2\omega)$ is directly proportional to the square of the non-linear polarization  $P_{\text{nl}}(2 \omega )$. This, in turn, is equal to the product of the near field on the surface \(E(\omega)\) and the surface second-order effective susceptibility \(\chi^{2}_{\text{eff}}\)\cite{mizrahi_phenomenological_1988}.
\begin{equation}\label{eq:ch4_amplitude-SH-Second_order}
P_{\text{nl}} (2\omega) = \varepsilon_0 \chi^{2}_{\text{eff}}:E_{\text{}}(\omega)E_{\text{}}(\omega) 
\end{equation}
Unlike SHG in non-centrosymmetric materials—where the response is governed by a bulk susceptibility—in a centrosymmetric solid--liquid system, $\chi^{2}_{\text{eff}}$ is an interfacial property jointly defined by both phases and can vary when either is modified (Fig.~\ref{fig:main_figure_1} B). In aqueous systems, liquid contributions arise from both the alignment of water molecules and surface groups within the Stern layer ($\chi^{2}_{\text{Stern}}$) and the third-order susceptibility of water, $\chi^{3}_{\text{H$_2$O}}$, through the electric-field-induced second-harmonic (EFISH) mechanism, where the static EDL field enters the nonlinear polarization together with the optical field. For oxide surfaces, the intrinsic solid contribution is typically weak and relatively invariant. In contrast, at metal--electrolyte and semiconductor--electrolyte interfaces, the solid surface contribution ($\chi^{2}_{\text{solid}}(\omega)$, Eq.~\ref{eq:SH_intensity_overlap_Integrals_form}) can become significant, particularly under resonant or near-resonant excitation conditions, and may vary with the electrostatic environment.  This variation stems from the polarizability of free and bound electrons in metals \cite{mizrahi_phenomenological_1988}, and from the space charge layer (band-bending) in semiconductors \cite{noauthor_water_nodate}. Thus, the total nonlinear polarization arises from both intrinsic surface contributions and electric-field-induced third-order polarization, whose relative importance depends on the material system and excitation photon energies.
 \begin{equation}\label{eq:ch4_amplitude-SH_Second_and_third_Order}
P_{\text{nl}} (2\omega)=P^{\text{(2)}} (2\omega)+P^{\text{(3)}} (2\omega) = \varepsilon_0 \chi^{2}_{\text{s}}:E_{\text{}}(\omega)E_{\text{}}(\omega)+\varepsilon_0 \chi^{3}_{\text{}}:E_{\text{}}(\omega)E_{\text{}}(\omega)E^{\text{DC}}
\end{equation}

\subsection{Nonlinear polarization and overlap formalism in nanostructured interfaces}\label{section:Overlap_formalism}

The SH intensity is governed by the total nonlinear polarization and is thus expressed as:
\begin{equation}
I(2\omega) \propto \left| \mathbf{P}^{(2)}(2\omega) + \mathbf{P}^{(3)}(2\omega) \right|^2
\end{equation}
In solid–liquid interfaces, the third-order contribution originates from the DC electric field within the EDL, which decays exponentially along the surface normal. Consequently, the effective third-order polarization is obtained by integrating along the normal direction, $\mathbf{P}^{(3)}(2\omega) = \int_{0}^{\infty} \mathbf{P}^{(3)}(2\omega; z)\, dz$. For planar interfaces, the incident field is typically assumed to be spatially uniform in the lateral ($x$–$y$) plane, with variation only along the surface normal ($z$). The depth dependence is captured through a phase factor $e^{i\Delta k_z z}$, where $\Delta k_z$ accounts for wavevector mismatch between the fundamental and SH fields \cite{gonella_second_2016, guyot-sionnest_local_1987}. This framework leads to interference between contributions from different depths, as established in prior studies of planar SHG.

In contrast, nanostructured interfaces break translational symmetry and give rise to strongly inhomogeneous optical fields (Fig.~\ref{fig:main_figure_1}C and \ref{fig:ch4_SH_exp_simulation}D) \cite{celebrano_mode_2015, smirnova_multipolar_2018},
\begin{equation}\label{eq:Electric_field_vector_inhomogeneous}
\mathbf{E}(\mathbf{r}) = E_x(x,y,z)\,\hat{\mathbf{x}} + E_y(x,y,z)\,\hat{\mathbf{y}} + E_z(x,y,z)\,\hat{\mathbf{z}}
\end{equation}
where $E_x, E_y,E_z$ are complex valued. Importantly, even under normally incident, linearly polarized excitation, all vector components of the electric field can be present due to near-field confinement and geometry-induced mode mixing (Fig.~\ref{fig:main_figure_1}C, right inset). As a result, the induced nonlinear polarization is inherently spatially varying and vectorial, with contributions from multiple tensor elements of the nonlinear susceptibility.

These considerations lead to two key generalizations beyond the planar case. First, the total SH response must be obtained by integrating the nonlinear polarization over the relevant regions: a surface integral for $\mathbf{P}^{(2)}$ and a volume integral for $\mathbf{P}^{(3)}$. Second, all symmetry-allowed tensor components of the susceptibility must be included, as the local field distribution couples different polarization channels irrespective of the incident polarization.

To account for these effects, we adopt an overlap-integral formalism commonly used in nanophotonics \cite{fan_electric-field-induced_2025, celebrano_mode_2015}. Once the fundamental electric field spatial distribution is obtained—typically via numerical methods such as finite-element simulations—the nonlinear polarization $\mathbf{P}^{(2)}$ and $\mathbf{P}^{(3)}$ can be evaluated pointwise across the nonlinear surface (e.g., nanodisk interface) and nonlinear volumes (e.g., EDL and space-charge layer). The total SH emission is then determined by the coherent overlap between the nonlinear polarization and the radiated SH field,
\begin{equation}
I(2\omega) \propto 
\left| 
\int_{\text{surface}} \left(\mathbf{E}^{2\omega}\right)^{*} \cdot \mathbf{P}^{(2)}\, dS
+
\int_{\text{volume}} \left(\mathbf{E}^{2\omega}\right)^{*} \cdot \mathbf{P}^{(3)}\, dV
\right|^2,
\end{equation}
where $\left(\mathbf{E}^{2\omega}\right)^{*}$ is the complex conjugate of the  electric field at SH frequency. 

Within this framework, SHG can be viewed as a coherent three-wave mixing process in which the efficiency is governed not only by the magnitude of the nonlinear polarization, but also by its spatial overlap and mode matching with the emitted SH field \cite{celebrano_mode_2015}. This highlights a key distinction from planar systems: in nanostructured interfaces, the nonlinear response is controlled by the interplay between geometry-induced field inhomogeneity, tensorial nonlinearities, and spatial mode overlap.

\subsection{Evaluating nonlinear polarization from electric field distributions}\label{section:non-linear polarisation_E-field_distribution}
We now outline how the nonlinear polarization is constructed from the fundamental field distribution, with particular emphasis on the active susceptibility components and their role in nanostructured geometries. The second-order and the electric field induced third-order nonlinear polarizations are given by
\begin{equation}
P_i^{(2)} = \chi_{ijk}^{2} E_j E_k
\qquad
P_i^{(3)} = \chi_{ijkl}^{3} E_j E_k E_l^{\mathrm{DC}}
\end{equation}
where the DC field $E^{\mathrm{DC}}$ is oriented along the surface normal \cite{zhao_contactless_2024, corn_optical_SHG, he_measurement_2012, fan_electric-field-induced_2025}. Owing to in-plane isotropy and broken inversion symmetry along the surface normal, the interface is well described by $C_{\infty v}$ symmetry, appropriate for amorphous silicon–liquid water systems. Under this symmetry, both the intrinsic surface susceptibility $\boldsymbol{\chi}^{2}$ and the effective third-order susceptibility associated with electric-field-induced second-harmonic generation (EFISH), $\boldsymbol{\chi}^{3}$, reduce to a small set of independent tensor components involving normal ($\perp$) and tangential ($\parallel$) directions \cite{heinz_chapter_1991} (Fig. \ref{fig:main_figure_1}B). For the second-order response, there are four  non-vanishing components, where intrinsic permutation symmetry further reduces the number of independent terms to three \(\chi^{2}_{\perp\perp\perp}, \
\chi^{2}_{\perp\parallel\parallel},\
\chi^{2}_{\parallel\perp\parallel} = \chi^{2}_{\parallel\parallel\perp}\)\cite{smirnova_multipolar_2018}. For the third-order susceptibility in isotropic media (water and amorphous silicon), we have $\chi^{3}_{\perp\perp\perp\perp}
, \ \chi^{3}_{\perp\parallel\parallel\perp}
, \ \chi^{3}_{\parallel\parallel\perp\perp}
=  \chi^{3}_{\parallel\perp\parallel\perp}$\cite{kielich_optical_1969, he_measurement_2012, sipe_phenomenological_1987}. As a result, the EFISH response is governed by the same reduced set of tensor elements as the surface second-order response. This provides a unified symmetry-based framework in which both $\chi^{2}$ and $\chi^{3}$ contributions can be treated on equal footing.

As discussed above, in nanostructured systems,  the local electric field is fully vectorial and spatially varying (Eq.~\ref{eq:Electric_field_vector_inhomogeneous}). Consequently, all symmetry-allowed tensor components can be excited irrespective of the incident polarization. To account for this, it is convenient to express the fields in a local normal–tangential basis $(n, t_1, t_2) $ aligned with the interface. For nanodisk geometries, cylindrical coordinates provide a natural description. Specifically, for the flat interface $(n, t_1, t_2) \rightarrow (z, r, \phi)$, whereas for curved sidewalls $(n, t_1, t_2) \rightarrow (r, \phi, z)$. The procedure is therefore as follows. Starting from the simulated fundamental fields $(E_x^{\omega}, E_y^{\omega}, E_z^{\omega})$, we transform to cylindrical components $(E_r^{\omega}, E_{\phi}^{\omega}, E_z^{\omega})$ and evaluate the nonlinear polarization components $(P_r^{(2)}, P_{\phi}^{(2)}, P_z^{(2)})$ and $(P_r^{(3)}, P_{\phi}^{(3)}, P_z^{(3)})$ using the active tensor elements. These are subsequently transformed back into Cartesian components $(P_x, P_y, P_z)$ for use in the overlap-integral formalism. This approach yields the full spatial distribution of the nonlinear polarization across the interfacial surface and adjacent nonlinear volumes (EDL and the SCL). When combined with the field distribution at the SH frequency $(E_x^{2\omega}, E_y^{2\omega}, E_z^{2\omega})$, it enables a rigorous evaluation of the SH response through spatial overlap integrals \cite{fan_electric-field-induced_2025, celebrano_mode_2015}, as described in the previous section. A detailed mathematical formulation is provided in the Supplementary Information (SI \ref{non-linear-polarisation_Second_Harmonic_intensity}). In the section \ref{section:effective_susceptibility_overlap_fomralism_intefacial_potential}, these spatially resolved contributions are incorporated into an effective susceptibility framework that captures the interplay between interfacial and bulk-mediated nonlinear processes.

\subsection{Second harmonic emission from nanostructured interfaces}\label{section:Second_harmonic_emission_nanostructured_interface}
Beyond the inhomogeneous near-field distribution discussed above, nanostructured interfaces fundamentally alter the \textit{outcoupling} of SH radiation. In contrast to planar interfaces, where SH emission is confined to specific directions by boundary conditions, nanostructures act as localized nonlinear scatterers that radiate over a broad polar and azimuthal angles $(\theta,\phi)$ (Fig.~\ref{fig:main_figure_1}A,D), and the detected signal is determined by the angular collection of the optical system (Fig.~\ref{fig:main_figure_1}A,D). In our setup, the objective numerical aperture (NA = 0.7) limits the collection to $\theta_{\text{max}} = \sin^{-1}(0.7)=44.4 ^{\circ}$. While this approach provides the total scattered signal (integrated over all angles), it does not capture the angular distribution of the emitted radiation, which contains critical information about the underlying emission. To access this information, we experimentally employ back focal plane (BFP) imaging (Fourier imaging), which directly maps the angular distribution of the scattered light \cite{cueff_fourier_2024}. Experimentally, BFP imaging yields the angular intensity distribution of the SH signal across $(\theta,\phi)$ space (Fig.~\ref{fig:main_figure_1}E, top panel).  Experimental results and full SHG-BFP simulations show that, for the nanodisk geometry, SH emission is predominantly directed toward higher angles, implying that a substantial fraction of the signal lies outside the collection cone. This is confirmed by simulations at different NAs: increasing the NA from 0.7 to 0.9 enhances the collected intensity by $\sim2\times$, while reducing it to 0.5 decreases the signal by nearly an order of magnitude (Fig.~\ref{fig:BFP_image_Polarisation_experiment_Simulation}). Thus, the measured SH response is strongly influenced by angular emission in addition to the near-field distribution. 

The polarization of the emitted SH field provides a second key distinction from planar systems. Due to the inhomogeneous local fields, multiple polarization components may be present even under a single polarization state of the incident light. Consequently, the emitted SH polarization reflects both the nonlinear susceptibility tensor and the geometry-dependent field distribution. We probe this by measuring the SH intensity as a function of analyzer angle $\alpha$, yielding $I(\alpha) \propto \left(E_s \cos\alpha + E_p \sin\alpha \right)^2$, where $E_s$ and $E_p$ are the s- and p-polarized SH components. The observed polarization mixing agrees well with full-wave simulations (Fig.~\ref{fig:main_figure_1}F). Polarization-resolved BFP images further reveal that different polarization components are emitted into distinct angular regions, indicating a strong preferred angular–polarization dependence of SH emission. For the nanodisks studied here, the total s-polarized emission is $\sim0.7$ times the p-polarized emission.

These results demonstrate that SHG from nanostructured interfaces is fundamentally governed by the coupled interplay of nonlinear generation, polarization mixing, and angle-dependent emission, in contrast to the more constrained response of planar systems. Importantly, the strong agreement between experiment and simulations across (i) total SH intensity (Fig.~\ref{fig:main_figure_3}A), (ii) polarization-resolved measurements (Fig.~\ref{fig:main_figure_1}F), and (iii) angle-resolved emission patterns (Fig.~\ref{fig:main_figure_1}E) demonstrates that the model accurately captures the underlying spatial field distributions and emission characteristics. Because the same nonlinear polarization is consistently incorporated in both the overlap-integral formalism and full-wave SHG simulations, this agreement validates the proposed framework as a generalized description of SHG in nanostructured systems, inherently accounting for both amplitude and phase through complex-valued local fields and their spatial interference.

\subsection{From overlap integrals to effective susceptibility and interfacial potentials}\label{section:effective_susceptibility_overlap_fomralism_intefacial_potential}
We focus on a semiconductor–oxide–electrolyte interface, specifically a silicon-silicon oxide-water system, due to its relevance for both hydrovoltaic and photo-electrochemical devices \cite{gao_dynamic_2023, anwar_salinity-dependent_2024, anwar_unified_2025, qin_constant_2020, noauthor_water_nodate}. In this case, the susceptibility consists of 4 different contributions (Fig. \ref{fig:main_figure_1} B). The overlap-integral formalism provides a natural framework for separating the distinct physical contributions to the second-harmonic (SH) response while preserving the underlying symmetry constraints. Within this description, the total nonlinear polarization can be decomposed into four distinct contributions \cite{zhao_contactless_2024, monch_semiconductor_2001}\cite{ma_new_2021, gonella_second_2016, ohno_beyond_2019}:
\begin{enumerate}
    \item the intrinsic second-order response of the silicon interface,
    \item the intrinsic second-order response of the electrolyte (Stern layer),
    \item the third-order EFISH contribution from the space-charge layer (SCL) in silicon,
    \item the third-order EFISH contribution from the electrical double layer (EDL) in the electrolyte.
\end{enumerate}
\begin{equation}\label{eq:Overlap_integral_fomalism_E2w_P2w}
\begin{split}
I_{\mathrm{SH}} \propto \Bigg| & \int_{S} \big(\mathbf{E}^{2\omega}\big)^{*} \cdot \left (\mathbf{P}_{\text{Si}}^{(2)}+\mathbf{P}_{\text{Stern}}^{(2)}\right ) \, dS \\
& + \int_{\text{SCL}} \big(\mathbf{E}^{2\omega}\big)^{*} \cdot \mathbf{p}_{\text{Si}}^{(3)}E_{\text{SCL}} \, dV + \int_{\text{EDL}} \big(\mathbf{E}^{2\omega}\big)^{*} \cdot \mathbf{p}_{\text{H$_2$O}}^{(3)}E_{\text{EDL}} \, dV \Bigg|^{2}
\end{split}
\end{equation}

where, $\mathbf{P}_{\text{Si}}^{(2)}, \ \mathbf{P}_{\text{Stern}}^{(2)}$ are the second-order polarizations, and $\mathbf{p}_{\text{Si}}^{(3)}, \ \mathbf{p}_{\text{H$_2$O}}^{(2)}$ are the third-order polarizations per-unit of the DC field. To connect this microscopic description to experimentally accessible quantities, we express the DC electric fields in the EDL and SCL as exponentially decaying profiles along the surface normal $\hat{n}$,
\begin{equation}\label{eq:EDL_and_SCL_fieldExpression decay_length}
E_{\text{EDL}} = \frac{\Delta\Phi_{\text{EDL}}}{\delta_{\text{EDL}}}
\exp\!\left(-\frac{r_{\text{n}}}{\delta_{\text{EDL}}}\right),
\qquad
E_{\text{SCL}} = \frac{\Delta\Phi_{\text{SCL}}}{\delta_{\text{SCL}}}
\exp\!\left(-\frac{r_{\text{n}}}{\delta_{\text{SCL}}}\right),
\end{equation}
where $r_{\text{n}}=(\mathbf{r}\cdot \hat{n})$ is the normal distance from the nanostructures surface, $\Delta\Phi_{\text{EDL}}$ and $\Delta\Phi_{\text{SCL}}$ denote the potential drops, and $\delta_{\text{EDL}}$ and $\delta_{\text{SCL}}$ are the corresponding decay lengths, all of which varies with electrolyte concentration as shown in Fig.~\ref{fig:EDL and SCL E-field_potential_Debye_Length}.

Substituting these fields into the overlap integrals yields a compact expression for the SH intensity (see SI \ref{non-linear-polarisation_Second_Harmonic_intensity} for detailed derivation),
\begin{equation}\label{eq:SH_intensity_overlap_Integrals_form}
I_{\text{SH}} \propto 
\left| \mathcal{I}_{\text{S}} \right|^2
\left|
\chi^{2}_{\text{Si,eff}} + \chi^{2}_{\text{Stern}}
+ \Delta\Phi_{\text{SCL}} \, \chi^{3}_{\text{Si,eff}} 
\frac{\mathcal{I}_{\text{SCL}}}{\delta_{\text{SCL}} \mathcal{I}_{\text{S}}}
+ \Delta\Phi_{\text{EDL}} \, \chi^{3}_{\text{H$_2$O}}
\frac{\mathcal{I}_{\text{EDL}}}{\delta_{\text{EDL}} \mathcal{I}_{\text{S}}}
\right|^2,
\end{equation}
where $\mathcal{I}_{\text{S}}$, $\mathcal{I}_{\text{SCL}}$, and $\mathcal{I}_{\text{EDL}}$ are the overlap integrals over the surface and the respective nonlinear volumes. 

The overlap factors $\mathcal{I}_{\text{EDL}}$ and $\mathcal{I}_{\text{SCL}}$ encode the spatial coupling within the respective nonlinear volumes of thickness $\delta_{\text{EDL}}$ and $\delta_{\text{SCL}}$ around the nanostructure. To account for finite regions extending along the surface normal, we introduce a depth-resolved overlap $\mathcal{I}_{\text{S}}(r_{\text{n}})$ at a distance $r_{\text{n}}$ from the interface ($r_{\text{n}}=0$ is on the surface of the nanostructure, see inset of Fig.~\ref{fig:main_figure_3}B) as:
\begin{equation}\label{eq:Surface_overlap_integral_non-linear_polarisation}
\mathcal{I}_{\text{S}}(r_{\text{n}}) \equiv \int_{S} 
\big(\mathbf{E}^{2\omega}\big)^{*} \cdot \mathbf{p}(r_{\text{n}}) E^{\text{DC}}/\chi_{\mathrm{}}^3 \, dS,
\end{equation}
and express volume contributions as weighted integrals. For the EDL, we have,
\begin{equation}\label{eq:EDL_overlap_integral_form}
\mathcal{I}_{\text{EDL}} = \int_{0}^{\infty} 
\mathcal{I}_{\text{S}}(r_{\text{n}})\, e^{-r_{\text{n}}/\delta_{\text{EDL}}} \, dr_{\text{n}},
\end{equation}
and similarly for the SCL. For three representative E-field distributions (with disk diameters of 510 nm, 520 nm, and 530 nm), we calculated from the spatial variation of $\mathcal{I}_{\text{S}}(r_{\text{n}})$ is shown in Fig.~\ref{fig:main_figure_3} B.\\

It is important to note that, although Eq.~\ref{eq:SH_intensity_overlap_Integrals_form} explicitly highlights the dominant susceptibility components ($\chi^{2}_{\text{Si,eff}}$, $\chi^{2}_{\text{Stern}}$, $\chi^{3}_{\text{Si,eff}}$, and $\chi^{3}_{\text{H$_2$O}}$), the full tensorial nature of the nonlinear response is retained in the overlap-integral formulation. In practice, the dominant tensor component is factored out to define an effective susceptibility, while the non-dominant components can be incorporated through relative weighting factors. For example, if $\chi^{(2)}_{\text{Si},\perp\perp\perp}$ is taken as the dominant component, the remaining tensor elements are normalized as $\chi^{2}_{r} = \chi^{2}_{\perp\parallel\parallel}/\chi^{2}_{\perp\perp\perp}$, with $\chi^{2}_{r} \sim 0.2$--$0.3$ for amorphous silicon \cite{falasconi_bulk_2001}. These relative contributions may not be neglected; as they can be implicitly included in the spatial overlap integrals. A detailed description of this normalization and its implementation is provided in the Supporting Information (section~\ref{non-linear-polarisation_Second_Harmonic_intensity} and Table \ref{table:SI:overlap_integral_implementation}).

This expression generalizes the well-established planar SHG formalism \cite{gonella_second_2016}, where the third-order contributions acquire phase factors due to wavevector mismatch, leading to interference between different probing depths. In nanostructured systems, this role is played by the spatial overlap integrals, which encode both amplitude and phase variations of the optical fields. A particularly important limiting case arises when the electrostatic screening lengths are much smaller than the optical field variation normal to the surface. This is typical for metals or highly doped semiconductors at high ionic strength. In this regime, the optical fields can be considered approximately constant across the EDL and SCL, and the volume integrals reduce to effective surface contributions. The SH intensity then simplifies to:
\begin{equation}\label{eq:SH_intensity_surface_limit_No_F_G}
I_{\text{SH}} \propto 
\left| \mathcal{I}_{\text{S}} \right|^2
\left|
\chi^{2}_{\text{Si,eff}} + \chi^{2}_{\text{Stern}}
+ \Delta\Phi_{\text{SCL}} \, \chi^{3}_{\text{Si,eff}} 
+ \Delta\Phi_{\text{EDL}} \, \chi^{3}_{\text{H$_2$O}}
\right|^2,
\end{equation}
which recovers the familiar planar-interface form, with the distinction that fundamental field intensity is replaced by the overlap integral $|\mathcal{I}_{\text{S}}|$. More generally, deviations from this limit can be captured through geometry-dependent complex factors,
\begin{equation}\label{eq:SH_intensity_F_G}
I_{\text{SH}} \propto 
\left| \mathcal{I}_{\text{S}} \right|^2
\left|
\chi^{2}_{\text{Si,eff}} + \chi^{2}_{\text{Stern}}
+ \Delta\Phi_{\text{SCL}} \, \chi^{3}_{\text{Si,eff}} \, \mathcal{G}
+ \Delta\Phi_{\text{EDL}} \, \chi^{3}_{\text{H$_2$O}} \, \mathcal{F}
\right|^2,
\end{equation}
where $\mathcal{F}$ and $\mathcal{G}$ are complex-valued overlap factors that account for the spatial distribution and phase of the optical fields within the EDL and SCL (SI \ref{non-linear-polarisation_Second_Harmonic_intensity}).
For convenience, we define normalized overlap factors
\begin{equation}
\mathcal{F} = \frac{\mathcal{I}_{\text{EDL}}}{\delta_{\text{EDL}} \mathcal{I}_{\text{S}}}, 
\qquad
\mathcal{G} = \frac{\mathcal{I}_{\text{SCL}}}{\delta_{\text{SCL}} \mathcal{I}_{\text{S}}},
\end{equation}
which capture the deviation from the surface-localized limit. This naturally leads to the definition of an effective susceptibility,
\begin{equation}\label{eq:chi2_effective_F_G}
\chi_{\text{eff}}^2 =
\chi^{2}_{\text{Si,eff}} + \chi^{2}_{\text{Stern}}
+ \chi^{3}_{\text{Si,eff}} \, \Delta\Phi_{\text{SCL}} \, \mathcal{G}
+ \chi^{3}_{\text{H$_2$O}} \, \Delta\Phi_{\text{EDL}} \, \mathcal{F},
\end{equation}
such that the measured SH response $(I_{\text{SH}},  P_{\text{SH}}$ are the intensity and non-linear polarization respectively) can be written as
\begin{equation}\label{eq:non-linear_polarisation_effective_chi2_EW}
P_{\text{SH}} \propto \sqrt{I_{\text{SH}}}
\propto \left| \mathcal{I}_{\text{S}} \, \chi_{\text{eff}}^2 \right|\approx A_{\text{S}}E(\omega)^2\left| \  \chi_{\text{eff}}^2 \right|
\end{equation}
 where the near-field factor $E(\omega)$, is contained within  the surface overlap integral as, $|\mathcal{I}_{\text{S}}| \propto A_{\text{S}} \langle E(\omega;\mathbf{r})^2 \rangle_{\text{S}}=A_{\text{S}}E(\omega)^2$  integrated over the surface of the nanostructure  with area $A_{\text{S}}$. The final expression (Eq.~\ref{eq:non-linear_polarisation_effective_chi2_EW}) will be used to interpret the variation in SH intensity under optical excitation in the sections \ref{section:charge_SHG_change} and \ref{section:Heat_SHG_change}. Altogether, this formulation establishes a direct link between the measured SH intensity and the physically relevant interfacial parameters $\Delta\Phi_{\text{EDL}}$ and $\Delta\Phi_{\text{SCL}}$. Crucially, in nanostructured systems, the quantities $\mathcal{F}$ and $\mathcal{G}$ are not determined by simple phase-matching considerations but by geometry-dependent overlap factors, which must be accurately accounted for to extract quantitative interfacial information.
 
\section{Discussion}

\subsection{Nanophotonic engineering of interfacial SHG sensitivity}\label{section:discussion_nanophotonic_engineering}

When employing SHG as a probe of interfacial phenomena, the central observable is the effective nonlinear susceptibility, $\chi^{2}_{\text{eff}}$  \cite{fischer_sensitivity_1994, nagy_surface_1993-1, guyot-sionnest_electronic_1990}. However, its intrinsically small magnitude, typically $\mathcal{O}(10^{-21}\ \text{m}^2\text{V}^{-2})$, often limits sensitivity. From Eq.~\ref{eq:non-linear_polarisation_effective_chi2_EW}, it is evident that the fundamental field $E(\omega)$ acts as a multiplicative amplifier of the SH response, enabling detection of subtle interfacial changes in a contactless and non-invasive manner, without perturbing the interface.

In planar geometries, this amplification is conventionally achieved by increasing the incident laser intensity. In contrast, metasurfaces—engineered arrays of optical nanoresonators—provide a fundamentally different route by locally enhancing the electromagnetic near field \cite{vabishchevich_nonlinear_2023,makarov_efficient_2017, tonkaev_even-order_2024, choudhury_surface_2015}. Our electromagnetic simulations of hydrogenated amorphous silicon (a-Si:H) nanodisks (SiNDs) reveal strong geometry-dependent variations in local field enhancement (SI~\ref{numerical_simulation}), resulting in substantial modulation of the SH response (Fig.~\ref{fig:main_figure_3}A and \ref{fig:ch4_SH_exp_simulation}A, B). For optimized geometries (diameter 520~nm, periodicity 800~nm, thickness 440~nm), we predict up to a fivefold enhancement of $E(\omega)$ at the interface.

Experimentally, these structures were fabricated using e-beam lithography (SI~\ref{ch4_fabrication}), and their linear optical response agrees well with simulations (Fig.~\ref{fig:ch4_SH_exp_simulation}A, B). Most strikingly, the SH intensity from the optimized nanodisk array immersed in a $10~\mu\text{M}$ KCl solution is enhanced by more than $200\times$ compared to a planar film (Fig.~\ref{fig:main_figure_1}C). This enhancement cannot be explained by increased surface area alone ($\sim$2–3$\times$), but instead arises from local field concentration, consistent with the $E(\omega)^4$ scaling of SHG. The dependence of SH intensity on nanodisk diameter further confirms that the enhancement follows the predicted near-field distribution (Fig.~\ref{fig:ch4_SH_exp_simulation}C, D).

\begin{figure}
    \centering
    \includegraphics[width=1\linewidth]{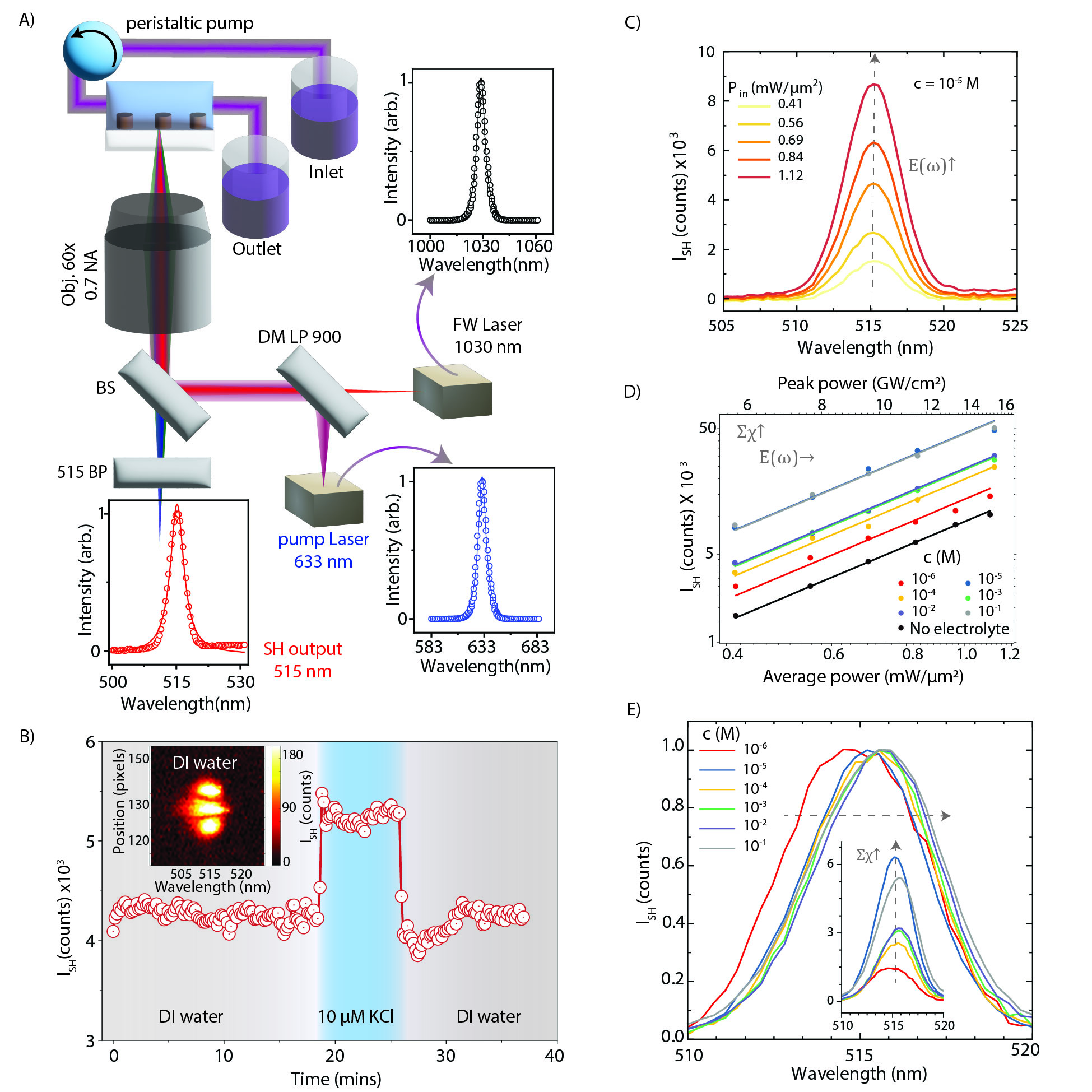}
 \caption{\textbf{In Situ Second-Harmonic Spectroscopy Reveals Electrolyte-Dependent Modulation of Interfacial Nonlinear Response .}
(A) Experimental configuration with the compression cell mounted on an inverted microscope. The excitation path includes an FW source (centered at 1030 nm, bandwidth 6 nm) and a pump laser (centered at 633 nm, bandwidth 10 nm). BS denotes a beam splitter, and BP denotes a band-pass filter. The lower panel shows a representative SH spectrum (centered at 515 nm, with a bandwidth of 4.3 nm) collected in the detection path.
(B) In situ SHG measurements (10 s acquisition time) showing reversible modulation of SH intensity during electrolyte exchange from deionized (DI) water (gray region) to 0.01 mM aqueous KCl (blue region) and back, demonstrating sensitivity to electrolyte concentration. The inset shows an SH emission image recorded in DI water, revealing localized enhancement of nonlinear polarization around the nanodisks.
(C) Second-harmonic (SH) spectra measured at 0.01 mM aqueous KCl concentration for different FW powers, showing constant spectral position with increasing peak intensity as the incident power increases.
(D) SH intensity as a function of FW intensity for various electrolyte concentrations, exhibiting a quadratic dependence (slope $\sim 2$), consistent with a second-order nonlinear process.
(E) Normalized SH spectra for different electrolyte concentrations, revealing a systematic red shift with increasing concentration. The inset shows the corresponding non-normalized spectra, highlighting simultaneous variations in both peak intensity and spectral position.}
    \label{fig:main_figure_2}
\end{figure}

\subsection{Concentration- and Nanostructure Geometry  Modulate Solid–Liquid Interfacial Susceptibility}\label{surface_potential_electrolyte_change}

Continuous SHG measurements were performed in a reflection configuration using a compression flow cell (Fig.~\ref{fig:main_figure_2}A and \ref{fig:ch4_operando_SHG_setup}D) that is connected to a peristaltic pump for continuous electrolyte circulation. As shown in Fig.~\ref{fig:main_figure_2}B, the spectroscopic signal is stable during electrolyte pumping at optimized flow rates, and the increased signal sensitivity (SI \ref{sensitivity_analysis}) enables continuous monitoring of SH intensity transitions as the bulk electrolyte environment is altered. In these measurements, no pump laser is used; only the FW beam is used for SHG. SH intensities were measured from SiND arrays with disk diameters ranging from 460 to 535~nm (pitch: 800~nm; height: 440~nm) fabricated on the same substrate. Measurements were first performed in air and subsequently repeated in electrolyte solutions with concentrations spanning six orders of magnitude (0.001 to 100 mM). Fig.~\ref{fig:main_figure_2}C shows the SH spectra for a 520~nm SiND array at 0.01~mM electrolyte concentration for varying FW intensities (0.41–1.12~mW$/\mu$m$^2$). As expected, the spectral position remains invariant with FW intensity, whereas the peak intensity increases in a quadratic fashion. In Fig.~\ref{fig:main_figure_2}D, SH intensity is plotted as a function of FW intensity for a wide range of electrolyte concentrations, as well as for the purely solid contribution (no electrolyte). The quadratic scaling (slope fixed at 2) confirms the second-order nonlinear origin of the signal (Eqs.  \ref{eq:ch4_amplitude-SH-Second_order},  \ref{eq:ch4_amplitude-SH_Second_and_third_Order}, and \ref{eq:non-linear_polarisation_effective_chi2_EW}) while systematic changes in the fitted intercept highlight electrolyte-dependent modifications of the effective susceptibility. Analytically, this can be easily appreciated because \(\log(I_{\text{SH}})\propto 2\log(\chi_{\text{eff}}^2)+2\log(I_{\text{FW}})\). Thus, when \(I_{\text{SH}}\)  is plotted as a function of \(I_{\text{FW}}\) in a log-log scale, it gives a slope of 2, while the intercept changes with electrolyte concentration, indicating that the effective susceptibility \(\chi_{\text{eff}}^2\) changes due to the surface contribution and the electric field induced bulk contributions (Eq.~\ref{eq:SH_intensity_F_G}) that is related to interfacial potentials, establishing the interfacial origin of the SH response in different electrolyte environments.

Amplifying the SH intensity enables the detection of very subtle changes in the measured SH spectra. Detailed spectral analysis reveals that the SH peak redshifts up to $\sim$1.3 nm with increasing electrolyte concentration (Fig.~\ref{fig:main_figure_2}E and Fig.~\ref{fig:SI_spectra_shift_electrolyte_concentration_diameter}). To the best of our knowledge, such shifts have not been reported previously, likely because most prior liquid--solid SHG studies focused on planar systems. These subtle spectral changes are detectable here due to the strong nanophotonic enhancement of the SHG signals. Notably, the observed spectral shift cannot be attributed to changes in the susceptibility magnitude of the liquid phase, which remains spectrally invariant under the non-resonant probing conditions (1030~nm excitation does not couple to water vibrational modes). Furthermore, the FW laser is non-invasive, as confirmed by the FW power-dependent measurements (fixed peak position for a given concentration, consistent quadratic power dependence, Fig.~\ref{fig:main_figure_2}C, D). Under the assumption that the overlap integral $(\mathcal{I}_{\text{S}})$ in Eq.~\ref{eq:SH_intensity_F_G} does not vary significantly with changes in the interfacial environment, the observed redshift is consistent with a concentration-dependent dispersion of the second- or third-order susceptibility of a-Si:H, $\chi^{2}_\text{Si,eff}(\omega)$ or $\chi^{3}_\text{Si,eff}(\omega)$, which is sensitive to interfacial surface charge density. Notably, first-principles DFT studies of metal--electrolyte junctions have reported analogous charge-induced spectral shifts in $\chi^{2}_{\text{solid}}(\omega)$ \cite{guyot-sionnest_electronic_1990, mizrahi_phenomenological_1988}. These observations therefore suggest that the effective silicon susceptibility may be modulated by electrolyte-induced changes in the interfacial potential within the framework of our model.

To extract the interfacial potential from the measured SH signal, an appropriate referencing scheme is required to isolate the intrinsic nonlinear response. In planar SHG, this is typically achieved using a nonlinear reference, such as $\alpha$-quartz, together with Fresnel factors \cite{dalstein_direct_2019, alghamdi_temperature_2025} that account for the wave propagation, boundary conditions, and refractive indices. In nanostructured systems, however, these effects are inherently embedded in the spatially resolved electromagnetic fields. Specifically, the local field obtained from full-wave simulations already captures all incoupling effects, including the dependence on refractive indices, illumination geometry (polarization, incidence angle). Therefore, in the nanostructured case, an analogous normalization must be performed, but with an important distinction: both incoupling and outcoupling factors are geometry-dependent. To this end, we normalize the nonlinear polarization of the SiND array in electrolyte to that in air, yielding $\eta= \sqrt{\frac{\eta_{\text{out, air}}^{2\omega}}{\eta_{\text{out}}^{2\omega}}}\left | \frac{\cplx{I}_{\text{S, air}}}{\cplx{I}_{\text{S}}} \right|$, which plays a role analogous to Fresnel corrections in planar systems. The resulting normalized SH response amplitude $(R_{\mathrm{SH}})$ and phase $\phi_{\text{SH}}$ can then be used to express the effective second-order susceptibility $\chi_s^2$ and the interfacial potential $\Delta \Phi_{\text{EDL}}$ (SI~\ref{surface_potential_derivation}). Combining Eqs.~\ref{eq:ch4_amplitude-SH-No-Electrolyte} and \ref{eq:ch4_amplitude-SH-Electrolyte}, we obtain an analytical expression for the SH amplitude ratio (Eq.~\ref{eq:ratio_SH_amplitude}), which can be solved explicitly to yield: 
\begin{equation}\label{eq:ch4_Phi0_expression_SH_intensity}
\Delta \Phi_{\text{EDL}}=
\frac{\chi_{\text{Si,0}}^2 \eta R_{\text{SH}}\sin{\phi_{\text{SH}}}}
{\chi^3_{\text{H$_2$O}} |\cplx{F}|\sin{(\arg(\mathcal{F}))}},
\end{equation}
\begin{equation}\label{eq:ch4_chi2_expression_SH_intensity}
\chi_s^2 =
\chi_{\text{Si,0}}^2\eta R_{\text{SH}}\cos{\phi_{\text{SH}}}
-\cot{(\arg(\mathcal{F}))}\,
\chi_{\text{Si,0}}^2\eta R_{\text{SH}}\sin{\phi_{\text{SH}}},
\end{equation}
where $R_{\text{SH}}$ and $\phi_{\text{SH}}$ are the amplitude and phase of the measured SH signals, $\chi_{\text{Si,0}}^2$ is the surface susceptibility of silicon without any electrolyte. Eqs.~\ref{eq:ch4_Phi0_expression_SH_intensity} and \ref{eq:ch4_chi2_expression_SH_intensity} are very similar to the ones reported in the literature for planar interfaces \cite{dalstein_direct_2019}, except the phase factor $\mathcal{F}$ is dependent on the nanostructure geometry (through near field distribution),  which will be discussed in more detail in the section~\ref{section:spatial_phase_engineering_nanostructuring}. In the above formulation, we have neglected the SH emission contribution from the SCL region. This approximation is valid under specific conditions, such as near flat-band (e.g., under external bias or sufficiently high illumination in the absence of bias), or when the spatial overlap of the optical fields within the SCL is significantly weaker than in the EDL. In such cases, the relative contribution from the SCL becomes negligible, satisfying 
\(\left|\frac{\chi^{3}_{\text{Si,eff}}\,\Delta \Phi_{\text{SCL}}\,\mathcal{G}}{\chi^{3}_{\text{H$_2$O}}\,\Delta \Phi_{\text{EDL}}\,\mathcal{F}}\right|^2 \ll 1\). As shown in Fig.~\ref{fig:ISH_EDL_SCL_comparsion_Ratio_FandG}, within the limit of thin oxide and absence of strong Fermi level pinning, we can self-consistently obtain the potential drops in the EDL and SCL. Under this conditions, the SCL contributions to the SH intensity is less than 10 percent for the three representative disk diameters. Overall, Eqs. \ref{eq:ch4_Phi0_expression_SH_intensity} and \ref{eq:ch4_chi2_expression_SH_intensity} highlight that both the amplitude and phase of the SH signal encode interfacial information, consistent with phase-resolved SHG approaches \cite{ohno_beyond_2019,dalstein_direct_2019}. While a complete separation of contributions from the Stern layer and intrinsic silicon response remains challenging, and may require additional spectroscopic dimensions such as wavelength dependence, the above framework already enables quantitative extraction of $\Delta \Phi_{\text{EDL}}$ and $\chi_s^2$ under varying electrolyte conditions. The non-trivial task of extending phase-sensitive SHG techniques to nanostructured interfaces and integrating them with the present formalism will be an important direction for future work.

\vspace{0.5em}
\noindent
\textbf{Role of the complex phase factor $\mathcal{F}$}

A key outcome of this analysis is that the determination of both $\Delta \Phi_{\text{EDL}}$ and $\chi_s^2$ depends explicitly on the \emph{complex-valued phase factor} $\mathcal{F}$, through both its magnitude $|\mathcal{F}|$ and phase $\arg(\mathcal{F})$. This represents a fundamental departure from planar SHG formulations, where the phase factor is typically known a priori from bulk optical constants and can be treated as a fixed quantity \cite{gonella_second_2016, dalstein_direct_2019}.

In nanostructured interfaces, however, $\mathcal{F}$ is not a simple propagation phase. Instead, it emerges from the spatial overlap between the nonlinear polarization and the optical fields within the electric double layer. As such, $\mathcal{F}$ depends sensitively on both the nanostructures' geometry (through the near field distribution) and the electrolyte properties (through the Debye screening length). Consequently, it must be determined explicitly within the overlap-integral formalism developed above.

To fully exploit the analytical expressions derived above, it is therefore essential to obtain an explicit form for $\mathcal{F}$ in terms of physically meaningful parameters. In the following section, we derive this complex phase factor by evaluating depth-resolved overlap integrals and show that it can be expressed in terms of the effective attenuation and phase constants of the nanostructured interface. This provides the missing link between the measurable SH amplitude and phase and the underlying interfacial potential and susceptibility.

\subsection{Beyond field enhancement: spatial and phase engineering of nonlinear interactions}\label{section:spatial_phase_engineering_nanostructuring}

\begin{figure}[h!]
    \centering
    \includegraphics[width=1\linewidth]{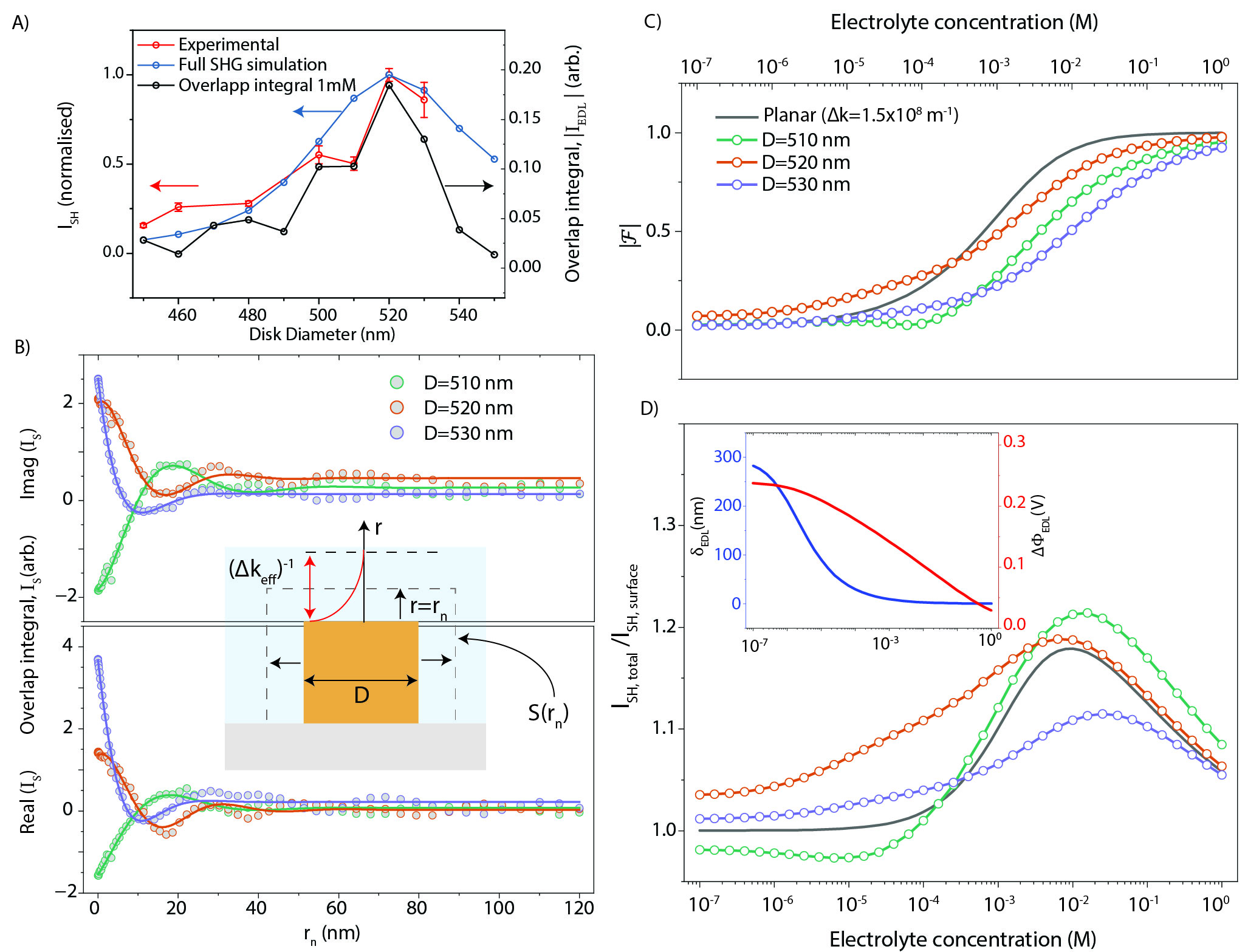}
   \caption{\textbf{Geometry-Dependent Overlap Integrals Govern Electrolyte-Tunable Second-Harmonic Generation .} 
(A) Left axis: experimentally measured and numerically simulated SH intensity as a function of disk diameter, normalized to the maximum value. Right axis: corresponding overlap integral evaluated within the EDL at 1 mM electrolyte concentration as a function of disk diameter. The error bars represent the standard deviation of $3\times5$ repeated measurements performed at 5 different FW laser powers, as shown in panel B.
(B) Real and imaginary parts of the depth-resolved surface overlap integrals as a function of the normal distance $r_{\text{n}}$ from the disk surface for three representative disk diameters. The points are obtained through numerical calculation of the overlap integral, while the solid lines are fits according to Eq.~\ref{eq:depth_resolved_surface_overlap_integrals_model}. The inset illustrates the cross-sectional geometry of the Si nanodisk, with water above and a fused-silica substrate below. The dashed gray lines denote surfaces $S(r_{\text{n}})$ at distance $r_{\text{n}}$ from the interface. The red curve in the inset shows the decay of the overlap integral, characterized by a decay length $\alpha=(\Delta k_{\mathrm{eff}})^{-1}$. 
(C) Magnitude of the complex factor $\cplx{F}$ as a function of electrolyte concentration for different disk diameters and a planar surface with an equivalent $\Delta k$ 
(D) Ratio of total SH intensity to surface SH intensity as a function of electrolyte concentration for the same disks as in (F). For calculation, we used $\chi_{\text{Si,eff}}^2=1,\ \chi_{\text{Stern}}^2=1,\ \chi_{\text{H$_2$O}}^2=1,\ \chi_r^2=\chi_r^3=0$ The inset shows the variation of the EDL thickness and surface potential with electrolyte concentration.}
    \label{fig:main_figure_3}
\end{figure}

While local field enhancement constitutes the first major advantage of nanostructuring, it is not the most fundamental one. A second, qualitatively distinct advantage emerges from the ability to \emph{engineer the spatial and phase distribution of the nonlinear polarization}. 

In planar interfaces, the nonlinear interaction is intrinsically constrained. The depth-dependent response (taking x-y as the planar surface, z is the distance normal to the surface) can be written as
\begin{equation}
\mathcal{I}_{\mathrm{S}}(z) = I_0 e^{i\Delta k\, z},
\end{equation}
where both the amplitude $I_0$ (set by the incident field) and the phase mismatch $\Delta k$ (determined by refractive indices and wavelength) are fixed by material properties and experimental configuration \cite{gonella_second_2016, dalstein_direct_2019}. As a result, the spatial profile of the nonlinear interaction is not tunable, and the probing depth is governed solely by intrinsic optical and electrostatic length scales.

In stark contrast, nanostructured interfaces break this constraint. The local optical field becomes strongly inhomogeneous, and the nonlinear response acquires a geometry-dependent spatial distribution. This can be captured through the depth-resolved overlap
\begin{equation}\label{eq:depth_resolved_surface_overlap_integrals_model}
\mathcal{I}_{\mathrm{S}}(r_{\text{n}})
=
\mathcal{I}_{\mathrm{S}}(0)\,e^{-\alpha r_{\text{n}}}e^{i\beta r_{\text{n}}},
\end{equation}
where $\alpha$ and $\beta$ represent the effective attenuation and phase accumulation constants, respectively. Crucially, all three quantities: $\mathcal{I}_{\mathrm{S}}(0)$, $\alpha$, and $\beta$—are no longer fixed, but can be tuned via nanophotonic design. As shown in Fig.~\ref{fig:main_figure_3}B, we obtain the real and imaginary part of the $\mathcal{I}_{\mathrm{S}}$  for three representative nanodisk diameters (510 nm, 520 nm, and 530 nm) through the overlap integral formalism presented in this work. To quantify the impact of nanostructure geometry on nonlinear interactions, we extract parameters from depth-dependent overlap integrals.  The fitted parameters are summarized below in Table \ref{tab:overlap_fit_parameters}:

\begin{table}[h]
\centering
\renewcommand{\arraystretch}{1.3}
\begin{tabular}{c c c c}
\hline
\textbf{Diameter} & $\mathcal{I}_{\mathrm{S}}(0)\;[\text{arb.}]$ & $\alpha\;[\text{m}^{-1}]$ & $\beta\;[\text{m}^{-1}]$ \\
\hline
510 nm & $-1.621 - 2.095i$ & $8.04\times10^{7}$ & $1.54\times10^{8}$ \\
520 nm & $1.324 + 1.554i$  & $8.44\times10^{7}$ & $2.02\times10^{8}$ \\
530 nm & $3.458 + 2.385i$  & $1.59\times10^{8}$ & $1.56\times10^{8}$ \\
\hline
\end{tabular}
\caption{Fitted parameters of the depth-resolved surface overlap integral for different nanodisk diameters. $\mathcal{I}_{\mathrm{S}}(0)$ denotes the complex overlap integral evaluated on the surface of the disk, while $\alpha$ and $\beta$ represent attenuation and phase constants.}
\label{tab:overlap_fit_parameters}
\end{table}

This tunability is directly demonstrated in our system: varying the nanodisk diameter modifies the near-field distribution, leading to distinct values of $\mathcal{I}_{\mathrm{S}}(0)$, $\alpha$, and $\beta$. Consequently, both the magnitude and phase of the nonlinear polarization can be engineered at the nanoscale.

The impact of this control becomes evident when evaluating the EDL contribution to the SH intensity through the overlap integral, as shown in Fig.~\ref{fig:main_figure_3}A, for 1 mM and Fig.~\ref{fig:Overlap_integral_different_concentrations} for 1-100 mM concentrations as a function of disk diameters, which is in quite good agreement with experimental and full SH simulation approaches. 
\begin{equation}\label{eq:EDl_overlap_integral_from_Surface_overlap_integras}
\mathcal{I}_{\mathrm{EDL}}
=
\frac{\mathcal{I}_{\mathrm{S}}(0)}
{\alpha+\delta_{\mathrm{EDL}}^{-1}-i\beta},
\end{equation}
which leads to the normalized factor
\begin{equation}\label{eq:complex_F_using_surface_overlap_integral_model}
\mathcal{F}
=
\frac{1}{1+\alpha\delta_{\mathrm{EDL}}-i\beta\delta_{\mathrm{EDL}}}.
\end{equation}
Unlike planar systems, where the phase factor depends solely on EDL thickness, $\mathcal{F}$ here depends sensitively on both nanostructures' geometry (through $\alpha$ and $\beta$) and electrolyte properties (through $\delta_{\mathrm{EDL}}$). This enables tunable interference between surface and EDL contributions.

\noindent
\textbf{Geometry-dependent tuning of $\mathcal{F}$ and SH intensity}

The influence of nanophotonic control on interfacial SHG is captured in Fig.~\ref{fig:main_figure_3}C, D, which show the variation of the overlap factor $|\mathcal{F}|$ and the normalized SH response ($I_{\text{SH, total}}$/$I_{\text{SH, surface}}$) $\propto \left|1+(\chi_{\text{H$_2$O}}^3/\chi_{\text{s}}^2)\Delta\Phi_{\mathrm{EDL}} \mathcal{F}\right|^2$ as a function of electrolyte concentration. In contrast to planar interfaces, nanostructured interfaces exhibit a qualitatively different behavior, arising from the independent control of the attenuation ($\alpha$) and phase accumulation ($\beta$).\\

A systematic analysis (SI \ref{section:_SI_phase_factor_formulation}, Figs.~\ref{fig:F_ISH_beta_fixed} and \ref{fig:F_ISH_alpha_fixed}) reveals distinct and complementary roles of these parameters. For a fixed $\alpha$, decreasing $\beta$ leads to (i) an increase in the peak SH intensity, (ii) a shift of the maximum toward lower concentrations (larger $\delta_{\mathrm{EDL}}$), and (iii) a broadening of the peak. Physically, a smaller $\beta$ reduces phase accumulation across the EDL, enabling more constructive interference over a larger spatial extent. This also results in an overall increase in $|\mathcal{F}|$, reflecting enhanced coupling to the EDL. Conversely, for a fixed $\beta$, decreasing $\alpha$ increases the peak SH intensity without significantly shifting its position. In this case, the reduced attenuation extends the effective interaction depth, thereby increasing the contribution from the EDL while preserving the phase-matching condition set by $\beta$. The peak correspondingly becomes narrower, and $|\mathcal{F}|$ increases due to the reduced suppression of deeper regions. These trends are directly reflected in Fig.~\ref{fig:main_figure_3}D. The 510 nm disk, characterized by relatively low $\alpha$ and moderate $\beta$, exhibits the highest and broadest peak, indicating optimal coupling to the EDL over an extended depth. The 520 nm disk, with larger $\beta$, shows a slightly reduced and shifted peak due to stronger phase accumulation. In contrast, the 530 nm disk, with significantly larger $\alpha$, exhibits a suppressed and broadened response, reflecting reduced interaction volume. Importantly, the position of the SH intensity maximum is primarily governed by $\beta$, following an approximate condition $\beta\,\delta_{\mathrm{EDL}}\sim 1$, while its magnitude is controlled by both $\alpha$ and $\beta$. This establishes a direct link between nanophotonic field engineering and electrolyte-dependent sensitivity.\\

Overall, this dual capability transforms SHG from a passive probe into a tunable spectroscopic tool, enabling selective amplification of specific interfacial contributions. As a result, we achieve more than two orders of magnitude improvement in sensitivity compared to planar films, \((\Delta \chi)_{\mathrm{SiND}} \approx 0.005\,(\Delta \chi)_{\mathrm{film}}\), allowing detection of minute variations in interfacial potential and charge density (SI~\ref{sensitivity_analysis}). Beyond signal enhancement, nanostructuring enables independent control of the attenuation ($\alpha$) and phase ($\beta$), providing deterministic tuning of the interaction volume and phase interference between surface and EDL contributions. Consequently, the position, magnitude, and width of the SH sensitivity maximum can be engineered for specific electrolyte regimes—an ability fundamentally inaccessible in planar geometries. Notably, the enhanced signal levels permit direct SH imaging with moderate laser powers (Fig.~\ref{fig:main_figure_2}B inset and Fig.~\ref{fig:SHG_image_SI}), opening new possibilities for operando and spatially resolved studies of interfacial processes. In the following, we leverage this platform to reveal light-driven modifications of interfacial chemical equilibria.
 
\subsection{Photocharging Shifts Solid–Liquid Equilibria }  \label{section:charge_SHG_change}

In semiconductor-based electrochemical systems, illumination induces photocharging at the solid–solid interface, thereby perturbing the charge distribution at the solid–liquid interface. A consistent description of this response therefore requires a unified treatment of the coupled electrostatic and chemical equilibria spanning the semiconductor, oxide, and electrolyte (SI~\ref{section:SI_EDL_SCL_field_capacitive _interface}).

Upon immersion, the native oxide layer on silicon undergoes surface dissociation (Fig.~\ref{fig:main_figure_4}A), generating a net negative surface charge density, $\sigma_{\text{ox-el}}$, primarily through deprotonation of silanol groups. This charge is compensated by counter-ions in the electrolyte, leading to the formation of an electrical double layer (EDL). The resulting electrostatic potential can, in general, distribute across the electrolytes EDL, oxide layer, and SCL in silicon, depending on the degree of electrostatic coupling across the oxide. In particular, for sufficiently thin oxides (less than 2-3 nm) and in the absence of strong Fermi level pinning, variations in interfacial charge can induce a coupled response in the semiconductor, leading to band bending and formation of an SCL. In this case, the corresponding electric fields and potential profiles are obtained self-consistently from electrostatic boundary conditions and are described in detail in SI~\ref{section:SI_EDL_SCL_field_capacitive _interface} and Fig.~\ref{fig:EDL and SCL E-field_potential_Debye_Length}.

The surface charge at the oxide–electrolyte interface is governed by interfacial chemical equilibrium. For SiO$_2$, this is described by the surface complexation reaction \(\ce{SiOH <=> SiO^- + H^+}\), leading to
\begin{equation}\label{eq:surface-charge-illumination}
    \sigma_{\text{ox-el}} = \frac{-e\Gamma}{1 + \frac{[\ce{H+}]_{\text{s}}}{K_{\text{a}}}}.
\end{equation}
Hydrogenation of silicon may influence interfacial charge, but that a detailed investigation of these effects is beyond the scope of the present work. The charge density thus depends on the surface groups density $(\Gamma)$, dissociation constant $(K_{\text{a}})$ local interfacial proton concentration $([\ce{H+}]_{\text{s}})$ which is directly coupled to the electrostatic potential. Increasing electrolyte concentration enhances screening (reducing $\delta_{\mathrm{EDL}}$), which facilitates deprotonation and increases the magnitude of $\sigma_{\text{ox-el}}$. This interplay between chemical equilibrium and electrostatics governs the profiles of the EDL potential and electric field.

Importantly, the electric fields in the EDL and SCL are not independent but are coupled through capacitive boundary conditions across the oxide layer (SI~\ref{section:SI_EDL_SCL_field_capacitive _interface}). Enforcing continuity of the normal component of the displacement field yields
\begin{equation}\label{eq:EDL_SCl_field_Surface_Charge_Si-ox_and_ox-el}
    \varepsilon_{\text{Si}}E_{\text{SCL}}-\varepsilon_{\text{el}}E_{\text{EDL}}=\sigma_{\text{Si-ox}}+\sigma_{\text{ox-el}},
\end{equation}
where $\varepsilon_{\text{Si}},\varepsilon_{\text{ox}}$, and $\varepsilon_{\text{el}}$ are the dielectric constants of silicon, oxide, and electrolyte, respectively, and $\sigma_{\text{Si-ox}}$ is the charge at the silicon–oxide interface. This relation highlights that variations in surface charge or EDL potential directly modify the field within the semiconductor.

Under low-irradiance conditions and in the absence of external heating, temperature variations can be neglected, and changes in equilibrium arise primarily from photocharging. In the dark, equilibration of the Fermi level establishes band bending at the silicon–oxide interface. Upon illumination, two processes occur simultaneously: (i) photogenerated carriers move towards the semiconductor surface, reducing the electrostatic potential within the SCL and entirely flattening the bands at higher intensities; and (ii) through capacitive coupling across the oxide, this electronic redistribution modifies the interfacial proton concentration, thereby altering $\sigma_{\text{ox-el}}$. In this way, light dynamically shifts the surface chemical equilibrium, generating a photovoltage that reflects a capacitive, rather than faradaic, response.

For p-type silicon, the initial downward band bending (Fig.~\ref{fig:main_figure_4}A, left panel) drives photogenerated electrons toward the semiconductor surface under illumination. In the limit of a thin oxide and absence of strong Fermi-level pinning, this carrier redistribution modifies the interfacial electrostatic boundary conditions and is accompanied by a corresponding rearrangement of ions in the electrolyte, with hydronium ions moving toward the solid--liquid interface. Together, these coupled responses lead to partial band flattening, a reduction in the electric field in the space-charge layer, and charge regulation in the EDL. Specifically, the increased proton concentration at the interface shifts the surface chemical equilibrium, thereby modifying the surface charge density in accordance with Eq.~\ref{eq:surface-charge-illumination},  and produces a negative photovoltage \cite{li_kinetic_2021, monch_semiconductor_2001}. Defining the open-circuit voltage (OCV) as the potential difference between bulk silicon and the electrolyte, the OCV decreases under illumination for p-type silicon \cite{anwar_enhancing_2026, li_kinetic_2021}.

To probe this effect experimentally, we monitored SH spectra in real time from two SiND $(\text{p-doped, }N_A =  10^{24}-10^{25}\ \mathrm{m^{-3}})$ in DI water (diameters: 490~nm and 510~nm), while periodically modulating a 633~nm pump laser (intensity: 7.5~$\mu$W/$\mu$m$^2$). SH spectra were acquired every 20~s, with alternating light on/off cycles of 100~s (Fig.~\ref{fig:Dynamic_SH_510_490nm_light_on_off}). After baseline normalization and analysis using Eq.~\ref{eq:non-linear_polarisation_effective_chi2_EW}, the response is expressed as the fractional change in susceptibility,
\begin{equation}
    \frac{\Delta \chi}{\chi_0}=\frac{(|\chi_{\text{eff}}^2|)^{\text{light}}-(|\chi_{\text{eff}}^2|)^{\text{dark}}}{(|\chi_{\text{eff}}^2|)^{\text{dark}}}.
\end{equation}
As shown in Fig.~\ref{fig:main_figure_4}B, both SiNDs exhibit a negative $\Delta \chi$, consistent with a reduction in surface potential and/or SCL contribution to the SH response.

Light-induced photocharging modifies the entire interfacial potential profile across the semiconductor–oxide–electrolyte system. While Eq.~\ref{eq:chi2_effective_F_G} provides a framework to relate $\Delta\chi$ to variations in EDL and SCL potentials, extracting absolute values requires detailed knowledge of the light-induced spectral dependence of $\chi_{\text{s}}^{2}$, which is not yet available. Nevertheless, our measurements provide direct, real-time tracking of changes in the total susceptibility. We observe a decrease in effective susceptibility of $\approx 5\%$ and $\approx 2.5\%$ for 490~nm and 510~nm SiNDs, respectively, under $7.5~\mu\text{W}/\mu\text{m}^2$ illumination. These results demonstrate that optical excitation provides a direct and reversible means to modulate interfacial charge and potential in SiND arrays, complementing conventional control via electrolyte composition or pH.

\subsection{Photothermal Effect Modulates Interfacial SHG  }\label{section:Heat_SHG_change}

\begin{figure}[h!]
    \centering
    \includegraphics[width=1\linewidth]{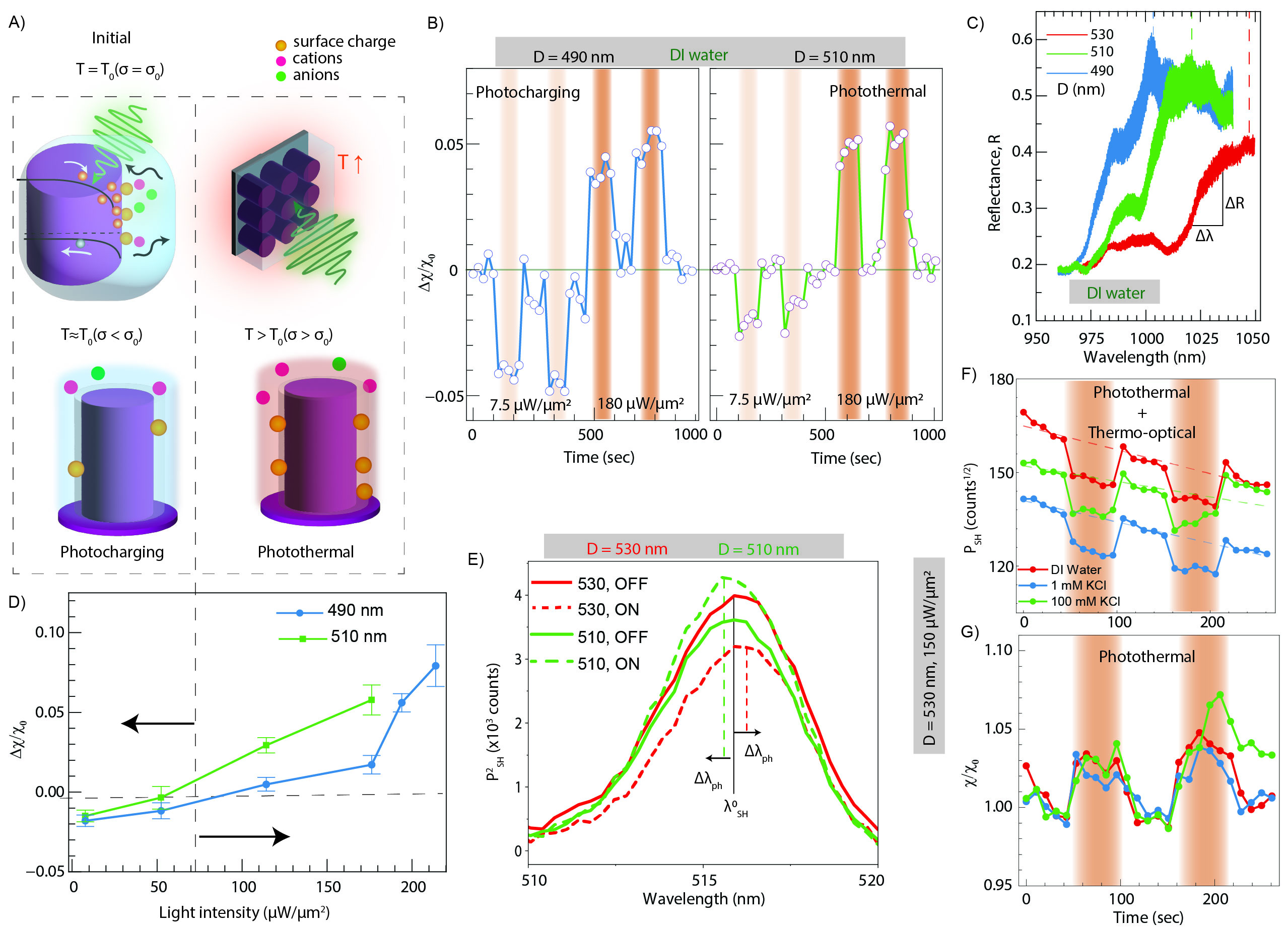}
    \caption{ \textbf{Photocharging and Photothermal Modulation of the Interfacial Susceptibility probed by Second-Harmonic Generation. }(A) Schematic illustration of the light-induced (right) heating and (left) changes in the capacitive interface, affecting surface charge and potential. The inset on the top highlights the initial surface charge conditions at ambient temperature $T_0$.  (B) Time traces of the light-induced fractional change in effective susceptibility (\(\Delta\chi\)), normalized to the initial value in the dark  (\(\chi_0\)), for SiNDs with D = 490 nm and D = 510 nm in DI water under 633 nm pump irradiation. Two distinct regimes, depending on the pump laser intensity, are observed. At low intensity, the photocharging effect is observed; at high power, the photothermal effect dominates due to light-induced heating of the structures. (C) The reflectance spectra for the SiND arrays in DI water with diameters of 490, 510, and 530 nm. (D). Fractional change is total susceptibility as a function of light intensity for the two SiNDs, showing photocharging and photothermal regimes, with transition points at distinct light intensities. (E)SH spectra recorded using deionized (DI) water for disk arrays with diameters D = 510 nm (red lines) and D = 530 nm (blue lines) under 633 nm pump laser irradiation (dashed lines) and subsequent pump turn-off (solid lines). (F) The time-trace of non-linear polarization \(P_{SH}\) during 633 nm excitation for a SiND with disk diameter D = 530 nm, showcasing variations under different electrolyte conditions, with dashed lines indicating the decay of the non-linear polarization baseline.  (G) Corresponding to the measurements in panels F,  the normalized susceptibility values obtained by decoupling  the TO effect induced change in $E(\omega)^2 \propto R$.}
    \label{fig:main_figure_4}
\end{figure}
At high irradiation intensities of the same 633 nm pump, silicon nanoresonators can heat up. This leads to localized temperature changes that influence chemical equilibrium and alter both the surface charge and interfacial chemistry (Fig. \ref{fig:main_figure_4}A, right panel). Notably, this heating can also affect the electrode's optical properties, as the refractive indices of silicon and water are strongly temperature-dependent. In Fig. \ref{fig:TO_coeffecients}, we show the ellipsometry data for the silicon's refractive indices across various wavelengths (from 280 to 2400 nm) at different temperatures (up to 200 $^\circ$C). These measurements allow us to calculate the thermo-optical (TO) coefficients ($dn/dT$ and $dk/dT$). At the fundamental wave excitation (1030 nm), the TO coefficient is positive, with a magnitude of 2x10$^{-4}~^\circ \text{C}^{-1}$. This coefficient is substantial enough to shift the resonance peak from 1017 nm at 25 $^\circ$C to 1022 nm at 75 $^\circ$C (Fig. \ref{fig:ref_spectra_temperature}). Moreover, as temperature increases, the refractive index of water decreases slightly, with a negative thermo-optic coefficient of around -10$^{-4}~^\circ \text{C}^{-1}$ in the 1030 nm spectral region. Overall, light-induced thermal excitation can lead to two main effects on the interfacial SH response of the solid-liquid system (Eq.  \ref{eq:ch4_amplitude-SH-Electrolyte}). First, it can change the effective susceptibility ($\chi_{\text{eff}}^2$), as the third-order susceptibility of water and surface charges (and $\Delta \Phi_{\text{EDL}}$) change with temperature (SI \ref{temperature_dependent_equilibrium_constant} and Eq. \ref{temp_equilibrium_constant}). Besides, the polarizability of silicon is related to the linear dielectric properties \cite{mendoza_exactly_1996} and is therefore temperature-dependent (Fig. \ref{fig:Second_order_polarizability_temperaure}). Importantly, at high irradiation, the band bending in the space-charge layer is reduced and approaches the flat-band condition ($\Delta \Phi_{\text{SCL}} \rightarrow 0$) \cite{monch_semiconductor_2001}. In this regime, the EFISH contribution from the SCL becomes negligible, and the effective susceptibility is governed by the surface and EDL contributions. This is consistent with the observed positive change in $\chi_{\text{eff}}^2$ at high irradiation, which cannot originate from SCL contributions, as band bending can be reduced to zero but does not invert in the absence of an external bias. Independently, the electric field factor \((E(\omega))\) in Eq.~\ref{eq:non-linear_polarisation_effective_chi2_EW} can also change due to TO effects, which induce shifts of the nanostructures' optical resonances. Since the SH response is constructed from the same local fields within the overlap-integral formalism, the impact of refractive-index changes on the nonlinear polarization can be linked to changes in the linear response. The field-factor contribution to the change in nonlinear polarization can be positive, negative, or nearly zero, depending on the initial reflectance peak's position, whether it blueshifts or redshifts, or if it appears broad and flat within the TO spectral transition, because reflectance is an experimental measure of near-field enhancements as discussed in section \ref{Nanostructuring_SHG_enhancement}. Thus, to explore the modulation of chemical equilibrium through changes in interfacial susceptibility induced by photothermal effects at the solid-liquid interface, it’s crucial to carefully separate the TO.

To understand the significant role of light-induced local heating on SH intensity, we performed a thorough SHG probing of three SiND arrays with diameters of 490, 510, and 530 nm. Their linear reflectance spectra around the FW, in deionized (DI) water, shown in Fig. \ref{fig:main_figure_4}C, reveal important differences. The  530 nm SiNDs display a pronounced reflectance peak at 1047 nm, to the right of the FW (red dashed line). In contrast, the 490 nm and 510 nm SiNDs exhibit almost flat spectra across the 1000$-$1050 nm range. As a result, we expect that the TO effect for the 490 nm and 510 nm SiNDs will be minimal within the temperature range of 25-100 $^\circ$C employed in our experiments, while the 530 nm SiND is expected to show a significant temperature-dependent change in  \(E(\omega)\). Quantitatively, the slopes of the reflectance curves around 1030 nm in DI water \((\Delta R/\Delta\lambda)\) are almost zero for 490 nm and 510 nm. Instead, a significant slope of 0.008 nm$^{-1}$ is obtained for the 530 nm SiNDs.  Since the temperature-induced variation in \(E(\omega)^2\) is approximately proportional to reflectance $(R)$, this analysis effectively separates the TO effects from the photothermal changes in susceptibility.

First, we analyze the SH response for the 490 nm and 510 nm SiNDs (Fig. \ref{fig:main_figure_4}D), as in these cases the modulation in interfacial SHG arises solely from the susceptibility term (negligible TO effects). As mentioned earlier, at low pump intensities, the surface potential tends to decrease, while at higher intensities, it increases. Using SHG as a probe, we can observe these effects via corresponding increases and decreases in interfacial susceptibility. Notably, by carefully tuning the geometry of the SiND, we can effectively control the threshold pump intensity required to switch between regimes dominated by photothermal or photocharging effects. The variation in interfacial susceptibility as a function of light intensity for the two SiNDs (Fig. \ref{fig:main_figure_4}D and \ref{fig:Dynamic_SH_510_490nm_light_on_off}) clearly shows that the threshold intensity is 100 $\mu \text{W}/\mu \text{m}^2$ for the 490 nm SiND, whereas it drops to just 65 $\mu \text{W}/\mu \text{m}^2$ for the 510 nm SiND. This difference stems from our ability to optimize light absorption through nanophotonic design, as supported by the reflectance spectra, which reveal an optical mode near the pump wavelength (Fig. \ref{fig:pump_wavelength_spectra_reflectance}). This indicates higher absorption for the 510 nm SiND, which correlates with the lower threshold pump intensity.

To explore the contribution of the TO effect in the interfacial SH response, we irradiated the SiNDs with the 633 nm pump, applying a high laser intensity of 150 $\mu \text{W}/\mu \text{m}^2$ for the 530 nm SiND and 180 $\mu \text{W}/\mu \text{m}^2$ for the 510 nm SiNDs. This power limitation was crucial in keeping local temperatures below 100 $^\circ$C, thereby avoiding any phase transition to a liquid-vapor state that could significantly affect SH intensity. The results, illustrated in Fig.~\ref{fig:main_figure_4}E, clearly show that, compared to the dark case (Pump OFF), under photoexcitation (pump ON), a change in SH intensity and a shift in the SH-peak position occur for both 510 nm and 530 nm SiNDs. Interestingly, the 510 nm SiND exhibits an increase in SH intensity, accompanied by a notable blueshift, which directly corresponds to the expected increase in surface potential (Fig.~\ref{fig:main_figure_2}E) resulting from temperature changes as discussed above. In contrast, the SH spectrum for the 530 nm SiNDs shows a decrease in peak amplitude and a redshift. When we examine the SH response under cyclic pump laser-on and off conditions, we consistently observe a decrease in non-linear polarization during irradiation (Fig. \ref{fig:main_figure_4}F). This trend holds across a range of electrolyte concentrations, underscoring that the behavior has a non-electrochemical origin. Therefore, as noted earlier, TO effects can be convoluted and alter the SH response in a non-intuitive manner.

To separate the thermo-optical effects, we estimate the change in  \(E(\omega)\) from reflectance spectra. We emphasize that the reflectance normalization is not intended to replace a rigorous Fresnel-type correction \cite{1984wi...book.....S, heinz_chapter_1991}. Rather, it serves as an experimental metric to monitor and partially compensate for temperature-induced changes in linear optical coupling. For instance, for the 530nm SiNDs, a spectral shift of approximately 4 nm and \((\Delta R/\Delta\lambda)=0.008 ~\text{nm}^{-1}\) was previously noted (Fig. \ref{fig:main_figure_4}C) and we can estimate \((\Delta R/R)\approx 0.09 \). As \(E(\omega)^2 \propto R\), we can normalize the SH signal with the reflectance, and extract the sole variation of \(\chi_{\text{eff}}^2\), which clearly shows the increase in \(\chi_{\text{eff}}^2\) with an increase in temperature (Fig. \ref{fig:main_figure_4}G), as expected based on the discussion of the temperature dependence of solid-liquid chemical equilibrium. 

Together, these results indicate that light can generate charge or heat, thereby influencing surface charges and the electrostatic landscape at the solid-liquid interface. When dealing with nanostructures exhibiting optical resonances, it is crucial to be aware of thermo-optical modulation, as it can produce complex effects on the SH signal. The insights from this work, based on nanostructuring and probing, enable us to effectively disentangle these effects by carefully controlling optical resonances at targeted wavelengths. Importantly, by fine-tuning the SiND geometry, we can achieve reversible switching between regimes in which the \emph{modulation} of the SH signal is governed by photocharging at low intensities and by photothermal effects at higher intensities, simply by adjusting the light intensity. This phenomenon, as evidenced by both intensity and spectral changes in the SH response, introduces a novel mechanism for all-optically driven interfacial control that goes beyond conventional methods of tuning via pH or electrolyte concentration. This dynamic response, which depends on geometry, opens up new possibilities for actively modulating electrochemical processes, enhancing nonlinear optical responses, and improving sensing capabilities at semiconductor-electrolyte interfaces.

\section{Conclusion}

Overall, this work establishes nanophotonic-driven electromagnetic design as a powerful approach for probing and controlling solid–liquid interfaces \textit{in situ}. By combining nanostructure-enabled near-field enhancement with SHG, we achieve more than a 200-fold increase in sensitivity, enabling detection of subtle variations in interfacial susceptibility and local electrostatic environment that are typically inaccessible in planar systems.

A central advance of this work is the development of a rigorous overlap-integral formalism that, for the first time, provides a general and quantitative framework for describing SHG at nanostructured solid-liquid interfaces. In contrast to conventional planar treatments, this approach explicitly accounts for spatially inhomogeneous and vectorial electromagnetic fields, and establishes a direct connection between the nonlinear response and geometry-dependent near-field distributions. This framework not only recovers the standard planar formalism in the appropriate limit, but also reveals fundamentally new degrees of freedom unique to nanostructures. Importantly, nanostructuring does not simply enhance the SHG signal; it enables \emph{tunability}—the ability to independently control both the attenuation and phase of the nonlinear interaction. As a result, the relative contributions of surface and electric-field-induced responses can be deterministically engineered through nanophotonic design, in line with emerging theoretical and experimental efforts toward structured-light control of nonlinear optical processes \cite{allayarov_strong_2025}. 

Experimentally, this enhanced sensitivity enables the observation of spectral shifts in the SH response (on the order of $\sim$1.3~nm) with electrolyte concentration, indicating a direct coupling between the EDL potential and the semiconductor's electronic polarizability. Furthermore, using a controlled optical pump, we observe reversible, intensity-dependent modulation of the interfacial susceptibility driven by two mechanisms. At low intensities, the SH signal decreases, consistent with photocharging, while at higher intensities it increases due to photothermal effects. The opposite sign of the modulation provides a clear signature of this transition. Notably, tailored optical resonances at the pump wavelength enable controlled light-induced modification of interfacial chemical equilibria and thermal effects. By tuning the nanodisk geometry, we control the threshold intensity separating these regimes, enabling reversible switching between photocharging- and photothermal-governed responses. These results highlight that nanostructuring transforms SHG from a passive probe into a tunable spectroscopic tool that can probe subtle interfacial phenomena.

Taken together, these results establish a unified, experimentally validated framework for understanding SHG at nanostructured solid–liquid interfaces, while opening new opportunities to actively control interfacial charge, potential, and chemical equilibria with light. Extending phase-resolved SHG techniques to nanostructured systems would provide additional insight and enable more direct separation of individual contributions. However, implementing such measurements remains experimentally challenging due to the complex angular emission and polarization mixing inherent to nanostructures, and thus represents an important direction for future work. More broadly, this work paves the way for integrating nanophotonics with electrochemical and catalytic systems, enabling new routes toward optically driven control of interfacial processes in energy conversion, photoelectrochemistry, and hydrovoltaic technologies.

\bibliography{references, references-2}
\section*{Acknowledgments}
We want to acknowledge Dr. Alan Bowman for his technical support during the initial stages of the setup development. We acknowledge the support of the Swiss National Science Foundation (Starting Grant 211695) and the Korean-Swiss Science and Technology Cooperation Fund (IZKSZ2\_188341). T.A. also acknowledges the support of the Swiss Government Excellence fellowship. We are thankful to Mr. Laurent Chevalley at the EPFL's mechanical workshop (ATME) for his assistance in fabricating the cell. We also acknowledge the support of the following experimental facilities at EPFL: Center of MicroNanoTechnology (CMi) and Interdisciplinary Center for Electron Microscopy (CIME).
\section*{Author contributions}
T.A. and G.T. conceptualized the study; T.A. developed the experimental platform, performed all the experiments, numerical calculations, and modeling, and investigated and visualized the results under the supervision of G.T.; D.D.A. fabricated the samples; M.S. and D.D.A. contributed to the setup development.  

\clearpage
\appendix
\renewcommand{\thesection}{S\arabic{section}}
\section*{Supplementary Information}

This section provides methods and additional figures.

\renewcommand{\thefigure}{S\arabic{figure}}
\renewcommand{\thetable}{S\arabic{table}}
\renewcommand{\theequation}{S\arabic{equation}}
\renewcommand{\thesubsection}{S\arabic{subsection}}
\setcounter{figure}{0}
\setcounter{table}{0}
\setcounter{equation}{0}
\subsection{Experimental}\label{experimental_setup}
\begin{figure}
    \centering
    \includegraphics[width=0.8\linewidth]{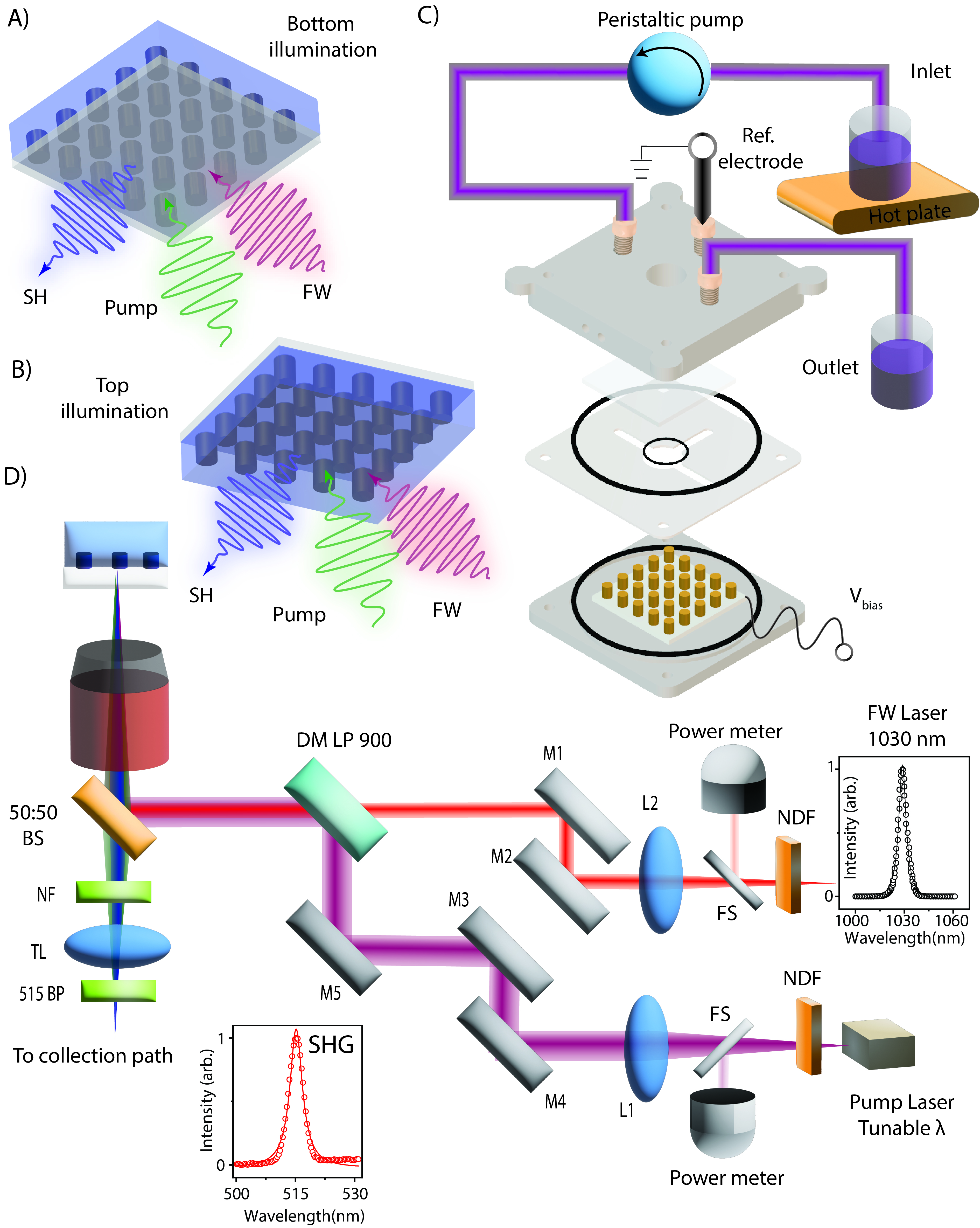}
    \caption{\textbf{The operando spectro-photo(electro)-chemical setup.}  A-B) Sample configuration depicting the (A) bottom and (B) top illumination configuration in the reflection mode. C) An external pump circulates the electrolyte through the designated inlet and outlet ports, and an additional port for a reference electrode is available for precise electrochemical measurements. The nanostructured sample can be arranged either on the top or bottom, based on whether front or back illumination is employed. A hot plate is used to heat the electrolyte for temperature-dependent measurements.  D)In the (bottom-right), we have the excitation path featuring the FW (centered at 1030 nm with a bandwidth of 6 nm),  and a tunable wavelength pump laser (wavelength, power, and bandwidth can be varied).  At the (top-left), an inverted microscope is mounted with the flow cell as shown in panel C. On the (bottom left), the 4F collection path facilitates the signal, which then travels through the spectrometer and is monitored via a CCD camera. The various critical elements in this setup are as follows: M1-M5 signify mirrors; L1, L2 are collimating lenses; TL denotes the tube lens; DM LP900 is the long pass dichroic mirror; FS consists of 0.17 mm thin fused silica substrates for reflecting a small percentage for real-time power monitoring; NDF is a neutral density filter for controlling the power level; BS is the beam splitter; NF is a notch filter designed to eliminate the pump excitation wavelength from the collection path; SP is the short pass filter; and BP is the band pass filter. The bottom graph shows an SH intensity (centered at 515 nm with a bandwidth of 4.3 nm) detected for a representative case. }
    \label{fig:ch4_operando_SHG_setup}
\end{figure}
We designed a compression flow cell, as shown in (Fig.  \ref{fig:ch4_operando_SHG_setup}A), which consists of three essential components: i) a top section with three ports for electrolyte inflow and outflow, along with a reference electrode situated within the bulk electrolyte; ii) a middle spacer available in thicknesses of 0.5 mm or 1 mm; and iii) a bottom section where the sample to be examined can be positioned for back illumination. This flow cell works in conjunction with an external peristaltic pump that continuously circulates the electrolyte during in-situ experiments. To ensure optimal observations, the cell is mounted on an inverted microscope (Nikon Eclipse T2), providing flexibility in sample placement—whether at the bottom for back illumination (Fig.  \ref{fig:ch4_operando_SHG_setup}B) or at the top for front illumination (Fig.  \ref{fig:ch4_operando_SHG_setup}C). When the sample is placed on top, a fused-silica substrate is set at the bottom, and this arrangement is reversed for front placement. The variable-coverslip correction of the 0.7 NA 60x objective enables precise focusing of light onto specific planes.

\begin{figure}[h!]
\centering 
\includegraphics[width=0.6\columnwidth]{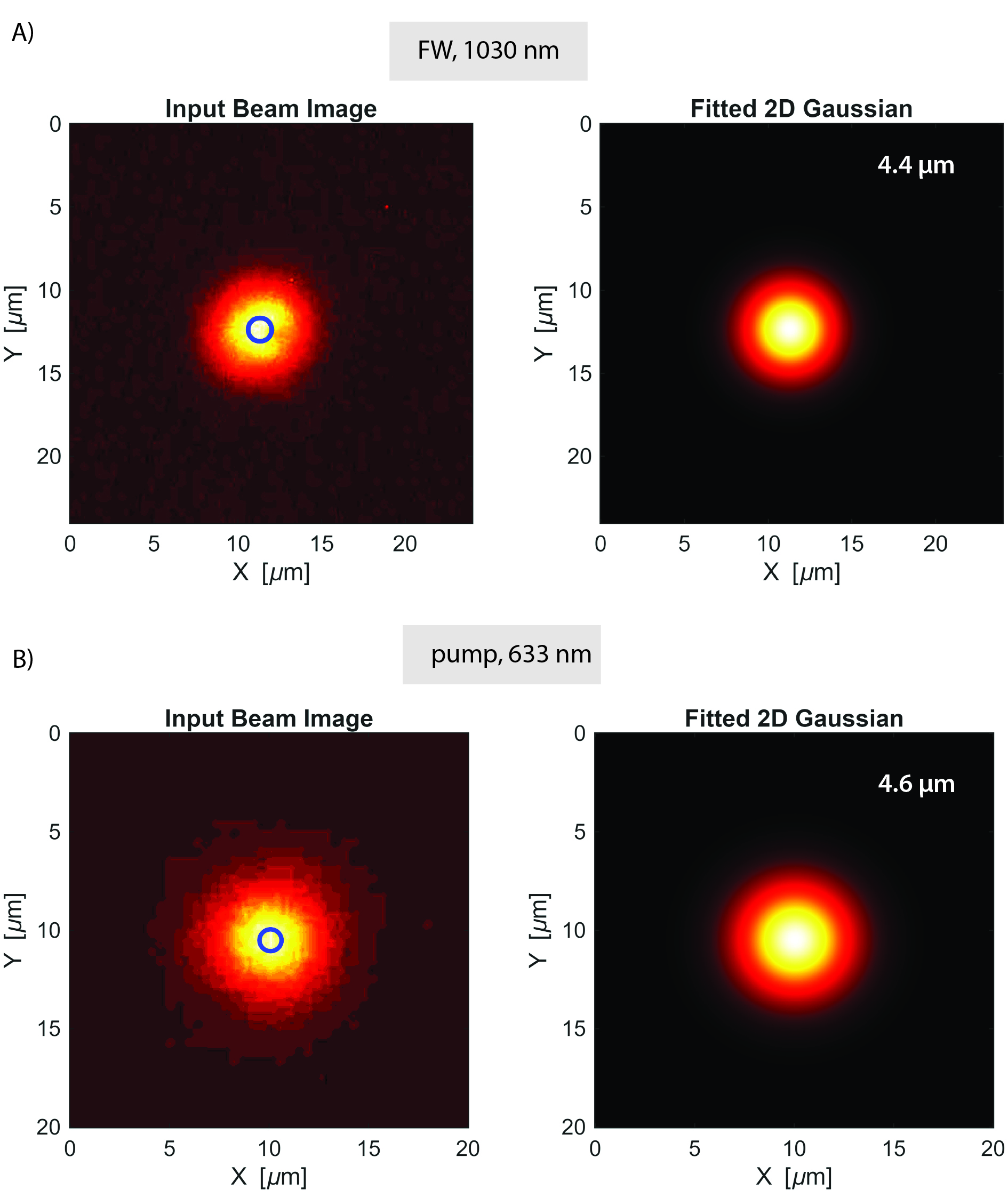} 
\caption{\footnotesize Beam diameter of the FW (1030 nm) and pump (633 nm) laser obtained by the corresponding 2D Gaussian fit.}
\label{fig:beam_diameter} 
\end{figure}

As illustrated in Fig.  \ref{fig:ch4_operando_SHG_setup}D, the inverted microscope is set up to simultaneously excite and capture optical signals. The excitation pathway features two lasers: i) a 1030 nm, 6 nm full-width-half-maximum, linearly polarized femtosecond laser (NKT origami, 200 fs, 40 MHz) serving as the fundamental beam (FW) for second harmonic generation, and ii) a supercontinuum white light laser (NKT Photonics) used for generating charge carriers and optically modulating the semiconductor. To fine-tune the wavelength and power of the NKT laser, we employ a tunable wavelength filter (SuperK VARIA) spanning 450-840 nm with a 10 nm bandwidth. Both laser beams are focussed and incident vertically to the sample and are carefully aligned to converge at the same point on the sample through independent paths, as shown in Fig.  \ref{fig:ch4_operando_SHG_setup}D. The beam diameter for the FW beam is fixed at 4.4 $\mu$m, slightly smaller than the pump laser beam's diameter of 4.6 $\mu$m (Fig. \ref{fig:beam_diameter}), with precision tuning achieved using collimating lenses.

We conduct SHG measurements in a reflection configuration, in which the emitted SH signal and the reflected fundamental wave (FW) are directed into the collection pathway. This pathway employs a 4F configuration to couple the signal to a grating spectrometer (Princeton Instruments Spectra Pro HRS-500), equipped with a Peltier-cooled 2D CCD detector (Princeton Instruments PIXIS 256). To characterize the system, we first perform SHG measurements without electrolyte in the cell, then introduce the electrolyte via the external pump. Notably, our spectroscopic signal remained stable during electrolyte pumping at optimal rates, enabling us to monitor SH intensity transitions as the bulk electrolyte environment changed. To monitor SH intensity under light stimuli, we switched the pump laser between on and off states using a shutter in the optical path. Notably, the substantial spectral separation between the excitation (1030 nm) and SH emission (515 nm), as well as the pump laser (633 nm) and an additional notch filter in the collection path, ensures that the measured SH intensity is not affected by residual signals from the pump and excitation. In addition, our in-situ setup is adapted for measurements that require an external electrical bias or an increase in bulk temperature, using the equipped reference electrode and hot plates beneath the inlet reservoir. 

\subsection{Non-linear polarization and effective susceptibility using overlap integral formalism}\label{non-linear-polarisation_Second_Harmonic_intensity}
The non-linear  polarization depends on the  second-order and effective third-order susceptibility, and the fundamental E-field according to:

\begin{equation}
P_{\text{nl}} (2\omega)=P^{\text{2}} (2\omega)+P^{\text{3}} (2\omega) = \varepsilon_0 \chi^{2}_{\text{s}}:E_{\text{}}(\omega)E_{\text{}}(\omega)+\varepsilon_0 \chi^{3}_{\text{}}:E_{\text{}}(\omega)E_{\text{}}(\omega)E^{\text{DC}}
\end{equation}

We need to write the normal and tangential components of the non-linear polarization in terms of the non-vanishing components of the susceptibility tensor. For the nanodisk geometry, it is convenient to transform the normal-tangential components into $(r,\phi,z)$. The fundamental electric field at the specific point in the non-linear region is given:
\begin{equation}
\mathbf{E}=
\begin{pmatrix}
E_r \\
E_\phi \\
E_z
\end{pmatrix}
=
\begin{pmatrix}
\cos\phi & \sin\phi & 0 \\
-\sin\phi & \cos\phi & 0 \\
0 & 0 & 1
\end{pmatrix}
\begin{pmatrix}
E_x(x,y,z) \\
E_y(x,y,z) \\
E_z(x,y,z)
\end{pmatrix}.
\end{equation}

The second-order polarization at the curved surface is given by:
\begin{equation}
\mathbf{P}^{(\mathrm{2})}_{(\mathrm{curved})}
=
\varepsilon_0
\begin{pmatrix}
\chi_{\perp\perp\perp} E_r^{2}
+ \chi_{\perp\parallel\parallel}\!\left(E_\phi^{2}+E_z^{2}\right)\\[6pt]
2\,\chi_{\parallel\perp\parallel}\,E_r E_\phi\\[6pt]
2\,\chi_{\parallel\perp\parallel}\,E_r E_z
\end{pmatrix}
\end{equation}
Similarly, the second-order polarization at the flat surface is given by:
\begin{equation}
\mathbf{P}^{(\mathrm{2})}_{(\mathrm{flat})}
=
\varepsilon_0
\begin{pmatrix}
2\,\chi_{\parallel\parallel\perp}\,E_z E_r\\[6pt]
2\,\chi_{\parallel\parallel\perp}\,E_z E_\phi\\[6pt]
\chi_{\perp\perp\perp} E_z^{2}
+ \chi_{\perp\parallel\parallel}\!\left(E_r^{2}+E_\phi^{2}\right)
\end{pmatrix}
\end{equation}
The third-order polarization requires the DC field $(E^{\text{DC}})$ oriented along the surface normal. So, the third-order polarization at the curved surface is given by:
\begin{equation}
\mathbf{P}^{(\mathrm{3})}_{(\mathrm{curved})}
=
\varepsilon_0 
\begin{pmatrix}
\chi_{\perp\perp\perp\perp} E_r^{2}
+ \chi_{\perp\parallel\parallel\perp}\!\left(E_\phi^{2}+E_z^{2}\right)\\[6pt]
2\,\chi_{\parallel\perp\parallel\perp}\,E_r E_\phi\\[6pt]
2\,\chi_{\parallel\perp\parallel\perp}\,E_r E_z
\end{pmatrix}E^{\text{DC}}= \mathbf{p}^{(\mathrm{3})}_{(\mathrm{curved})} E^{\text{DC}}
\end{equation}
 Similarly, the third-order polarization at the flat surface is given by:
\begin{equation}
\mathbf{P}^{(\mathrm{3})}_{(\mathrm{flat})}
=
\varepsilon_0
\begin{pmatrix}
2\,\chi_{\parallel\parallel\perp\perp}\,E_z E_r\\[6pt]
2\,\chi_{\parallel\parallel\perp\perp}\,E_z E_\phi\\[6pt]
\chi_{\perp\perp\perp\perp} E_z^{2}
+ \chi_{\perp\parallel\parallel\perp}\!\left(E_r^{2}+E_\phi^{2}\right)
\end{pmatrix}E^{\text{DC}}=\mathbf{p}^{(\mathrm{3})}_{(\mathrm{flat})} E^{\text{DC}}
\end{equation}

We transform the normal-tangential components back into Cartesian components as follows:
\begin{equation}
\begin{pmatrix}
P_x \\
P_y \\
P_z
\end{pmatrix}
=
\begin{pmatrix}
\cos\phi & -\sin\phi & 0 \\
\sin\phi & \cos\phi & 0 \\
0 & 0 & 1
\end{pmatrix}
\begin{pmatrix}
P_r \\
P_\phi \\
P_z
\end{pmatrix}
\end{equation}

Once we have the point-wise non-linear polarization $(P_x^{2\omega}, P_y^{2\omega}, P_z^{2\omega})$ and near field at SH frequency $(E_x^{2\omega}, E_y^{2\omega}, E_z^{2\omega})$, we can use the overlap integral formalism to derive the SH intensity (see Table \ref{table:SI:overlap_integral_implementation} for detailed implementation in terms of electric field vector and susceptibility tensor) as follows:

\begin{equation}
I_{\mathrm{SH}}
\propto
\left|
\int_{S}
\big(\mathbf{E}^{2\omega}\big)^{*} \cdot \left (\mathbf{P}_{\text{Si}}^{2}+\mathbf{P}_{\text{Stern}}^{2}\right ) \, dS
+
\int_{V}
\big(\mathbf{E}^{2\omega}\big)^{*} \cdot \left (\mathbf{p}_{\text{Si}}^{(3)}E_{\text{SCL}}+\mathbf{p}_{\text{H$_2$O}}^{(3)}E_{\text{EDL}}\right ) \, dV
\right|^{2}
\end{equation}
We now substitute an explicit expression for the electric double layer (EDL) and space charge layer (SCL) fields varying along the surface normal $(\hat{n})$ as follows:

\begin{equation}
    E_{\text{EDL}}=\frac{\Delta\Phi_{\text{EDL}}}{\delta_{\text{EDL}}}e^{-\frac{\mathbf{r}\cdot \hat{n}}{\delta_{\text{EDL}}}}
\end{equation}

\begin{equation}
    E_{\text{SCL}}=\frac{\Delta\Phi_{\text{SCL}}}{\delta_{\text{SCL}}}e^{-\frac{\mathbf{r}\cdot \hat{n}}{\delta_{\text{SCL}}}}
\end{equation}
where $\Delta \Phi_{\text{EDL}}$, $\Delta \Phi_{\text{SCL}}$ are the electrical potential difference between the surface and bulk of the EDL and SCL, respectively. $\delta_{\text{EDL}}$ and $\delta_{\text{SCL}}$ are the decay lengths of the electric field in the EDL and SCL, respectively. Notably, the electromagnetic field is discontinuous at the interface between silicon and the electrolyte due to the different dielectric constants. Thus, for the silicon surface contribution, the integral is evaluated just inside the disk, ${\text{S$^-$}}$ $(r=0^-)$ \cite{reddy_revisiting_2017}, whereas for the electrolyte, it is evaluated just outside the disk, ${\text{S$^+$}}$ $(r=0^+)$. To visualize how different E-field components and susceptibilities are involved in the overlap integral formalism, we first write the overlap integral by considering only the dominant tensor elements for the different contributions. Here we have considered that the tensor elements $\chi_{\perp\perp\perp}^2$ and $\chi_{\perp\perp\perp\perp}^3$ are the dominant components \cite{falasconi_bulk_2001, kielich_optical_1969}. We can denote the dominant component as follows: $\chi_{\text{Si, }\perp\perp\perp}^2\equiv\chi_{\text{Si}}^2$, $\chi^3_{\text{Si, }\perp\perp\perp\perp}\equiv \chi_{\text{Si}}^3$ ,  $\chi_{\text{Stern, }\perp\perp\perp}^2\equiv\chi_{\text{Stern}}^2$, $\chi^3_{\text{H$_2$O, }\perp\perp\perp\perp}\equiv \chi_{\text{H$_2$O}}^3$. 
\begin{equation}\label{eq:SI_dominant_component_ISH_overlap_integral}
I_{\text{SH}} \propto
\left|
\begin{aligned}
&\chi^2_{\text{Si}}
\int_{\text{S$^-$}} \big(E_n^{2\omega}\big)^* E_n E_n \, dS+\chi^2_{\text{Stern}}
\int_{\text{S$^+$}} \big(E_n^{2\omega}\big)^* E_n E_n \, dS \\
&\quad + \frac{\Delta\Phi_{\text{EDL}}}{\delta_{\text{EDL}}}\chi^3_{\text{Si}}
\int_{V,\text{SCL}} \big(E_n^{2\omega}\big)^* E_n E_n
\, e^{-\frac{\mathbf{r}\cdot \hat{n}}{\delta_{\text{SCL}}}} \, dV \\
&\quad + \frac{\Delta\Phi_{\text{SCL}}}{\delta_{\text{SCL}}}\chi^3_{\text{H$_2$O}}
\int_{V,\text{EDL}} \big(E_n^{2\omega}\big)^* E_n E_n
\, e^{-\frac{\mathbf{r}\cdot \hat{n}}{\delta_{\text{EDL}}}} \, dV
\end{aligned}
\right|^2
\end{equation}
In modeling the second-harmonic response of a silicon–electrolyte interface, the total effective surface susceptibility can be expressed as a sum of contributions from the silicon surface and the electrolyte (Stern layer). Because the normal component of the electric displacement field must remain continuous across the interface, the electric field itself is discontinuous; this discontinuity can be absorbed into an \textbf{effective susceptibility for silicon}, allowing both silicon and electrolyte contributions to be evaluated using a common reference electric field \cite{heinz_chapter_1991}. 

The SH intensity can be written in terms of surface and volume overlap integrals,  $\mathcal{I_S}$ (evaluated just outside the disk, i.e., $r=0^+$ ), $\mathcal{I}_{\text{EDL}}$, and $\mathcal{I}_{\text{SCL}}$ as follows:
\begin{equation}\label{eq:SI_general_ISH_overlap_integral}
    I_{\text{SH}}\propto \left | \left(\chi^2_{\text{Si,eff}}+\chi^2_{\text{Stern}}\right) \cplx{I}_{\text{S}} + \frac{\Delta\Phi_{\text{SCL}}}{\delta_{\text{SCL}}}\chi^3_{\text{Si,eff }} \cplx{I}_{\text{SCL}} + \frac{\Delta\Phi_{\text{EDL}}}{\delta_{\text{EDL}}}\chi^3_{\text{H$_2$O}} \cplx{I}_{\text{EDL}}\right |^2
\end{equation}
where, $\chi_{\text{Si, eff}}^2$ and $\chi_{\text{Si, eff}}^3$ are the effective second and third-order susceptibility of silicon; $\chi_{\text{Stern}}^2$ and $\chi_{\text{H$_2$O}}^3$ are the second and third order susceptibility of the Stern layer and interfacial water. 

The overlap integral formalism in Eq.~\ref{eq:SI_dominant_component_ISH_overlap_integral} assumes the normal component is the dominant one, and the other component does not contribute to the SH intensity. However, the Eq.~\ref{eq:SI_general_ISH_overlap_integral} is a general one and contains all the active susceptibility components and not just the dominant components. The dominant one is taken out of the overlap integral, so the susceptibility component is normalized by it. For example, if $\chi_{\text{Si, }\perp\perp\perp}^2$ is the dominant one, we define $\chi_{r}^2=\chi_{\text{Si, }\perp\parallel\parallel}^2/ \chi_{\text{Si, }\perp\perp\perp}^2$, then the normalized susceptibility becomes, 1 and $\chi_r^2$ for the dominant and non-dominant components. For silicon $\chi^2_r \approx 0.2-0.3$ \cite{falasconi_bulk_2001}.The table \ref{table:SI:overlap_integral_implementation} shows how the non-linear polarization component can be evaluated by considering different electric components and all the non-vanishing components of the susceptibility tensor, both for the second and third order terms. 

\begin{table}[h]
\centering
\renewcommand{\arraystretch}{1.4}
\begin{tabular}{c l}
\hline
\textbf{Variable} & \textbf{Expression} \\
\hline

$\cos\phi$ & $\cos\!\bigl(\operatorname{atan2}(y,x)\bigr)$ \\
$\sin\phi$ & $\sin\!\bigl(\operatorname{atan2}(y,x)\bigr)$ \\

$E_r$ & $E_x \cos\phi + E_y \sin\phi$ \\
$E_\phi$ & $-E_x \sin\phi + E_y \cos\phi$ \\
$E_z$ & $E_z$ \\

$\mathrm{p2c}_r$ & $\chi_{\perp\perp\perp} E_r^2 + \chi_{\perp\parallel\parallel}(E_\phi^2 + E_z^2)$ \\
$\mathrm{p2c}_\phi$ & $2\chi_{\parallel\perp\parallel} E_r E_\phi$ \\
$\mathrm{p2c}_z$ & $2\chi_{\parallel\perp\parallel} E_r E_z$ \\

$\mathrm{p2f}_r$ & $2\chi_{\parallel\perp\parallel} E_z E_r$ \\
$\mathrm{p2f}_\phi$ & $2\chi_{\parallel\perp\parallel} E_z E_\phi$ \\
$\mathrm{p2f}_z$ & $\chi_{\perp\perp\perp} E_z^2 + \chi_{\perp\parallel\parallel}(E_r^2 + E_\phi^2)$ \\

$\mathrm{E}_f^{\text{DC}}$ & $\Delta \Phi_{\text{EDL}}/\delta_{\text{EDL}} \exp\!\left(-\frac{z-H}{\delta_{\text{EDL}}}\right)$ \\
$\mathrm{E}_c^{\text{DC}}$ & $\Delta \Phi_{\text{EDL}}/\delta_{\text{EDL}} \exp\!\left(-\frac{\sqrt{x^2+y^2}-R}{\delta_{\text{EDL}}}\right)$ \\

$\mathrm{p3c}_r$ & $\left(\chi_{\perp\perp\perp\perp} E_r^2 + \chi_{\perp\parallel\parallel\perp}(E_\phi^2 + E_z^2)\right)\mathrm{E}_c^{\text{DC}}$ \\
$\mathrm{p3c}_\phi$ & $\left(2\chi_{\parallel\perp\parallel\perp} E_r E_\phi\right)\mathrm{E}_c^{\text{DC}}$ \\
$\mathrm{p3c}_z$ & $\left(2\chi_{\parallel\perp\parallel\perp} E_r E_z\right)\mathrm{E}_c^{\text{DC}}$ \\

$\mathrm{p3f}_r$ & $\left(2\chi_{\parallel\perp\parallel\perp} E_z E_r\right)\mathrm{E}_f^{\text{DC}}$ \\
$\mathrm{p3f}_\phi$ & $\left(2\chi_{\parallel\perp\parallel\perp} E_z E_\phi\right)\mathrm{E}_f^{\text{DC}}$ \\
$\mathrm{p3f}_z$ & $\left(\chi_{\perp\perp\perp\perp} E_z^2 + \chi_{\perp\parallel\parallel\perp}(E_r^2 + E_\phi^2)\right)\mathrm{E}_f^{\text{DC}}$ \\

$\mathrm{p2c}_x$ & $\mathrm{p2c}_r \cos\phi - \mathrm{p2c}_\phi \sin\phi$ \\
$\mathrm{p2c}_y$ & $\mathrm{p2c}_r \sin\phi + \mathrm{p2c}_\phi \cos\phi$ \\

$\mathrm{p3c}_x$ & $\mathrm{p3c}_r \cos\phi - \mathrm{p3c}_\phi \sin\phi$ \\
$\mathrm{p3c}_y$ & $\mathrm{p3c}_r \sin\phi + \mathrm{p3c}_\phi \cos\phi$ \\

$\mathrm{p2f}_x$ & $\mathrm{p2f}_r \cos\phi - \mathrm{p2f}_\phi \sin\phi$ \\
$\mathrm{p2f}_y$ & $\mathrm{p2f}_r \sin\phi + \mathrm{p2f}_\phi \cos\phi$ \\

$\mathrm{p3f}_x$ & $\mathrm{p3f}_r \cos\phi - \mathrm{p3f}_\phi \sin\phi$ \\
$\mathrm{p3f}_y$ & $\mathrm{p3f}_r \sin\phi + \mathrm{p3f}_\phi \cos\phi$ \\

\hline
\end{tabular}
\caption{Definitions of field components and nonlinear polarization terms. Here, x,y,z are the spatial coordinates, $R$ and $H$ are the disk radius and height; c and f stand for curved and flat surface, respectively. E is the electric field, and p is the non-linear polarization (2:second order, 3:third-order).}
\label{table:SI:overlap_integral_implementation}
\end{table}

To write an expression for the effective susceptibility of the four contributions, we do some rearrangements as follows: 
\begin{equation}
    I_{\text{SH}}\propto \left | \cplx{I}_{\text{S}} \right |^2\left | \chi^2_{\text{Si,eff}}+\chi^2_{\text{Stern}}  + \Delta\Phi_{\text{SCL}}\chi^3_{\text{Si,eff }} \frac{\cplx{I}_{\text{SCL}}}{\delta_{\text{SCL}}\cplx{I}_{\text{S}}} + \Delta\Phi_{\text{EDL}}\chi^3_{\text{H$_2$O}}\frac{\cplx{I}_{\text{EDL}}}{\delta_{\text{EDL}}\cplx{I}_{\text{S}}} \right |^2
\end{equation}

\begin{equation}
    I_{\text{SH}}\propto \left | \cplx{I}_{\text{S}} \right |^2\left | \chi^2_{\text{Si,eff}}+\chi^2_{\text{Stern}}  + \Delta\Phi_{\text{SCL}}\chi^3_{\text{Si,eff }} \ \cplx{G} + \Delta\Phi_{\text{EDL}}\chi^3_{\text{H$_2$O}}\ \cplx{F} \right |^2
\end{equation}

\begin{equation}
    \chi_{\text{eff}}^2 = \chi^2_{\text{Si,eff}}+\chi^2_{\text{Stern}} + \chi^3_{\text{Si,eff }}\Delta\Phi_{\text{SCL}} \ \cplx{G} + \chi^3_{\text{H$_2$O}}\Delta\Phi_{\text{EDL}}\ \cplx{F}
\end{equation}
where the factor $\mathcal{F}$ and $\mathcal{G}$ are defined as follows:
\begin{equation}
\mathcal{F} = \frac{\mathcal{I}_{\text{EDL}}}{\delta_{\text{EDL}} \mathcal{I}_{\text{S}}}, 
\qquad
\mathcal{G} = \frac{\mathcal{I}_{\text{SCL}}}{\delta_{\text{SCL}} \mathcal{I}_{\text{S}}},
\end{equation}
Therefore, the SH intensity can be written in terms of the effective susceptibility.
\begin{equation}
    I_{\text{SH}}\propto \left| \cplx{I}_{\text{S}}* \chi_{\text{eff}}^2 \right|^2
\end{equation}
where, $|\mathcal{I}_{\text{S}}| \propto A_{\text{S}} \langle E(\omega;\mathbf{r})^2 \rangle_{\text{S}}=A_{\text{S}}E(\omega)^2$

\subsection{Numerical simulation}\label{numerical_simulation}

All simulations were performed using the finite element method (FEM) in COMSOL Multiphysics 6.3. The primary objective was to optimize the total SH emission of silicon nanodisks supported on a fused silica substrate, with either water or air as the superstrate. A three-dimensional unit-cell model was constructed to represent the periodic array, with the lateral dimensions set by the array periodicity \(p = 800~\text{nm}\). The nanodisk height and diameter were systematically varied to identify geometries that maximize the SH response.

\vspace{0.5em}
\noindent
\textbf{Linear simulation.}  
The simulation workflow begins with a linear electromagnetic analysis to obtain the fundamental-wave (FW) electric field distribution. To accurately represent an infinite array, periodic boundary conditions were applied to the lateral faces of the unit cell.

\begin{figure}[h!]
\centering 
\includegraphics[width=1\columnwidth]{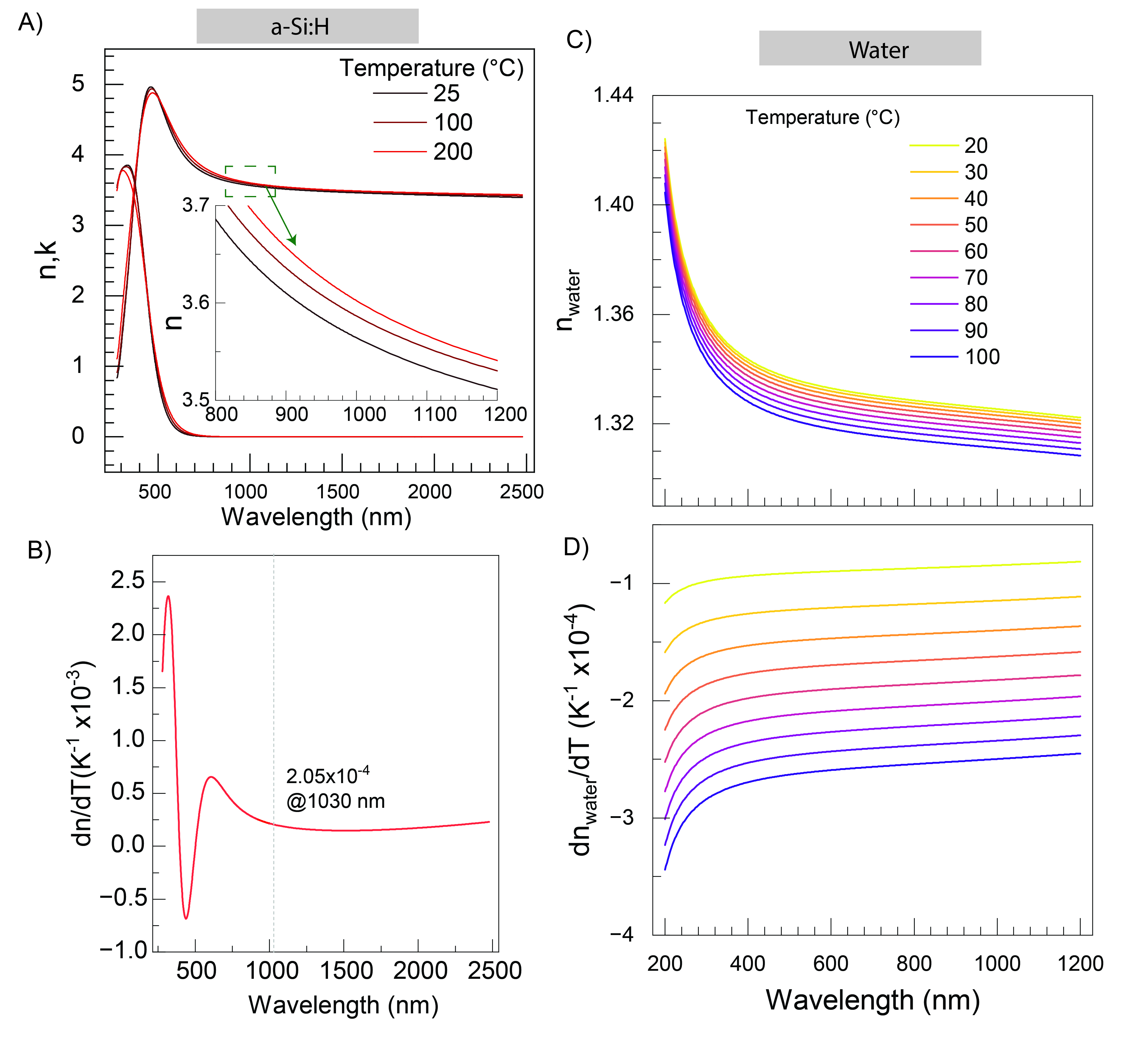} 
\caption{\footnotesize \textbf{Thermo-optical coefficients:} A) Measured n and k values of a-Si: H films with a thickness of 440 nm under different temperatures. Ellipsometry measurements at different temperatures were carried out using an external heater. B) The real part of the thermo-optical coefficient obtained from the measurements in panel A. C)Refractive index of water at different temperatures. Reproduced from: International Association for the Properties of Water and Steam (IAPWS). Release on the Refractive Index of Ordinary Water Substance as a Function of Wavelength, Temperature, and Pressure; Erlangen, Germany, 1997. Available at: http://www.iapws.org/relguide/rindex.pdf D)Corresponding thermo-optical coefficients.}
\label{fig:TO_coeffecients} 
\end{figure}

\begin{enumerate}
    \item \textit{Excitation configuration:} A port boundary condition was defined at the bottom boundary to simulate back-illumination by a normally incident plane wave. The polarization of the electric field was aligned along a fixed axis to define the excitation conditions.

    \item \textit{Boundary conditions:} The top boundary was assigned a second port without excitation, allowing transmitted fields to exit the simulation domain. Perfectly matched layers (PMLs) were implemented at both top and bottom boundaries to suppress spurious reflections.

    \item \textit{Reflectance and transmittance:} The optical response was quantified by integrating the Poynting vector over planes parallel to the substrate and superstrate, enabling accurate evaluation of reflected and transmitted power.

    \item \textit{Material properties:} The refractive index of silicon was obtained from experimental ellipsometry measurements (Fig.~\ref{fig:TO_coeffecients}A), ensuring realistic optical constants.
\end{enumerate}

\vspace{0.5em}
\noindent
\textbf{Nonlinear simulation.}  
The FW electric field distribution obtained from the linear simulation is used to compute the nonlinear surface polarization, which serves as the source of SH radiation. The second-order nonlinear polarization is expressed as

\begin{equation}\label{eq:non-linear-polarisation}
\textbf{P}^{2}_{\text{nl}}(2\omega)=\varepsilon_0 \left[\chi^{2}_{\perp \perp\perp}E_n(\omega)^2+\chi^{2}_{\perp\parallel\parallel}E_t(\omega)^2\right]\hat{\textbf{n}}+\varepsilon_0 \left[\chi^{2}_{\parallel\perp\parallel}E_t(\omega)E_n(\omega)\right]\hat{\textbf{t}},
\end{equation}

where \(E_n(\omega)\) and \(E_t(\omega)\) denote the normal and tangential components of the FW electric field at the nanodisk surface, respectively. The tensor components \(\chi^{2}_{\perp\perp\perp}\), \(\chi^{2}_{\perp\parallel\parallel}\), and \(\chi^{2}_{\parallel\perp\parallel}\) correspond to the surface nonlinear susceptibility of silicon, while \(\hat{\textbf{n}}\) and \(\hat{\textbf{t}}\) are the unit vectors normal and tangential to the surface. We use the notation of $\perp \text{and} \parallel$ for denoting susceptibility to follow the standards in the literature \cite{heinz_chapter_1991}

\begin{enumerate}
    \item \textit{Nonlinear source construction:} The computed FW fields are used to evaluate the nonlinear surface polarization, which is implemented as a source term for SH radiation.

    \item \textit{Second-harmonic simulation:} A frequency-domain simulation at the SH wavelength is performed, with the nonlinear polarization acting as the excitation source. This enables direct computation of the generated SH fields.

    \item \textit{SH power extraction:} The SH signal is quantified by integrating the Poynting vector over planes parallel to the substrate, analogous to the linear case. This provides a consistent measure of SH reflectance and transmission.
\end{enumerate}

\vspace{0.5em}
\noindent
\textbf{Back focal plane calculation for SH emission.}  
Once the near field and far field are obtained from the non-linear step of the simulation as described above, we can relate them to the BFP. From COMSOL, we can obtain the direction-dependent fields $(\theta,\phi)$ in the front-focal plane, which we can relate to positions $(u,v)$ in its BFP, as:
\begin{equation}
u = n \sin\theta \cos\phi,  \quad v = n \sin\theta \sin\phi
\end{equation}
The total intensity collected by the objective $I_{\mathrm{det}}$ is the integral of the angular field distribution $|E(\theta,\phi)|^2$ over the detection cone $\Omega_{\mathrm{det}}$:
\begin{equation}
I_{\mathrm{det}} = \int_{\Omega_{\mathrm{det}}} |E(\theta,\phi)|^2\, d\Omega,
\end{equation}
Assuming no losses, the same power is spread over the detection area $A_{\mathrm{det}}$ in the objective BFP, defined by coordinates $(u,v)$, which is limited by the objective numerical aperture:
\begin{equation}
\int_{\Omega_{\mathrm{det}}} |E(\theta,\phi)|^2\sin\theta\, d\theta\, d\phi 
= \int_{A_{\mathrm{det}}} I_{\mathrm{BFP}}(u,v)\, du\, dv,
\end{equation}
which holds if the integrands are equivalent:
\begin{equation}
|E(\theta,\phi)|^2\sin\theta\, d\theta\, d\phi 
= I_{\mathrm{BFP}}(u,v)\, du\, dv.
\end{equation}

To eliminate the differential elements, we can express BFP coordinates in a polar reference frame, where $u = r\cos\phi$ and $v = r\sin\phi$.
\begin{equation}
|E(\theta,\phi)|^2\sin\theta\, d\theta\, d\phi 
= I_{\mathrm{BFP}}(\theta,\phi)\,(n\sin\theta)\,(n\cos\theta\, d\theta)\, d\phi,
\end{equation}
eventually resulting in
\begin{equation}
I_{\mathrm{BFP}}(\theta,\phi) 
= \frac{|E(\theta,\phi)|^2}{n^2 \cos\theta}
\end{equation}

The simulated BFP intensity maps are shown in Fig.\ref{fig:main_figure_1}E (bottom panel), F, and this is observable directly in our experiment by imaging the BFP onto a CCD camera as shown in Fig.~\ref{fig:main_figure_1}E (top panel). Note also that, with these definitions, $u = k_x$ and $v = k_y$, with $(k_x, k_y)$ components of the transverse momentum as represented in Fig.~\ref{fig:main_figure_1}D, E, and F.\\

\textbf{Polarization resolved SH emission}  

For calculating the polarisation resolved intnsity, we decompose the $E_{\theta}(\theta, \phi)$ and  $E_{\phi}(\theta, \phi)$ into $E_{\text{s}}$ and $E_{\text{p}}$ components (see Fig.~\ref{fig:main_figure_1}D) as follows:
\begin{equation}
E(\alpha) = 
\left(E_\theta \cos\theta \cos\phi - E_\phi \sin\phi \right)\sin \alpha 
+ \left(E_\theta \cos\theta \sin\phi + E_\phi \cos\phi \right)\cos \alpha
\end{equation}

yielding $I(\alpha) \propto \left(E_s \cos\alpha + E_p \sin\alpha \right)^2$, where $E_s$ and $E_p$ are the s- and p-polarized SH components. The observed polarization mixing agrees well with the simulation results (Fig.~\ref{fig:main_figure_1}F)

Overall this approach, combining linear-field calculations with nonlinear source-driven simulations, provides a rigorous and self-consistent framework for evaluating SHG at nanostructured interfaces. It enables direct correlation between geometry-dependent field distributions and the resulting SH emission, forming the basis for optimizing nanodisk designs and interpreting experimental observations.

\subsection{Second harmonic emission from nanostructured interface}\label{Section:SI_Nanostructured_SHG_emission}

\begin{figure}[h!]
    \centering
    \includegraphics[width=1\linewidth]{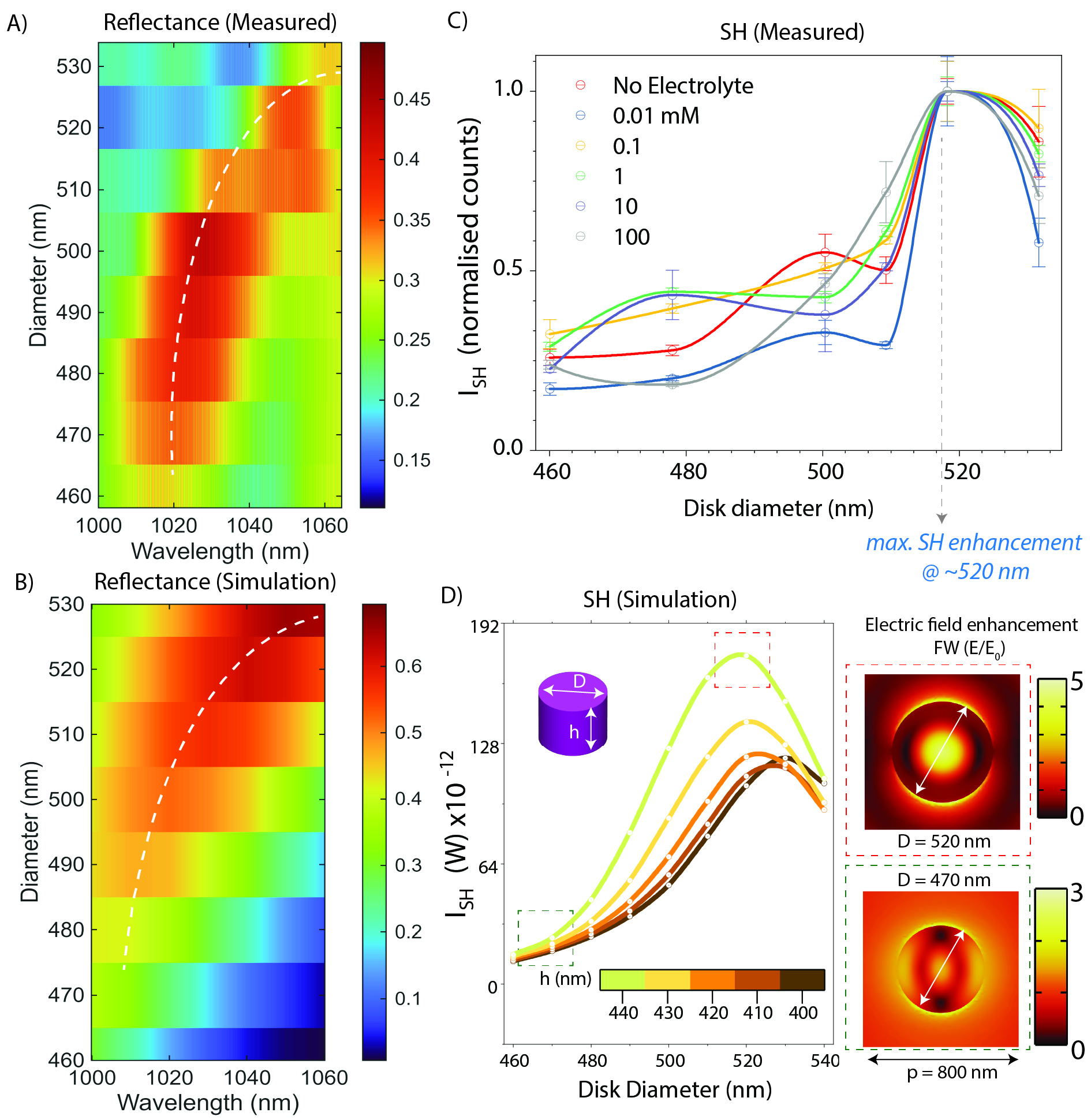}
    \caption{ \textbf{Characterization of SH intensity and linear reflectance in the nanodisk array.}(A) Linear optical characterization showcasing measured reflectance spectra for different nanodisk diameters at a fixed pitch of 800 nm, with Ag mirror reflectance as a reference. (B) Corresponding Comsol linear reflectance simulation for a square lattice with a pitch of 800 nm and height of 440 nm across various nanodisk diameters. (C) Measured second harmonic (SH) intensity as a function of nanodisk diameters under varying KCl electrolyte concentrations (0.01 mM to 100 mM) in DI water, normalized to the peak intensity observed for the ~520 nm diameter disks. (D) COMSOL simulation of SH intensity as a function of disk diameter, focusing on surface contributions to nonlinear polarization, revealing a peak enhancement at a diameter closely matching experimental trends as a function of disk diameter for a disk height of 440 nm. The red and green rectangles show the electric field enhancement for disk diameters of 520 nm and 470 nm, respectively.}
    \label{fig:ch4_SH_exp_simulation}
\end{figure}

Using the full SHG simulation framework described in Section~\ref{numerical_simulation}, we compute the near-field distribution at the second-harmonic frequency arising from the nonlinear polarization source (Eq.~\ref{eq:non-linear-polarisation}). The total emitted SH power is then obtained by integrating the Poynting vector over planes parallel to the substrate. The resulting SH intensities show good agreement with experimental measurements of second-harmonic scattering (Fig.~\ref{fig:main_figure_3}A and Figs.~\ref{fig:ch4_SH_exp_simulation}C,D), validating the overall accuracy of the numerical model.

While this approach provides the total scattered signal, it does not capture the angular distribution of the emitted radiation, which contains critical information about the underlying emission and scattering mechanisms. As discussed in Section~\ref{section:Second_harmonic_emission_nanostructured_interface} of the main text, SH emission from nanostructured interfaces occurs over a broad range of polar and azimuthal angles $(\theta,\phi)$ (Fig.~\ref{fig:main_figure_1}D). To access this information, we employ back focal plane (BFP) imaging (Fourier imaging), which directly maps the angular distribution of the scattered light \cite{cueff_fourier_2024}.

Experimentally, BFP imaging yields the angular intensity distribution of the SH signal across $(\theta,\phi)$ space (Fig.~\ref{fig:main_figure_1}E, top panel). To further validate the numerical model and the underlying nonlinear polarization formalism, we compute the corresponding BFP images from simulations (Fig.~\ref{fig:main_figure_1}E, bottom panel). The excellent agreement between experiment and simulation confirms that the model accurately captures both the magnitude and angular characteristics of the SH emission.

Taken together, the consistency between (i) total scattered intensity and (ii) angle-resolved emission provides strong validation of the combined numerical and theoretical framework developed in this work. In particular, the ability to reproduce the full angular emission pattern demonstrates that the model reliably captures the spatial distribution of nonlinear sources and their coupling to the far field.

\begin{figure}[h!]
    \centering
    \includegraphics[width=0.6\linewidth]{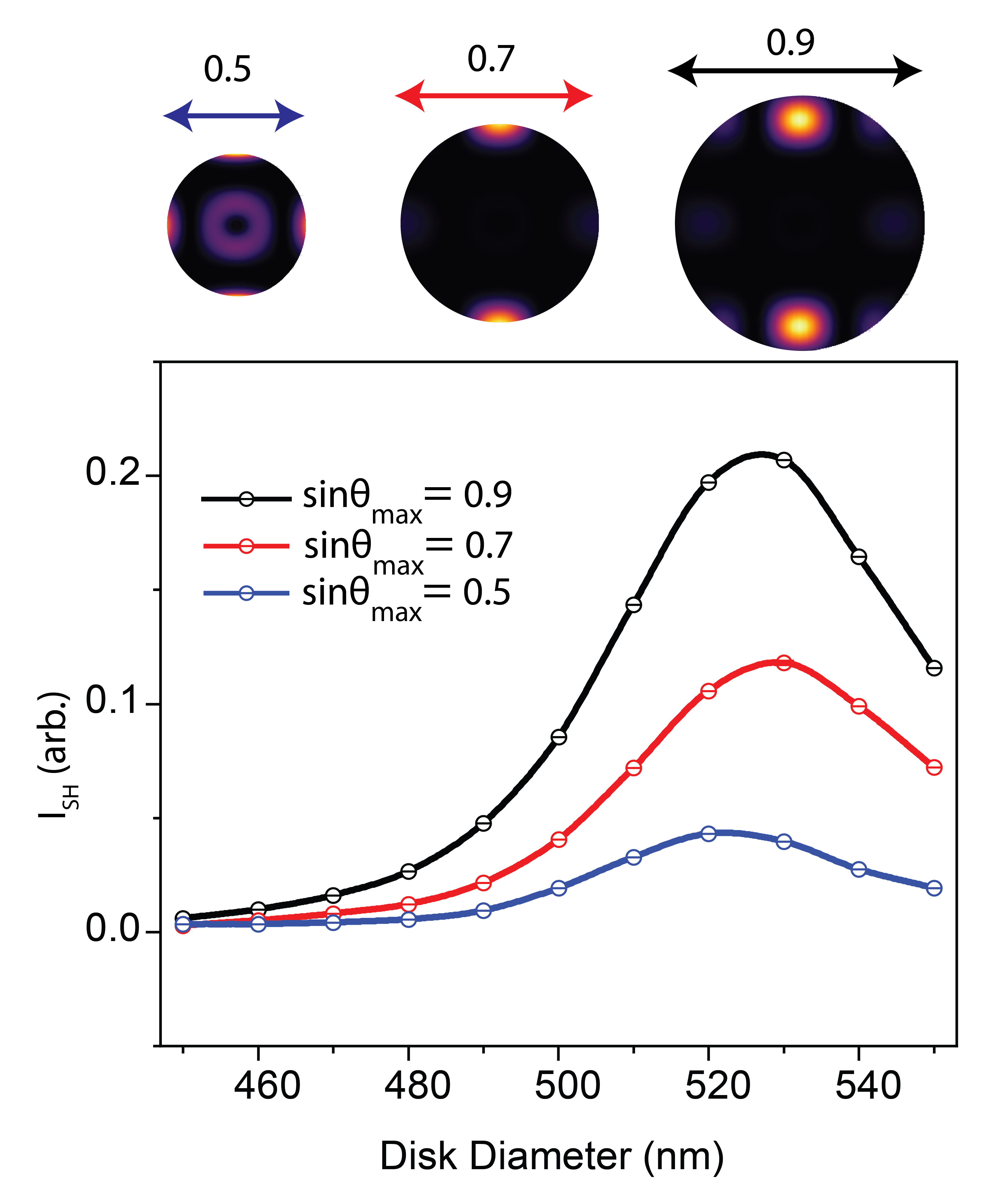}
    \caption{ \textbf{Second harmonic emission from nanostructured interface.}  Simulated SH intensity as a function of disk diameter for different values of NA. The corresponding back focal plane images. The data for NA=0.5 is multiplied by 5.}
    \label{fig:BFP_image_Polarisation_experiment_Simulation}
\end{figure}

\subsection{Nanofabrication of a-Si:H Nanopillar Arrays}\label{ch4_fabrication}

Hydrogenated amorphous silicon (a-Si:H) nanopillar arrays were fabricated on fused silica substrates using electron-beam lithography (EBL) and reactive-ion etching. Fused silica wafers were first cleaned in piranha solution, followed by deposition of a 430 nm-thick a-Si:H layer via plasma-enhanced chemical vapor deposition (PECVD, Corial D250L) at a substrate temperature of 280 $^\circ$C.
The wafers were diced into 20 $\times$20 mm chips, each of which was plasma-cleaned (Tepla 300, 2 min) and hard-baked at 180 $^\circ$C (5 min). A bilayer resist stack was spin-coated using a Sawatec SM-150 system: ZEP 520A (50 \%, 2000 rpm, ~150 nm thickness) was applied and baked, followed by Electra 92 conductive polymer at 1000 rpm, which was left to air dry without post-bake. Nanopillar arrays were patterned by exposing the surrounding regions to negative exposure using a Raith EBPG5000+ EBL system operating at 100 keV, with exposure doses between 150 and 300 $\mu$C/cm$^2$.
The resist was developed in amyl acetate at room temperature (90 s), rinsed in 90:10 methyl isobutyl ketone (MIBK):isopropanol, and dried under nitrogen. Pattern transfer into the a-Si:H layer was performed by inductively coupled plasma etching (AMS 200, 0  $^\circ$C, 3.5 min). The etch selectivity was $>$10:1 between ZEP and a-Si:H, and $ >$20:1 between a-Si:H and fused silica, enabling complete etching of the silicon layer. The residual resist was removed by high-power plasma ashing (Tepla Gigabatch, 5 min).
The final structures consisted of nanopillars arranged in a square lattice with 800 nm pitch, 500 nm diameter, and 430 nm height. The dose sweep during EBL allowed systematic tuning of the pillar diameter. Structural quality and fidelity were verified using scanning electron microscopy (SEM), with samples grounded by surrounding copper tape to minimize charging effects.
\subsection{Enhancing sensitivity for monitoring interfacial susceptibility}\label{sensitivity_analysis}

The second-harmonic (SH) intensity can be expressed as
\begin{equation}
I_{\text{SH}} \propto E(\omega)^4 \chi^2
\end{equation}
where $E(\omega)$ is the fundamental electric field at the interface and $\chi$ is the effective nonlinear susceptibility. Taking the logarithm of both sides yields
\begin{equation}
\log I_{\text{SH}} = 4 \log E(\omega) + 2 \log \chi
\end{equation}

Assuming that the fundamental field $E(\omega)$ remains unchanged for small variations in susceptibility, the fractional change in SH intensity can be written as
\begin{equation}
\frac{\Delta I_{\text{SH}}}{I_{\text{SH}}} = 2 \frac{\Delta \chi}{\chi}
\end{equation}
This relation provides a direct link between measurable intensity changes and variations in interfacial susceptibility.

Rearranging, the change in susceptibility is given by
\begin{equation}
\Delta \chi = \frac{\chi}{2 I_{\text{SH}}} \Delta I_{\text{SH}} 
= \frac{1}{2 E(\omega)^4 \chi} \Delta I_{\text{SH}}
\end{equation}

To quantify the sensitivity enhancement enabled by nanostructuring, we compare the susceptibility changes for a silicon nanodisk (SiND) array and a planar film:
\begin{equation}
\frac{(\Delta \chi)_{\text{SiND}}}{(\Delta \chi)_{\text{film}}}
=
\frac{\left( \frac{1}{E(\omega)^4 \chi} \Delta I_{\text{SH}} \right)_{\text{SiND}}}
{\left( \frac{1}{E(\omega)^4 \chi} \Delta I_{\text{SH}} \right)_{\text{film}}}
\end{equation}

For a given detection system, the minimum measurable change in intensity $\Delta I_{\text{SH}}$ is fixed by the detector sensitivity. Furthermore, the intrinsic susceptibility is comparable for the two cases, i.e., $(\chi)_{\text{SiND}} \sim (\chi)_{\text{film}}$. Under these conditions, the sensitivity ratio simplifies to
\begin{equation}
\frac{(\Delta \chi)_{\text{SiND}}}{(\Delta \chi)_{\text{film}}}
\approx
\left( \frac{E(\omega)_{\text{film}}}{E(\omega)_{\text{SiND}}} \right)^4
\end{equation}

Thus, the sensitivity is governed by the fourth power of the local field enhancement. In our system, the maximum local field enhancement in the SiND array is $\sim 5$, with a slightly lower surface-averaged value. Consistent with the experimentally observed $\sim 200$-fold increase in SH intensity, we obtain
\begin{equation}
(\Delta \chi)_{\text{SiND}} \approx \frac{1}{200} (\Delta \chi)_{\text{film}}
\end{equation}

This analysis demonstrates that nanostructuring significantly enhances the sensitivity for detecting changes in interfacial susceptibility. Specifically, the same detectable change in SH intensity corresponds to a $\sim 200$-times smaller variation in $\chi$ for the nanostructured system compared to a planar interface, enabling the resolution of subtle interfacial processes that would otherwise remain undetectable.

\subsection{Linear optical characterisation}\label{reflectance_measurement}

\begin{figure}
    \centering
    \includegraphics[width=1\linewidth]{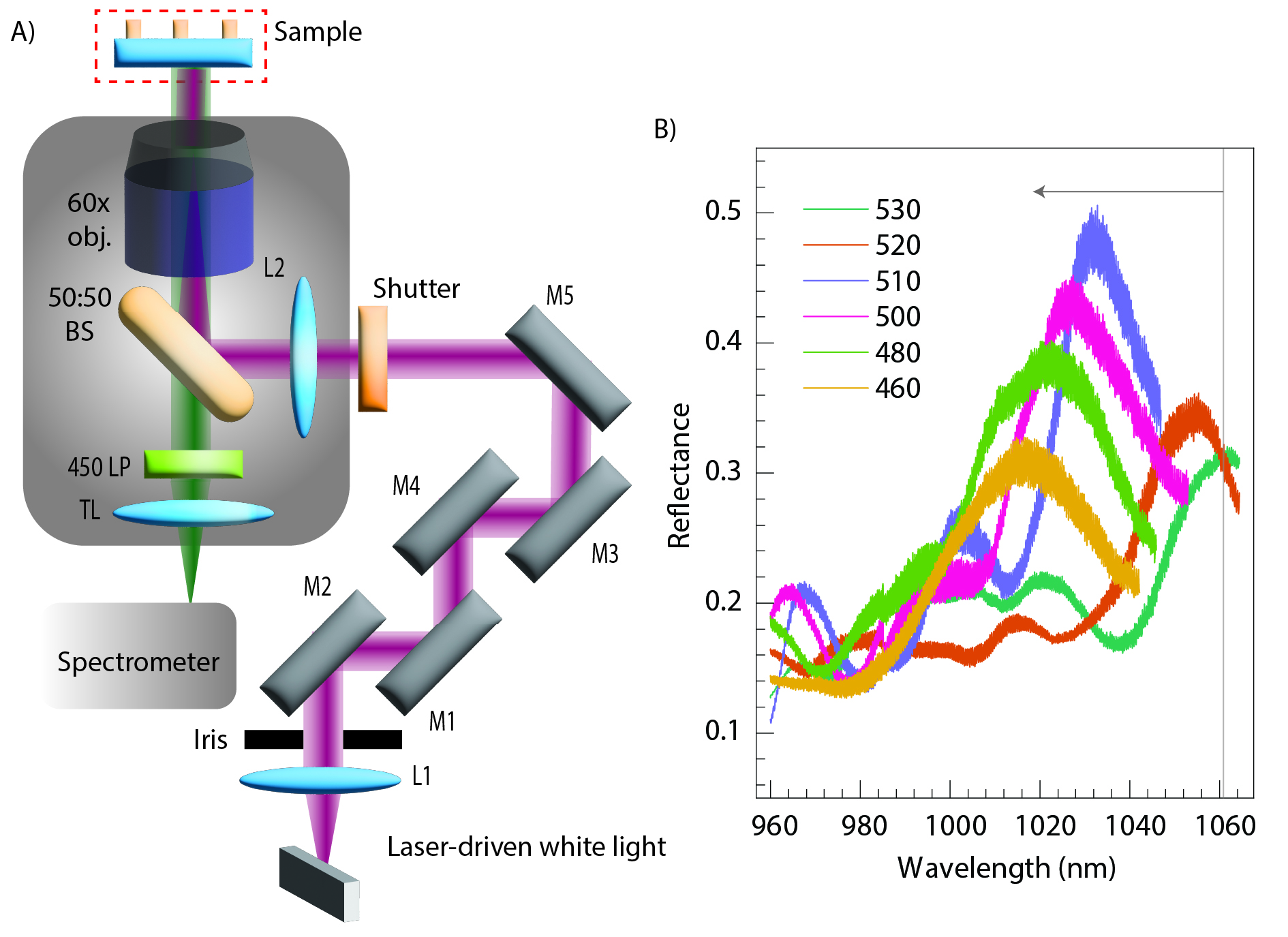}
    \caption{\textbf{Optical setup with details of the illumination and the detection paths for reflectance measurements.} A) In the setup, the red-shaded beam represents the incident light, and the green-shaded beam represents the reflected light from the sample. BS = BeamSplitter, L = Lens, M=Mirror, LP= Longpass filter. B) Reflectance spectra of the silicon nanodisk array for different diameters. The arrow indicates the blue shift in the spectra with increasing disk diameter.}
    \label{fig:ch4_linear_ref_Tr}
\end{figure}

The linear reflectance and transmittance spectra of the nanostructured samples were measured using the same inverted microscope and spectrometer employed for the nonlinear experiments. Illumination was provided by a fiber-coupled, broadband laser-driven white light source (Energetiq LDLS), which was focused onto the back focal plane (BFP) of a long working distance, high numerical aperture (NA) objective (Nikon 60$\times$, NA = 0.7). This configuration enables collimated illumination of the sample from below, ensuring uniform excitation over the region of interest (Fig.~\ref{fig:ch4_linear_ref_Tr}A).

The reflected light collected by the objective was directed to the spectrometer for spectral analysis. Absolute reflectance was obtained by normalizing the measured spectra to a calibrated silver mirror reference (Thorlabs, PF10-03-P01). A systematic background correction was applied to all measurements to remove instrumental noise and spurious signals, thereby improving the accuracy and reproducibility of the spectra.

Representative reflectance spectra for nanodisk arrays with fixed periodicity ($p = 800$~nm) and height (440~nm), and varying disk diameters, are shown in Fig.~\ref{fig:ch4_linear_ref_Tr}B.

\subsection{Estimating interfacial potential and susceptibility}\label{surface_potential_derivation}

In this section, we derive analytical expressions to extract the interfacial potential from measured SH intensities. Two experimentally relevant scenarios are considered: (i) when only the amplitude of the SH signal is available, and (ii) when both amplitude and phase information are accessible. In the former case, the extraction of the interfacial potential requires prior knowledge (or an assumption) of the surface susceptibility, whereas in the latter case, both the interfacial potential and the susceptibility can be determined independently.

\vspace{0.5em}
\noindent
\textbf{Normalization and reference scheme.}  
Quantitative extraction of interfacial properties requires an appropriate referencing scheme to isolate the intrinsic nonlinear response. In planar SHG, this is typically achieved using a nonlinear reference (e.g., $\alpha$-quartz) together with Fresnel corrections \cite{alghamdi_temperature_2025, speelman_quantifying_2025}. 

In nanostructured systems, however, both incoupling and outcoupling factors are geometry-dependent and must be explicitly accounted for. To address this, we normalize the SH response of the SiND array in electrolyte to that in air:
\begin{equation}
    I_{\text{SH}} \propto \eta_{\text{out}}^{2\omega} |\cplx{I}_{\text{S}}|^2 
    \left| \chi^2_{\text{Si,eff}} + \chi^2_{\text{Stern}} 
    + \Delta\Phi_{\text{SCL}}\chi^3_{\text{Si,eff}} \cplx{G} 
    + \Delta\Phi_{\text{EDL}}\chi^3_{\text{H$_2$O}} \cplx{F} \right|^2
\end{equation}

In air (absence of electrolyte), this reduces to
\begin{equation}
    I_{\text{SH, air}} \propto \eta_{\text{out, air}}^{2\omega} |\cplx{I}_{\text{S, air}}|^2 
    \left| \chi^2_{\text{Si,0}} 
    + \frac{\Delta\Phi_{\text{SCL}}}{\delta_{\text{SCL}}}\chi^3_{\text{Si}} 
    \frac{\cplx{I}_{\text{SCL,air}}}{\cplx{I}_{\text{S,air}}} \right|^2
\end{equation}

Here, $\eta_{\text{out}}^{2\omega}$ accounts for geometry-dependent outcoupling factor, while incoupling factor differences (similar to Fresnel factors) are captured through the overlap integrals $|\cplx{I}_{\text{S}}|$ and $|\cplx{I}_{\text{S,air}}|$. Note that this normalization scheme is not the only way, and one could also employ a non-linear crystal, such as quartz, with known $\chi^2$ values, as in SHG studies of planar systems. But in that case, one has to carefully consider the near-field distribution in the SH signal from the nanostructured case via the overlap integral. 

\begin{figure}[h!]
    \centering
    \includegraphics[width=0.6\linewidth]{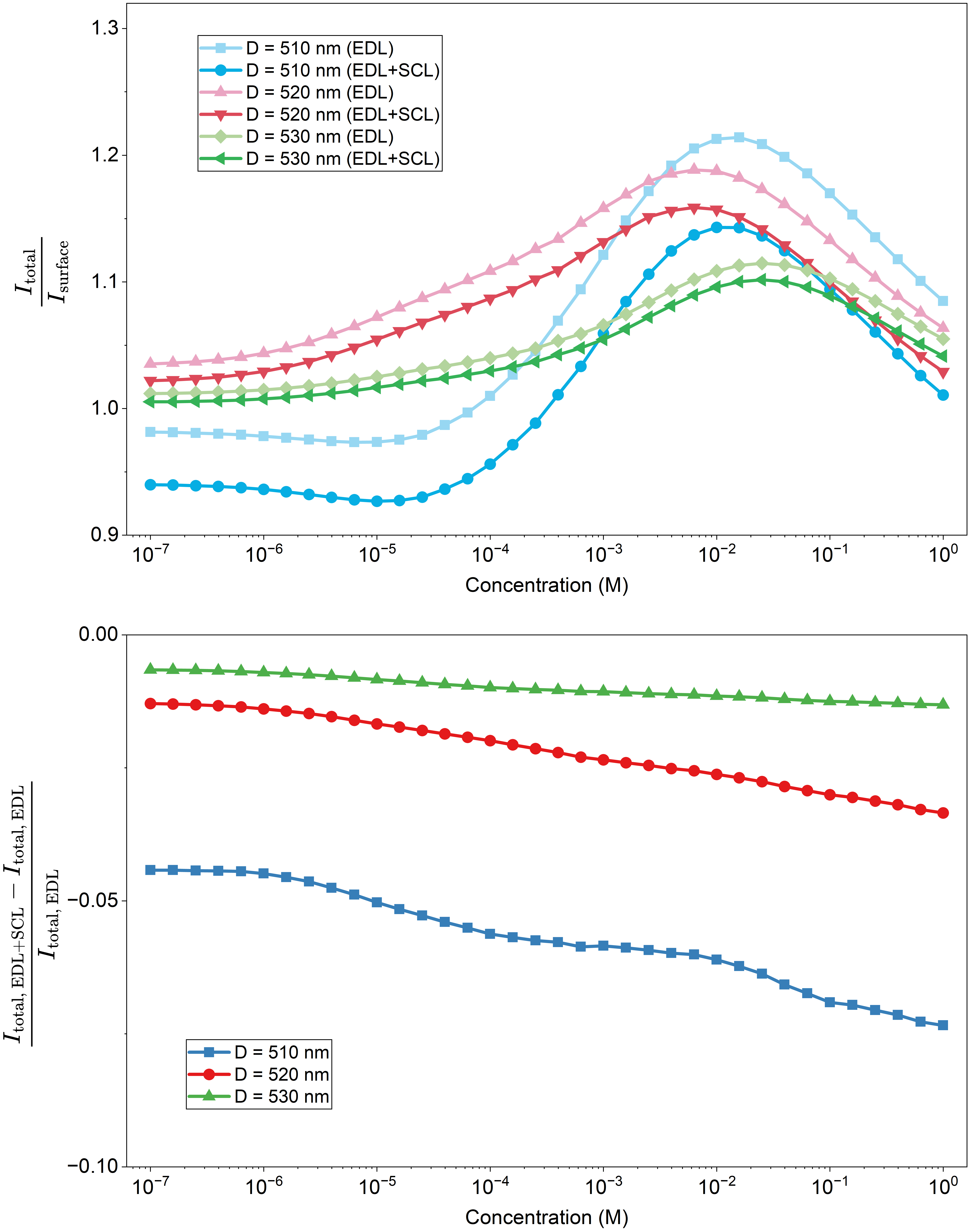}
    \caption{ \textbf{Normalized SH intensity comparison with and without the space charge layer contribution for the three representative disk diameters.}\\
$I_{\text{total, EDL+SCL}} \propto 
\left|
\mathcal{I}_{\text{S}}\ \chi^{2}_{\text{Si,eff}} + \mathcal{I}_{\text{S}}\  \chi^{2}_{\text{Stern}}
+\mathcal{I}_{\text{SCL}}  \, \chi^{3}_{\text{Si,eff}} 
\frac{\Delta\Phi_{\text{SCL}}}{\delta_{\text{SCL}}}
+\mathcal{I}_{\text{EDL}} \, \chi^{3}_{\text{H$_2$O}}
\frac{\Delta\Phi_{\text{EDL}} }{\delta_{\text{EDL}}}
\right|^2$ \\ 
$I_{\text{total, EDL}} \propto 
\left|
\mathcal{I}_{\text{S}}\ \chi^{2}_{\text{Si,eff}} + \mathcal{I}_{\text{S}}\  \chi^{2}_{\text{Stern}}
+\mathcal{I}_{\text{EDL}} \, \chi^{3}_{\text{H$_2$O}}
\frac{\Delta\Phi_{\text{EDL}} }{\delta_{\text{EDL}}}
\right|^2$ \\ 
$I_{\text{surface}} \propto 
\left|
\mathcal{I}_{\text{S}}\ \chi^{2}_{\text{Si,eff}} + \mathcal{I}_{\text{S}}\ \chi^{2}_{\text{Stern}}
\right|^2$\\
For calculation we have taken $\chi_{\text{Si,eff}}^2 =1,\ \chi_{\text{Stern}}^2=1,\ \chi_{\text{H$_2$O}}^3=1,\  \chi_{\text{Si,eff}}^3/\chi_{\text{Si,eff}}^2=2$ \cite{zhao_contactless_2024}. For computing the overlap integrals, we have taken $\chi_{\text{r}}^2=0,\ \chi_{\text{r}}^3=0$}
\label{fig:ISH_EDL_SCL_comparsion_Ratio_FandG}
\end{figure}

\vspace{0.5em}
\noindent
\textbf{Simplified regime.}  
Under flatband conditions ($\Delta\Phi_{\text{SCL}} \approx 0$), or when the SCL contribution is negligible compared to the EDL contribution, the term involving $\cplx{G}$ can be ignored (Fig.~\ref{fig:ISH_EDL_SCL_comparsion_Ratio_FandG}). The expressions simplify to:
\begin{equation}\label{eq:ch4_amplitude-SH-No-Electrolyte}
    I_{\text{SH, air}} \propto \eta_{\text{out, air}}^{2\omega} |\cplx{I}_{\text{S, air}}|^2 (\chi^2_{\text{Si,0}})^2
\end{equation}
\begin{equation}\label{eq:ch4_amplitude-SH-Electrolyte}
    I_{\text{SH}} \propto \eta_{\text{out}}^{2\omega} |\cplx{I}_{\text{S}}|^2 
    \left| \chi^2_{\text{s}} + \Delta\Phi_{\text{EDL}}\chi^3_{\text{H$_2$O}} \cplx{F} \right|^2
\end{equation}

We define the normalized SH response as
\begin{equation}
    R_{\text{SH}} = \sqrt{\frac{I_{\text{SH}}}{I_{\text{SH, air}}}}
\end{equation}

\vspace{0.5em}
\noindent
\textbf{Case 1: Amplitude-only measurements.}  
When only the amplitude of the SH signal is available,
\begin{equation}\label{eq:ratio_SH_amplitude}
    R_{\text{SH}} =
    \sqrt{\frac{\eta_{\text{out}}^{2\omega}}{\eta_{\text{out, air}}^{2\omega}}}
    \left|\frac{\cplx{I}_{\text{S}}}{\cplx{I}_{\text{S, air}}}\right|
    \left|
    \frac{\chi^2_{\text{s}}}{\chi^2_{\text{Si,0}}}
    + \frac{\chi^3_{\text{H$_2$O}}}{\chi^2_{\text{Si,0}}}\cplx{F}\,\Delta\Phi_{\text{EDL}}
    \right|
\end{equation}

Defining
\begin{equation}
\eta =
\sqrt{\frac{\eta_{\text{out, air}}^{2\omega}}{\eta_{\text{out}}^{2\omega}}}
\left|\frac{\cplx{I}_{\text{S, air}}}{\cplx{I}_{\text{S}}}\right|,
\end{equation}
we obtain a quadratic equation in $\Delta\Phi_{\text{EDL}}$, which can be solved analytically. 

 \begin{equation}
\left(\frac{\chi^3_{\text{H$_2$O}}}{\chi^2_{\text{Si,0}}}\right)^2 |\cplx{F}|^2 \, (\Delta\Phi_{\text{EDL}})^2
+ 2\left(\frac{\chi^2_{\text{s}}}{\chi^2_{\text{Si,0}}}\right)
\left(\frac{\chi^3_{\text{H$_2$O}}}{\chi^2_{\text{Si,0}}}\right)
\Re(\cplx{F})\, \Delta\Phi_{\text{EDL}}
+ \left[\left(\frac{\chi^2_{\text{s}}}{\chi^2_{\text{Si,0}}}\right)^2
- (\eta R_{\text{SH}})^2\right] = 0 .
\end{equation}

where
\begin{equation}
\cplx{F}= \Re(\cplx{F}) + i \Im(\cplx{F}), \qquad \cplx{F} \in \mathbb{C}
\end{equation}
The solution of the above equation
\begin{equation}
\Delta\Phi_{\text{EDL}}
=
\frac{
-\left(\frac{\chi^2_{\text{s}}}{\chi^2_{\text{Si,0}}}\right)\Re(\cplx{F})
\ \pm\
\sqrt{
(\eta R_{\text{SH}})^2 |\cplx{F}|^2
-
\left(\frac{\chi^2_{\text{s}}}{\chi^2_{\text{Si,0}}}\right)^2 \Im(\cplx{F})^2
}
}{
\left(\frac{\chi^3_{\text{H$_2$O}}}{\chi^2_{\text{Si,0}}}\right) |\cplx{F}|^2
}.
\end{equation}

Further simplified to:
\begin{equation}
\Delta\Phi_{\text{EDL}}
=\frac{\chi^2_{\text{s}}}{\chi^3_{\text{H}_2{\text{O}}}}
\frac{
-\Re(\cplx{F})
\ \pm\
\sqrt{
(\eta R_{\text{SH}}\ \chi^2_{\text{Si,0}}/\chi^2_{\text{s}})^2 |\cplx{F}|^2
-
\Im(\cplx{F})^2
}
}{
 |\cplx{F}|^2
}
\end{equation}

This approach provides a route to extract the interfacial potential; however, it relies on prior knowledge or assumptions regarding $\chi^2_{\text{s}}$, since only a single observable (intensity) is available.

\vspace{0.5em}
\noindent
\textbf{Case 2: Amplitude and phase measurements.}  
If both amplitude ($R_{\text{SH}}$) and phase ($\phi_{\text{SH}}$) are measured, the SH response becomes
\begin{equation}\label{eq:ratio_SH_amplitude_phase}
\eta R_{\text{SH}} e^{i\phi_{\text{SH}}} =
\frac{\chi^2_{\text{s}}}{\chi^2_{\text{Si,0}}}
+\frac{\chi^3_{\text{H$_2$O}}}{\chi^2_{\text{Si,0}}} \Delta\Phi_{\text{EDL}} \cplx{F}
\end{equation}

Separating real and imaginary parts yields two independent equations:
\begin{equation}
    \Delta \Phi_{\text{EDL}} =
    \frac{\chi_{\text{Si,0}}^2 \eta R_{\text{SH}}\sin{\phi_{\text{SH}}}}
    {\chi^3_{\text{H$_2$O}} \Im(\cplx{F})}
\end{equation}
\begin{equation}
    \chi_s^2 =
    \chi_{\text{Si,0}}^2 \eta R_{\text{SH}}\cos{\phi_{\text{SH}}}
    - \frac{\Re(\cplx{F})}{\Im(\cplx{F})}
    \chi_{\text{Si,0}}^2 \eta R_{\text{SH}}\sin{\phi_{\text{SH}}}
\end{equation}

Thus, phase-resolved SHG enables independent determination of both interfacial potential and susceptibility.

\vspace{0.5em}
\noindent
\textbf{Discussion.}  
Here, $\chi_{\text{Si,0}}^2$ represents the intrinsic surface susceptibility in the absence of electrolyte, while $\chi_{\text{s}}^2$ includes contributions from both the solid and interfacial liquid (Stern layer). Changes in electrolyte conditions can modify $\chi_{\text{s}}^2$ through surface reactions such as protonation/deprotonation. 

While the amplitude-only approach provides limited information, phase-resolved measurements offer a complete description of the interfacial response. Extending phase-sensitive SHG techniques to nanostructured interfaces and integrating them with the present formalism will be an important direction for future work.

\subsection{Temperature dependent equilibrium constant}\label{temperature_dependent_equilibrium_constant}
In our system, the observed linear relationship between temperature and surface charge arises from the thermally driven shift in the equilibrium of surface ionization reactions. Specifically, temperature increases promote the dissociation of surface hydroxyl groups, for example:  
\begin{equation}\nonumber
    \equiv \mathrm{SiOH} \;\rightleftharpoons\; \equiv \mathrm{SiO}^- + \mathrm{H}^+
\end{equation}

This thereby increases the surface charge density. The equilibrium constant can be expressed as  
\begin{equation}\nonumber
    K_a = \frac{[\mathrm{SiO}^-][\mathrm{H}^+]}{[\mathrm{SiOH}]} 
     = \frac{\sigma_{\text{ox-el}} [\mathrm{H}^+]}{\Gamma - \sigma_{\text{ox-el}}}
\end{equation}
The expression for surface charge in terms of surface proton concentration is given by  
\begin{equation}\nonumber
    \sigma_{\text{ox-el}} = \frac{-e \Gamma}{1 + \frac{[\mathrm{H}^+]_s}{K_a}}
\end{equation}
We note that the temperature dependence of the ionization equilibrium is central to our analysis. In our study, we relate the temperature-dependent equilibrium constant \(K(T)\) to the thermodynamic parameter enthalpy \(\Delta H\) via the Van’t Hoff equation:  
\begin{equation}\nonumber
    \ln K_1 - \ln K_2 = -\Delta H \left(\frac{1}{RT_1} - \frac{1}{RT_2}\right)
\end{equation}
Equivalently, this can be written in the form: 
\begin{equation}\label{temp_equilibrium_constant}
    K = K_0 \exp \left[ -\frac{\Delta H}{R} \left( \frac{1}{T} - \frac{1}{T_0} \right) \right]
\end{equation}

\subsection{Phase factor formulation in planar and nanostructured surface}\label{section:_SI_phase_factor_formulation}
The effective second-order susceptibility $\chi^{2}_{\text{eff}}$ at a charged interface is the sum of the intrinsic surface contribution and the potential-dependent bulk contribution:
\begin{equation}
    \chi^{2}_{\text{eff}} = \chi^{2}_{\text{s}} + \chi^{3}_{\text{H$_2$O}} \Delta \Phi_{\text{EDL}} \mathcal{F}_{\text{planar}}(\kappa, \Delta k)
\end{equation}

\textbf{The Phase Factor}
Following the work of \textit{Gonella and Roke} \cite{gonella_second_2016}, the interference between the surface and the bulk is modulated by the phase factor $F$. This factor accounts for the mismatch between the coherence length and the Debye length:
\begin{equation}
    \mathcal{F}_{\text{planar}}(\kappa, \Delta k) = \frac{\kappa}{\kappa + i \Delta k}
\end{equation}
Where:
\begin{itemize}
    \item $\kappa$ is the inverse Debye length (reciprocal of the EDL thickness).
    \item $\Delta k = k_{2\omega} - 2k_{\omega}$ is the wave vector mismatch.
\end{itemize}

\textbf{Second Harmonic Intensity}
The measured SHG intensity $I(2\omega)$ is proportional to the square of the total susceptibility:
\begin{equation}
    I(2\omega) \propto \left| \chi^{2}_{\text{s}} + \chi^{3}_{\text{H$_2$}O} \Delta \Phi_{\text{EDL}} \left( \frac{\kappa}{\kappa + i \Delta k} \right) \right|^2 I(\omega)^2
\end{equation}

Expanding the intensity to highlight the interference term:
\begin{equation}
    I(2\omega) \propto |\chi^{2}_{\text{s}}|^2 + |\chi^{3}_{\text{H$_2$}O} \Delta \Phi_{\text{EDL}} \mathcal{F}_{\text{planar}}|^2 + 2 \text{Re} \left[ \chi^{2}_{\text{s}} (\chi^{3}_{\text{H$_2$O}} \Delta \Phi_{\text{EDL}} \mathcal{F}_{\text{planar}})^* \right]
\end{equation}
The normalized SH intensity is obtained  by dividing the total SH intensity by the SH intensity contributed only by surface i.e when $\mathcal{F}_{\text{planar}}=0$ \cite{gonella_second_2016}
\begin{equation}
    I_{\mathrm{SH, norm}} \propto \left|1+(\chi_{\text{H$_2$O}}^3/\chi_{\text{s}}^2)\Delta\Phi_{\mathrm{EDL}} \mathcal{F}\right|^2
\end{equation}
For nanostructures, within the framework proposed in this work, the factor $\mathcal{F}$ is tunable by $\alpha$ and $\beta$. For understanding how the factor $\mathcal{F}$ and $I_{\text{SH, norm}}$ evolves under varying $\alpha$ and $\beta$, we start with the depth-resolved overlap
\begin{equation}
\mathcal{I}_{\mathrm{S}}(r_{\text{n}})
=
\mathcal{I}_{\mathrm{S}}(0)\,e^{-\alpha r_{\text{n}}}e^{i\beta r_{\text{n}}},
\end{equation}
where $\alpha$ and $\beta$ represent the effective attenuation and phase accumulation constants, respectively.
which leads to the normalized factor
\begin{equation}
\mathcal{F}
=
\frac{1}{1+\alpha\delta_{\mathrm{EDL}}-i\beta\delta_{\mathrm{EDL}}}.
\end{equation}
Unlike planar systems, where the phase factor $F$ depends solely on EDL thickness, $\mathcal{F}$ here depends sensitively on both nanostructures geometry (through $\alpha$ and $\beta$) and electrolyte properties (through $\delta_{\mathrm{EDL}}$). This enables tunable interference between surface and EDL contributions.

\begin{figure}[h!]
    \centering
    \includegraphics[width=0.8\linewidth]{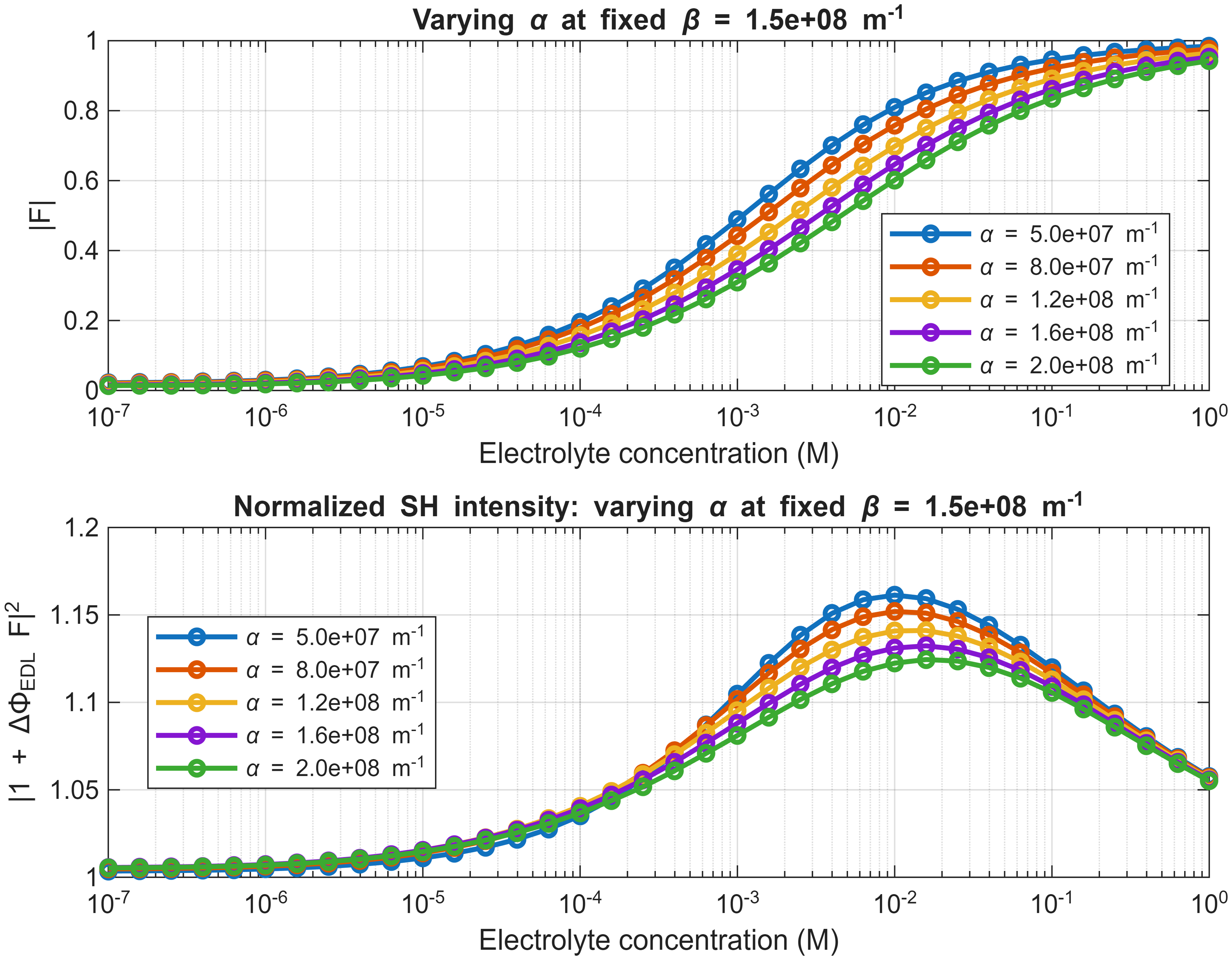}
    \caption{ \textbf{Variation of $|\mathcal{F}|$ and $I_{\text{SH,norm}}$ at varying $\alpha$ and fixed $\beta$.} For calculation, we have taken $\chi_{\text{s}}^2=\chi_{\text{H$_2$O}}^3=1$}
    \label{fig:F_ISH_beta_fixed}
\end{figure}
\begin{figure}[h!]
    \centering
    \includegraphics[width=0.8\linewidth]{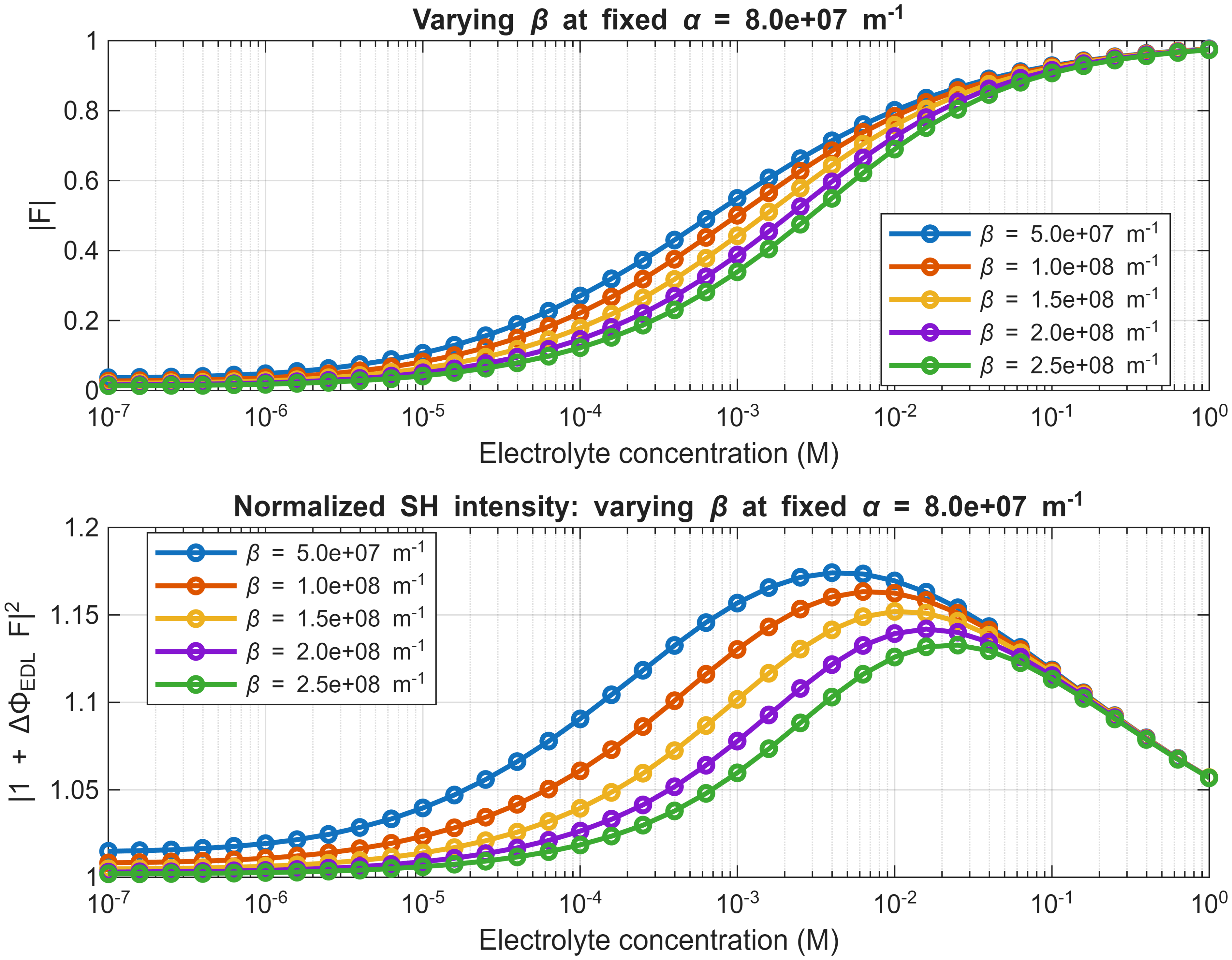}
    \caption{ \textbf{Variation of $|\mathcal{F}|$ and $I_{\text{SH,norm}}$ at varying $\beta$ and fixed $\alpha$.} For calculation, we have taken $\chi_{\text{s}}^2=\chi_{\text{H$_2$O}}^3=1$}
    \label{fig:F_ISH_alpha_fixed}
\end{figure}
\begin{figure}[h!]
    \centering
    \includegraphics[width=0.6\linewidth]{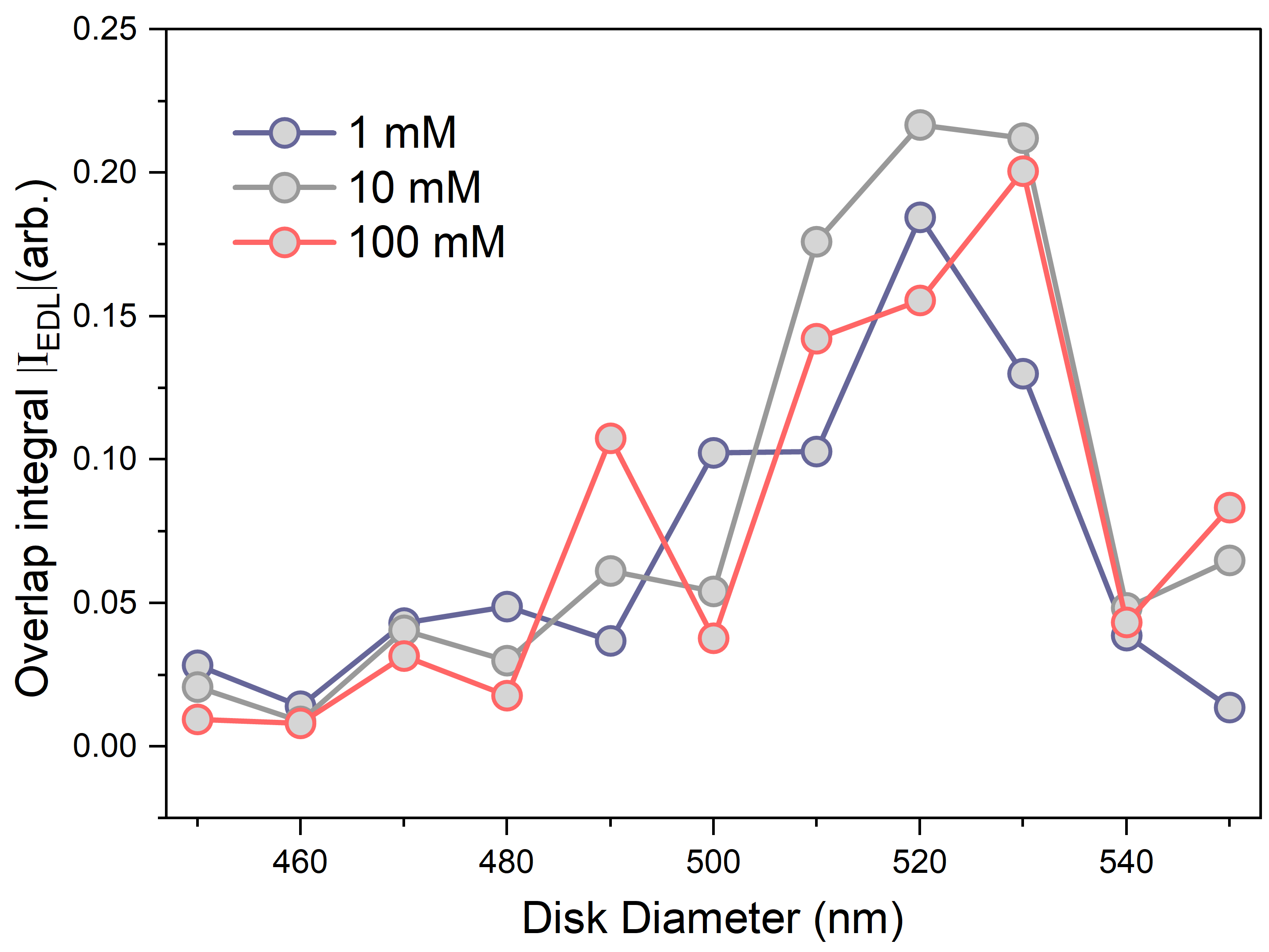}
    \caption{ \textbf{Overlap integral $\mathcal{I}_{\text{EDL}}$ with diameter and for different concentrations} obtain from the simulation. The EDL thickness $\delta_{\text{EDL}}$ was varied as a function of electrolyte concentration according to Eq.~\ref{eq:Debye_Length_expression}. For calculation, we have taken $\chi_{\text{H$_2$O}}^3$=1 and $\chi_{\text{r}}^3=0$}.
    \label{fig:Overlap_integral_different_concentrations}
\end{figure}

\begin{figure}[h!]
    \centering
    \includegraphics[width=0.6\linewidth]{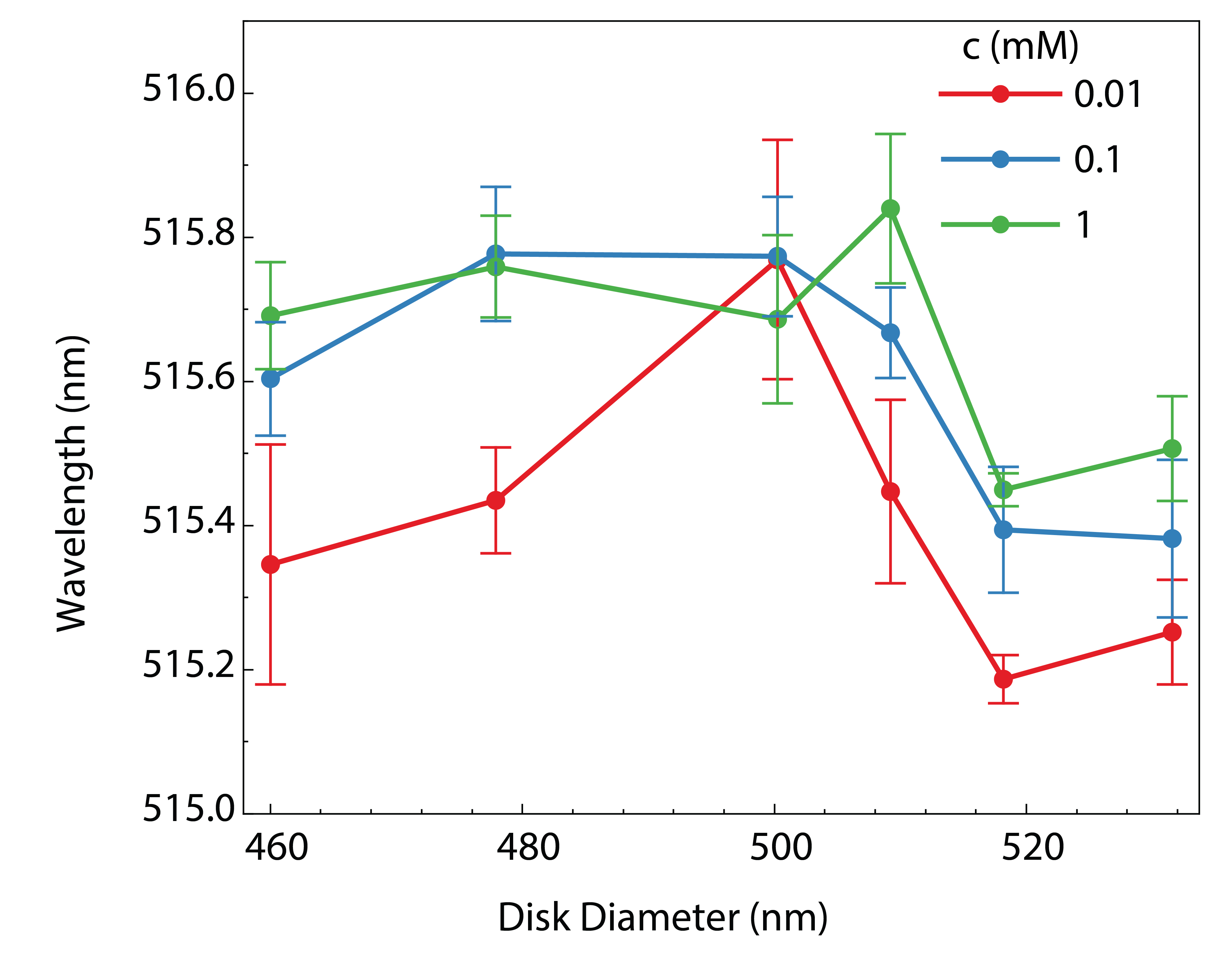}
    \caption{ \textbf{Spectral position of the SH spectra} obtained by the excitation of FW laser centered at 1030 nm with varying electrolyte concentration and diameter. The error bars represent the standard deviation of $3\times5$ repeated measurements performed at 5 different FW laser powers, as shown in Fig.~\ref{fig:main_figure_2}C and D. }
    \label{fig:SI_spectra_shift_electrolyte_concentration_diameter}
\end{figure}

\subsection{Mathematical framework for calculating EDL and SCL field}\label{section:SI_EDL_SCL_field_capacitive _interface}
\begin{figure}[h!]
    \centering
    \includegraphics[width=0.75\linewidth]{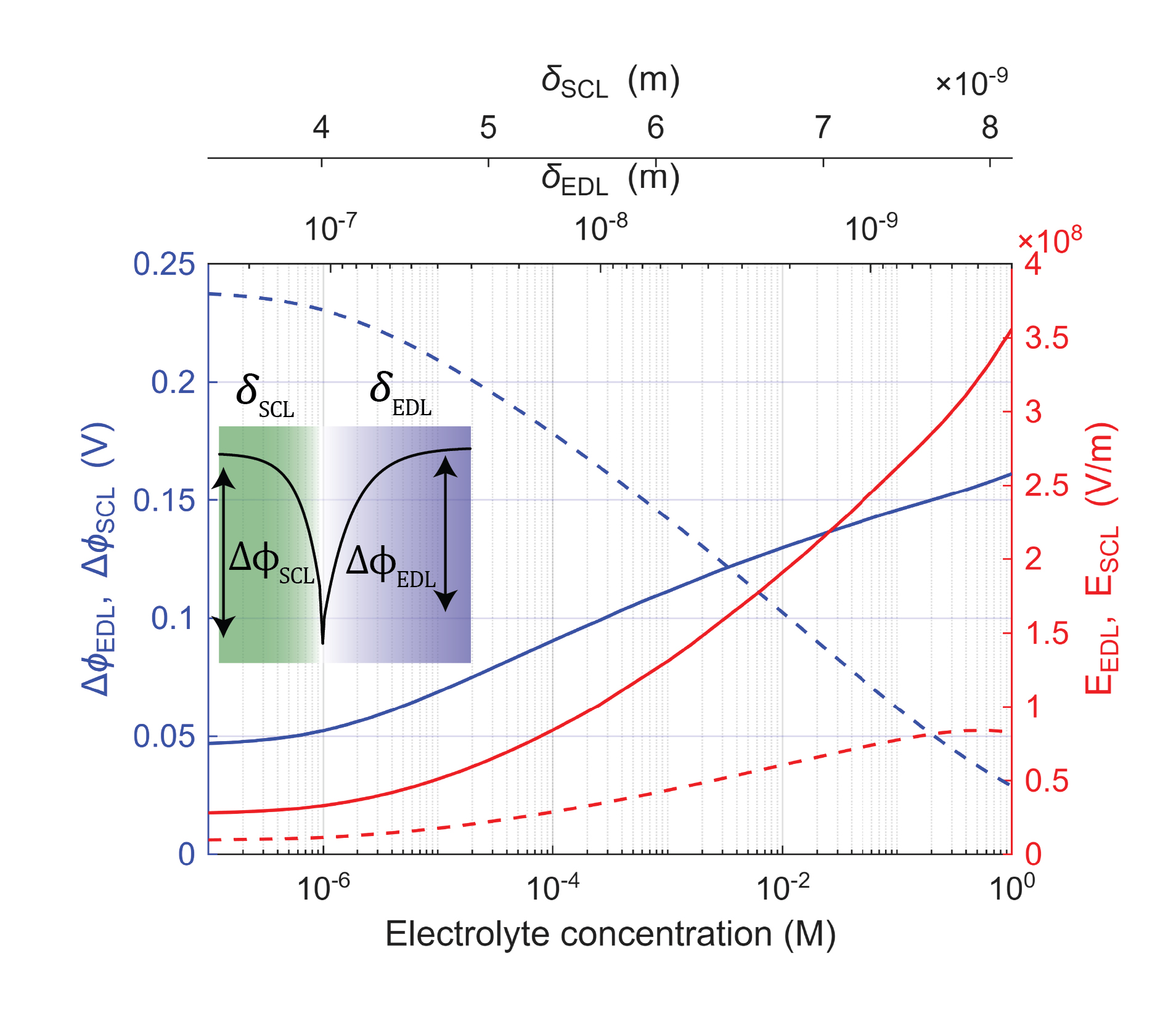}
    \caption{ \textbf{Electric field, potential and  field-decay length in the EDL and SCL as a function of electrolyte concentration.} Blue dashed line: EDL potential, blue solid line: SCL potential. Red dashed line: EDL electric field at the oxide-electrolyte interface , red solid line: electric field at the Si-oxide interface. The decay lengths are represented on the top axis.  For the above calculation we assumed $\sigma_{\text{Si-ox}}=0$. The inset shows a potential distribution for a representative case of 1 mM aqueous KCl concentration. }
    \label{fig:EDL and SCL E-field_potential_Debye_Length}
\end{figure}

As reported in our prior works \cite{anwar_salinity-dependent_2024, anwar_enhancing_2026}, we first obtained the electrical double layer potential $(\Delta \Phi_{\text{EDL}})$, E-field $(E_{\text{EDL}})$ and decay length $(\delta_{\text{EDL}})$ ,  as a function of electrolyte concentration  shown in Fig.~\ref{fig:EDL and SCL E-field_potential_Debye_Length}. 

We developed a 3D numerical model to solve the Nernst--Planck--Poisson equation to determine the equilibrium distribution of ions and resulting electrostatic potentials. For modeling, we considered different modes of ion transport, using the Nernst--Planck equation for dilute species and the Poisson--Boltzmann equation for the equilibrium distribution of ions. 
\paragraph{Governing equations:}

\begin{equation}
\nabla \cdot J_i + \mathbf{U} \cdot \nabla c_i = 0
\end{equation}

\begin{equation}
J_i = -D \nabla c_i - z_i \mu_i^{m} F c_i \nabla \Phi
\end{equation}

\begin{equation}
\nabla^2 \Phi
=
-\frac{1}{\varepsilon_0 \varepsilon_{\text{r}}}
\sum_i
F z_i c_i
\exp\!\left(
-\frac{e z_i \Phi}{k_B T_S}
\right)
\end{equation}

We used surface charge density as the boundary condition, which is not a fixed value but is instead governed by the above equilibrium reaction. This results in a variable surface charge density that depends on the surface potential at the stern plane (and therefore the concentration-dependent change in Debye length), given by the following equation:
\begin{equation}
\sigma_{\text{ox-el}} =
\frac{-e\Gamma}{1 + \dfrac{[H^+]_S}{K_a}},
\quad \text{and}
\end{equation}
\begin{equation}
[H^+]_S = [H^+]_{\textit{bulk}}
\exp\!\left(
-\frac{e\Phi_S}{k_B T_S}
\right)
\end{equation}
In the above set of equations, $\Phi$ is the potential in the EDL region and $\Phi_S$ is the surface potential. $J_i, c_i, \mu_i^m, z_i$ are the ionic flux, ionic concentration, ionic mobility, and ionic valency, respectively, of the different ionic constituents in the electrolyte.

Using the boundary condition across the semiconductor-oxide electrolyte interface, in particular, the continuity of displacement field at the interface, taking into account the surface charges, we have,  

\begin{equation}
\varepsilon_{\text{Si}} E_{\text{Si}}(0^-) - \varepsilon_{\text{ox}} E_{\text{ox}}(0^+) = \sigma_{\text{Si-ox}}
\end{equation}

\begin{equation}
\varepsilon_{\text{ox}} E_{\text{ox}}(0^+) - \varepsilon_{\text{ox}} E_{\text{ox}}(t_{\text{ox}}^-) = 0
\end{equation}

\begin{equation}
\varepsilon_{\text{ox}} E_{\text{ox}}(t_{\text{ox}}^-) - \varepsilon_{\text{el}} E_{\text{el}}(t_{\text{ox}}^+) = \sigma_{\text{ox-el}}
\end{equation}
the above three relations allows one to link the $E_{\text{EDL}}$ and $E_{\text{SCL}}$ in terms of surface charges  as follows: 
\begin{equation}
    \varepsilon_{\text{Si}}E_{\text{SCL}}-\varepsilon_{\text{el}}E_{\text{EDL}}=\sigma_{\text{Si-ox}}+\sigma_{\text{ox-el}}
\end{equation}
where $\varepsilon_{\text{Si}},\varepsilon_{\text{ox}}$, and $\varepsilon_{\text{el}}$ are the dielectric constants of silicon, oxide, and electrolyte. $\sigma_{\text{Si-ox}} \text{ and }\sigma_{\text{ox-el}}$ are the surface charge at the two interfaces. The E-field at the the Si-oxide interface can be used to obtain $\Delta \Phi_{\text{SCL}}$ and $\delta_{\text{SCL}}$ \cite{memming_semiconductor_2015} in terms of doping $(N_A =  10^{24}-10^{25}\ \mathrm{m^{-3}}$)  of the semiconductor,
\begin{equation}\label{eq:SI_E-field_Potential_SCL}
E_{\text{SCL}}
=
\frac{k_B T}{e L_{D\text{,eff}}}
\,
\sqrt{
\exp\left(\frac{e\Delta\Phi_{\text{SCL}}}{k_B T}\right)
-1
-\frac{e\Delta\Phi_{\text{SCL}}}{k_B T}
}
\end{equation}
\begin{equation}\label{eq:SI_Decay_length_potential_SCL}
\delta_{\text{SCL}}
=
2 L_{D,\mathrm{eff}}
\sqrt{
\frac{e\Delta\Phi_{\text{SCL}}}{k_B T}
-1
}
\end{equation}
where, the effective Debye length is defined as:$L_{D,\mathrm{eff}}=\sqrt{\frac{\varepsilon_{\text{Si}}\varepsilon_0 k_B T}{2 e^2 N_A}}$. In the absence of external bias, the variation of space charge layer potential $(\Delta \Phi_{\text{SCL}})$, E-field $(E_{\text{SCL}})$ and decay length $(\delta_{\text{SCL}})$,  as a function of electrolyte concentration shown in Fig.~\ref{fig:EDL and SCL E-field_potential_Debye_Length}.

\textbf{Expression for Debye length in EDL}
\begin{equation}\label{eq:Debye_Length_expression}
\delta_{\text{EDL}}
=
\sqrt{
\frac{\varepsilon_0 \varepsilon_{\text{r}} k_{\text{B}}T}
{2eF\left(c_0 + 10^{-\mathrm{pH}} + 10^{\mathrm{pH}-14}\right)}
}
\end{equation}
where $\varepsilon_0, \ \varepsilon_{\text{r}},\ e,\ F, \ k_{\text{B}}, T, \ c_0, \text{pH}$ are the free space permittivity, relative permittivity of the medium, electronic charge, Faraday's constant, Boltzmann constant, bulk electrolyte concentration, and pH, respectively. The factor 14 arises in the formula because we have taken pH + pOH = 14.

\begin{figure}[h!]
\centering 
\includegraphics[width=0.5\columnwidth]{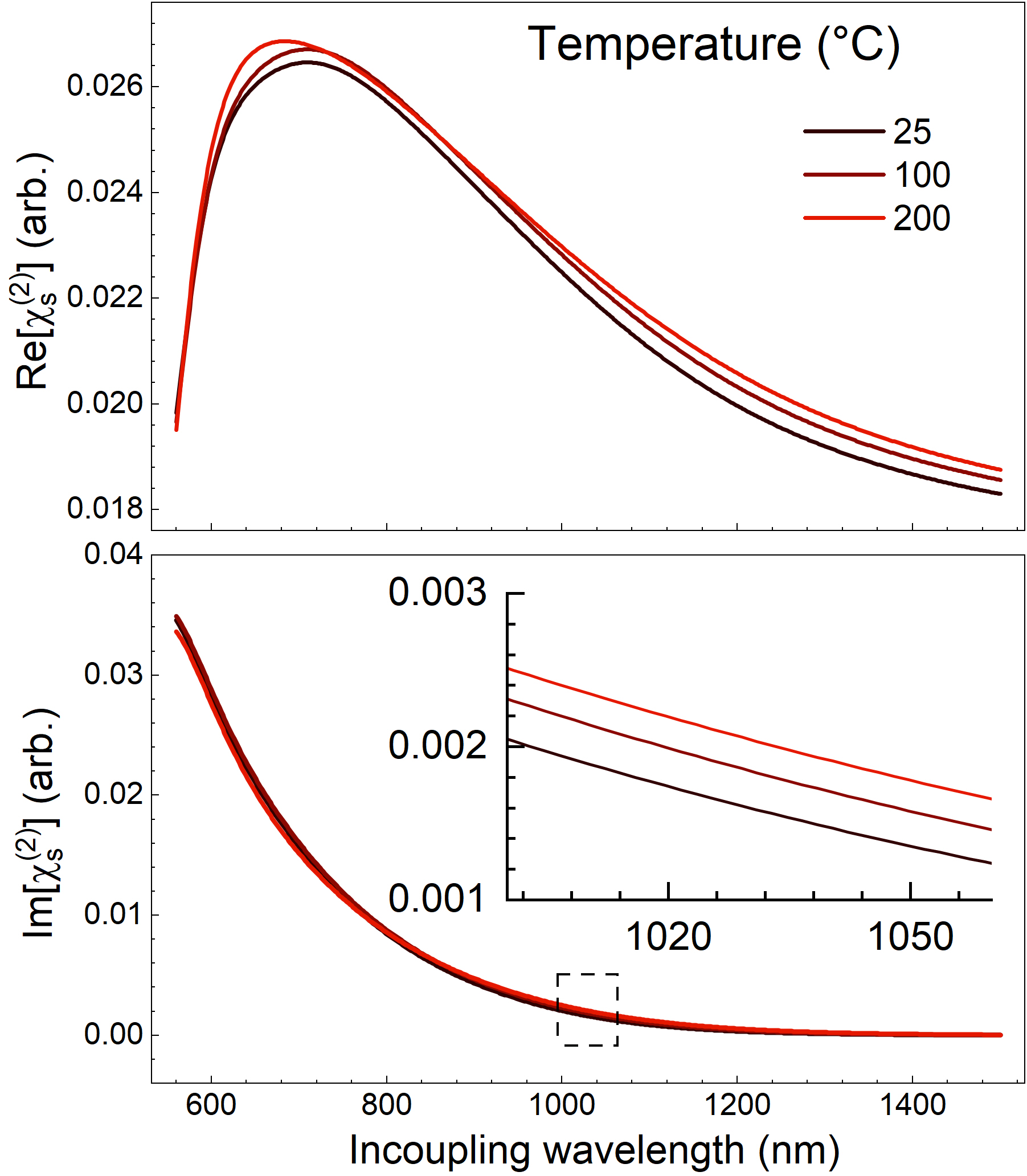} 
\caption{\footnotesize Real and imaginary parts of the surface second-order susceptibility for different temperatures obtained from the linear refractive index shown in Fig. \ref{fig:TO_coeffecients} based on the analytical formulation of surface second harmonic generation \cite{mendoza_exactly_1996}.}
\label{fig:Second_order_polarizability_temperaure} 
\end{figure}

\begin{figure}[h!]
\centering 
\includegraphics[width=1\columnwidth]{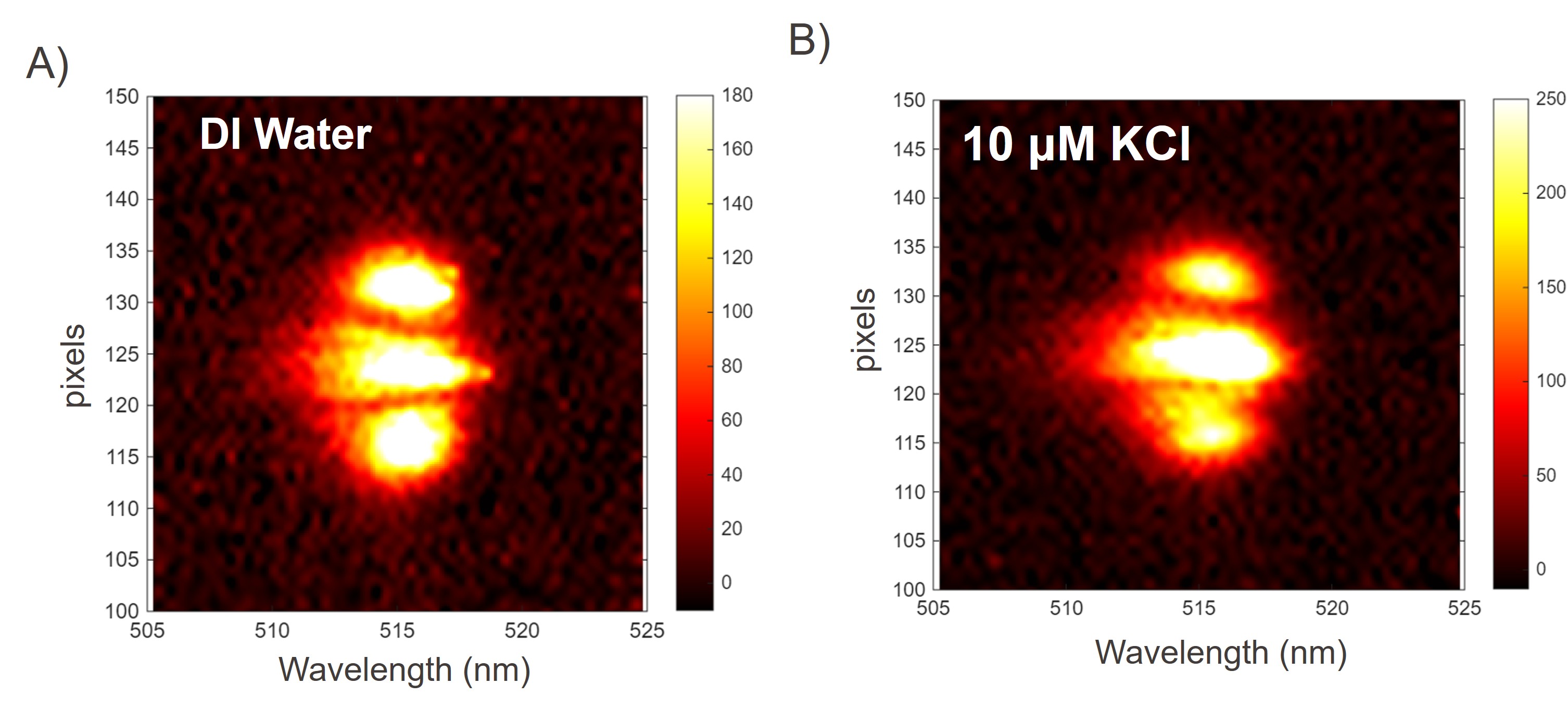} 
\caption{\footnotesize Second harmonic emission image taken from the spectrometer-CCD camera. The x-axis is wavelength, centered at 515 nm, while the y-axis is the real-space distance in pixels. The measurements are part of a continuous measurement series on a disk array sample placed in the cell. The electrolyte was changed in situ, and the SH intensity was measured. The time-trace of the SH intensity is shown in Fig. \ref{fig:main_figure_2}B}
\label{fig:SHG_image_SI} 
\end{figure}


\begin{figure}
    \centering
    \includegraphics[width=0.6\linewidth]{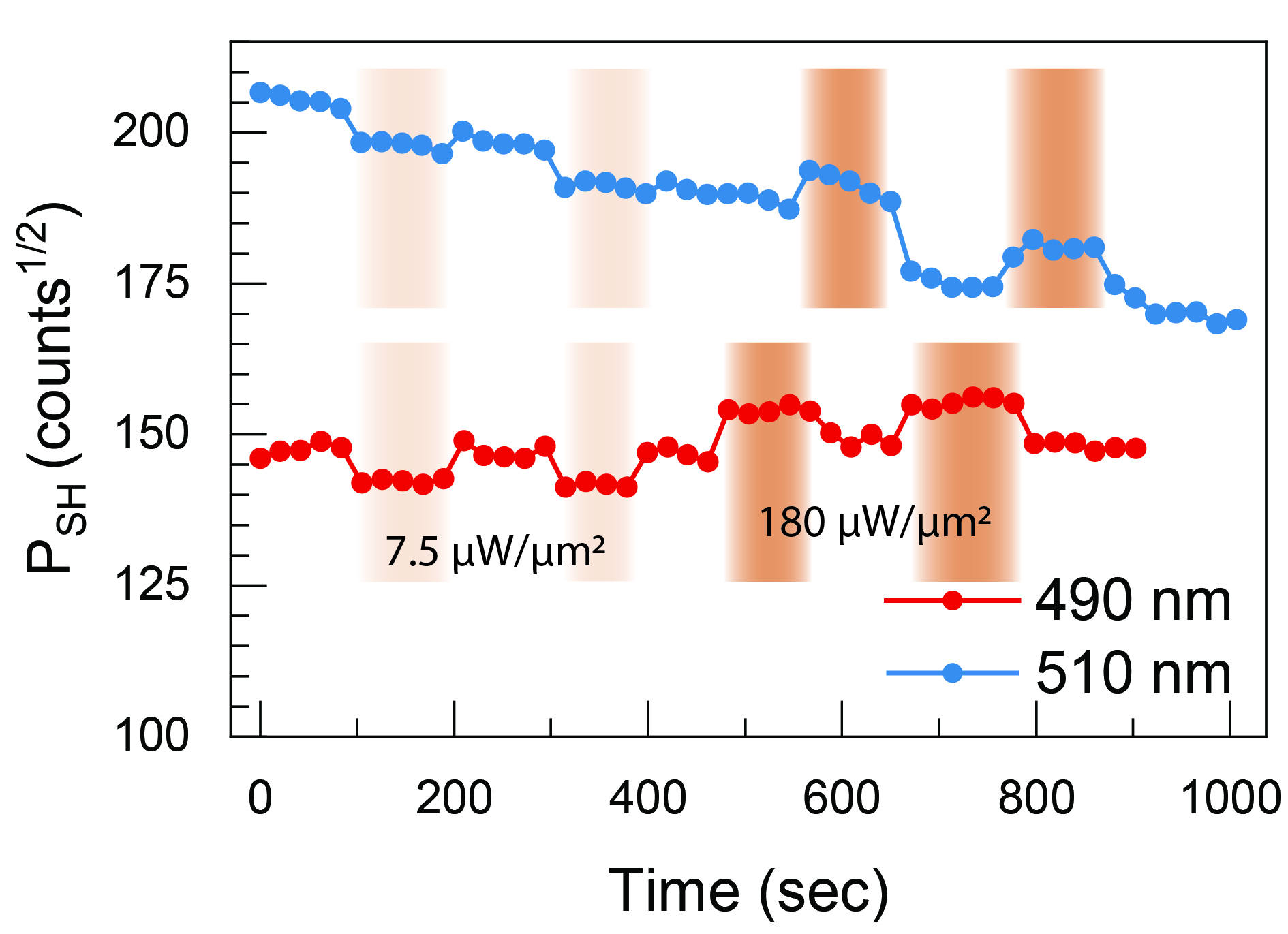}
    \caption{Non-linear polarization (as reported in the manuscript, it is equal to the $\sqrt{I_{\text{SH}}}$) during dynamic light on and off test for two SiNDs with 490 nm and 510 nm diameters under low and high pump (633 nm) irradiation, showing a decrease and an increase in SH amplitude, respectively.}
    \label{fig:Dynamic_SH_510_490nm_light_on_off}
\end{figure}

\begin{figure}[h!]
\centering 
\includegraphics[width=0.5\columnwidth]{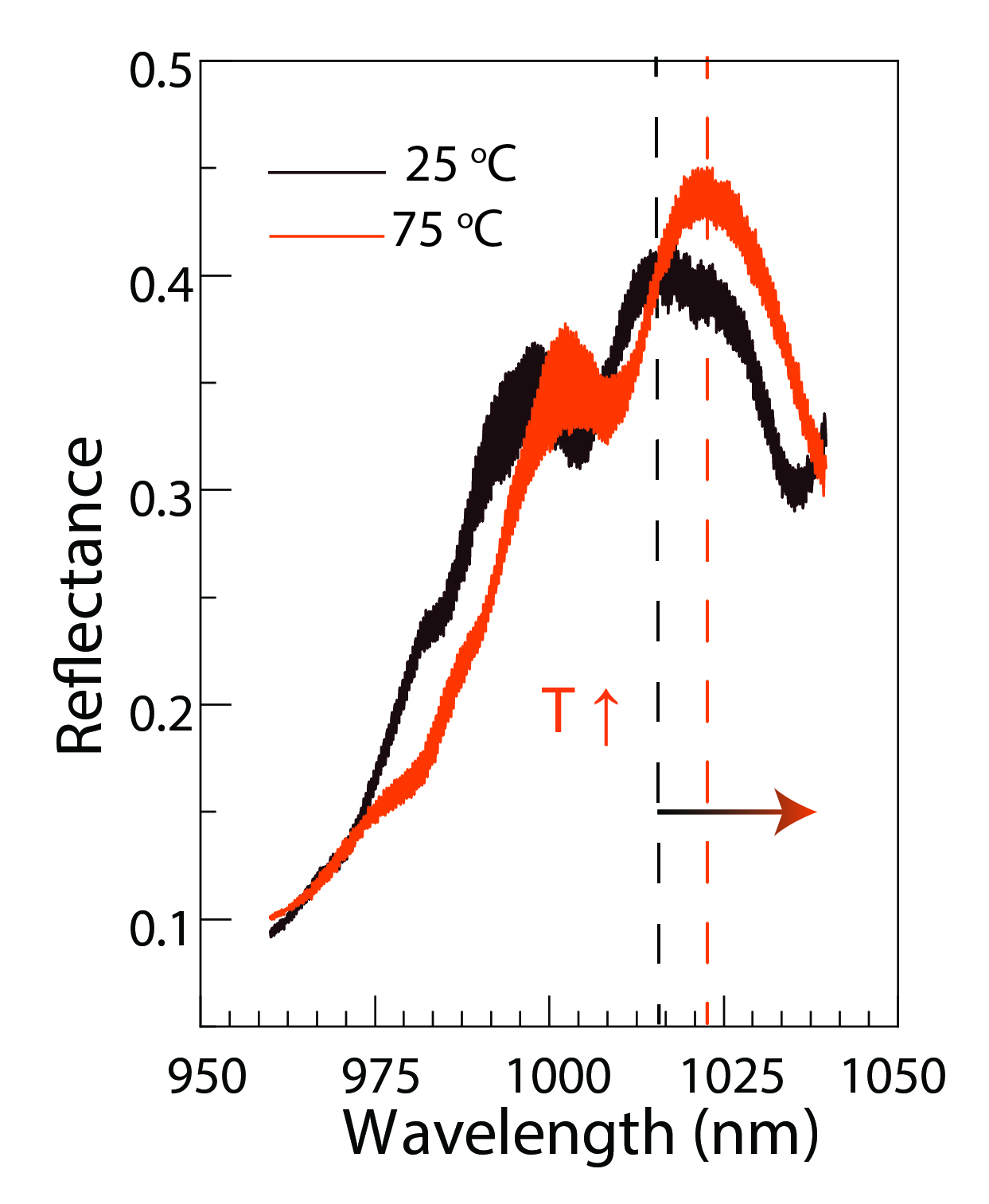} 
\caption{\footnotesize Reflectance spectra for the 510 nm Silicon Nanodisk (SiND) in Air measured at different temperatures according to SI \ref{reflectance_measurement}. The temperature was varied using an external heater.}
\label{fig:ref_spectra_temperature} 
\end{figure}

\begin{figure}
    \centering
    \includegraphics[width=0.5\linewidth]{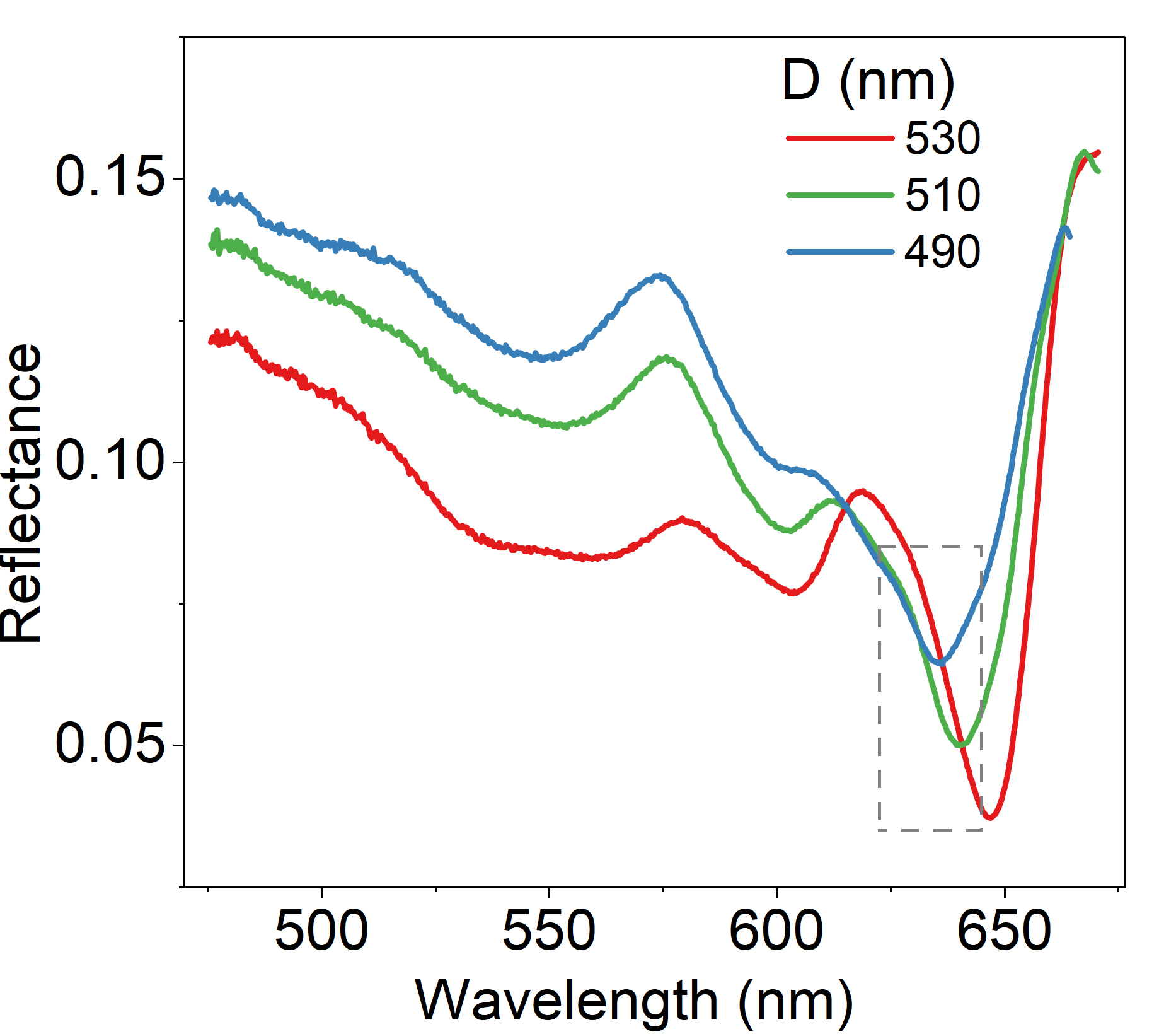}
    \caption{The reflectance spectra reveal an optical mode near the pump wavelength (633 nm), indicating a higher absorption. The reflectance peak decreases with increasing diameter. The measurements were performed according to SI \ref{reflectance_measurement} and Fig. \ref{fig:ch4_linear_ref_Tr}.
}
    \label{fig:pump_wavelength_spectra_reflectance}
\end{figure}

\end{document}